# AI-Powered Surrogate Modelling for Multiscale Combustion: A Critical Review and Opportunities

Amirali Shateri[1], Zhiyin Yang[1], Yuying Yan[2,*], Manosh C. Paul[3,*], Jianfei Xie[1,*]

[1] School of Engineering, University of Derby, Derby DE22 3AW, UK

[2] Faculty of Engineering, University of Nottingham, Nottingham NG7 2RD, UK

[3] Systems, Power & Energy Research Division, James Watt School of Engineering, University of Glasgow, Glasgow G12 8QQ, UK

**Abstract**

Recent advances in combustion science have led to the generation of large volumes of data from high-fidelity simulations, detailed chemical-kinetic calculations and engine-relevant measurements and create new opportunities for data-driven modelling across interacting physical and chemical scales. Among these approaches, artificial intelligence has emerged as a promising framework for constructing surrogate models that reduce computational costs, deliver substantial speed-up and support prediction in complex reacting systems. This review provides a state-of-the-art assessment of AI-powered surrogate modelling for multiscale combustion, spanning chemical kinetics, mechanism reduction, turbulent flames, combustors, engines, and emissions prediction. Supervised, unsupervised, and hybrid or physics-guided learning approaches are examined and compared in terms of predictive accuracy, physical consistency, computational efficiency, and generalizability across conditions and scales. The review further discusses key challenges, including limited transferability across fuels and operating regimes, extrapolation errors, inconsistency in datasets and benchmarks, and the difficulty of building robust and trustworthy models for practical combustion workflows. Future opportunities are identified in the development of more reliable, scalable, and physically grounded surrogate frameworks for next-generation combustion research.

| **Nomenclature** | | | |
|---|---|---|---|
| **Abbreviations** | | | |
| AI | Artificial intelligence | LES | Large-eddy simulation |
| ML | Machine learning | MD | Molecular dynamics |
| ANN | Artificial neural network | DFT | Density functional theory |
| MLP | Multilayer perceptron | PINN | Physics-informed neural network |

*Corresponding authors: yuying.yan@nottingham.ac.uk (Y.Y); Manosh.Paul@glasgow.ac.uk (M.C. Paul); j.xie@derby.ac.uk (J.X)

| | | | |
|---|---|---|---|
| DNN | Deep neural network | DeepONet | Deep operator network |
| CNN | Convolutional neural network | FNO | Fourier neural operator |
| GNN | Graph neural network | NODE | Neural ordinary differential equation |
| RL | Reinforcement learning | GPR | Gaussian process regression |
| CFD | Computational fluid dynamics | RANS | Reynolds-averaged Navier-Stokes |
| DNS | Direct numerical simulation | TCI | Turbulence-chemistry interaction |
| FPV | Flamelet/progress-variable | POD | Proper orthogonal decomposition |
| FGM | Flamelet generated manifold | PCA | Principal component analysis |
| VAE | Variational autoencoder | OH-PLIF | OH planar laser-induced fluorescence |
| CH-PLIF | CH planar laser-induced fluorescence | $CH_2$O-PLIF | Formaldehyde planar laser-induced fluorescence |
| PIV | Particle image velocimetry | FES | Flame emission spectroscopy |
| NNP | Neural network potential | MLIP | Machine-learned interatomic potential |
| ReaxFF | Reactive force field | DPMD | Deep Potential molecular dynamics |
| $\mathbf{NO}_x$ | Nitrogen oxides | $NH_3$ | Ammonia |
| CA50 | Crank angle at 50% heat release | RFR | Random forest regression |
| SVM | Support vector machine | LSTM | Long short-term memory network |
| U-Net | Encoder-decoder convolutional neural network with skip connections | CombML | Combustion machine learning |
| **Symbols** | | | |
| $\boldsymbol{f_\theta(x)}$ | Parameterized model/surrogate mapping | $k(T)$ | Temperature-dependent reaction-rate coefficient |
| $\boldsymbol{x}$ | Input state or feature vector | $T$ | Temperature |
| $\boldsymbol{\theta}$ | Trainable model parameters | $\tilde{T}$ | Filtered or resolved temperature |
| $\boldsymbol{E_a}$ | Activation energy | $\tilde{Y}_{CO}$ | Filtered carbon monoxide mass fraction |
| $\boldsymbol{\Delta H}$ | Reaction enthalpy | $\tilde{Z}$ | Filtered mixture fraction |
| $\boldsymbol{\nu}$ | Characteristic frequency | $\dot{\omega}$ | Reaction or burning rate |
| $\boldsymbol{\phi}$ | Equivalence ratio | $\overline{\dot{\omega}}_{NN}$ | Neural-network-predicted filtered burning rate |
| $\mathbf{R^2}$ | Coefficient of determination | $\overline{\dot{\omega}}_F$ | Filtered tabulated-chemistry burning-rate estimate |
| DSF | Downsampling factor | $\overline{\dot{\omega}}_{FC}$ | Alternative filtered tabulated-chemistry burning-rate estimate |
| SNR | Signal-to-noise ratio | $\overline{\dot{\omega}}^*$ | Reference filtered burning rate |

| **Subscripts** | | | |
|---|---|---|---|
| $\boldsymbol{a}$ | Activation-related quantity in $E_a$ | $F$ | Filtered/tabulated reference quantity |
| $\boldsymbol{CO}$ | Carbon monoxide | $FC$ | Alternative filtered/tabulated combustion quantity |
| $\boldsymbol{g}$ | Global quantity, where used | $NN$ | Neural-network prediction |
| *0* | Initial condition, where used | ref or $*$ | Reference value |

## Contents

## 1. Introduction

Decarbonizing combustion-based energy conversion remains a central engineering and societal challenge because combustion systems are deeply embedded in aviation, marine transport, heavy-duty mobility, industrial heat, and dispatchable power generation. In the near to medium term, material emissions reduction is therefore expected to come from a combination of (i) low-carbon and carbon-free fuels (e.g., hydrogen, ammonia, e-fuels, and oxygenated blends), (ii) device-level efficiency improvements, and (iii) robust emissions control strategies spanning in-cylinder measures, combustor staging, and aftertreatment. Recent syntheses underscore both the promise and complexity of these pathways: hydrogen introduces new safety and operability constraints [1], ammonia offers carbon-free storage advantages but raises ignition and NOx trade-offs [2-3], and oxygenated e-fuels such as oxymethylene ethers can shift soot/NOx behaviour while challenging legacy modelling assumptions [4]. These transitions do not eliminate modelling difficulties; they often amplify them by widening the operating envelope (i.e., pressure, temperature, dilution, and stratification) and introducing additional kinetic pathways for nitrogen chemistry, low-temperature oxidation, and pollutant formation.

At the same time, the modelling problem is becoming more urgent. Regulatory pressure on NOx and particulate matter, which is coupled with climate-driven limits on lifecycle $CO_2$, pushes designers toward multi-objective decisions under uncertainty: stable ignition versus flashback, efficiency versus knock, and low NOx versus low unburned-fuel slip (especially for ammonia). For engines and gas turbines, these objectives are tightly coupled to turbulence–chemistry interaction, heat loss and wall effects, and transient operation. The literature has long emphasized that “single-physics” or purely steady analyses are rarely sufficient at device scale, motivating multiscale and multiphysics simulation strategies spanning detailed kinetics, turbulence modelling, sprays, and wall interactions [5-6]. The implication is straightforward: if the community cannot accelerate the credible predictive simulation and validation pipelines, the design loop will remain too slow to explore the fuel-device-control co-design space that decarbonization demands.

Fig. 1 provides a concrete policy-facing illustration of why emissions-oriented combustion research remains a live problem even in mature economies. Based on the latest UK National Atmospheric Emissions Inventory (NAEI) sectoral statistics (2020–2023), it highlights that $NO_2$, particulate matter (PM10 and PM2.5), and CO remain distributed across multiple source categories, reinforcing why emissions control must be addressed alongside fuel switching rather than treated as a secondary detail.

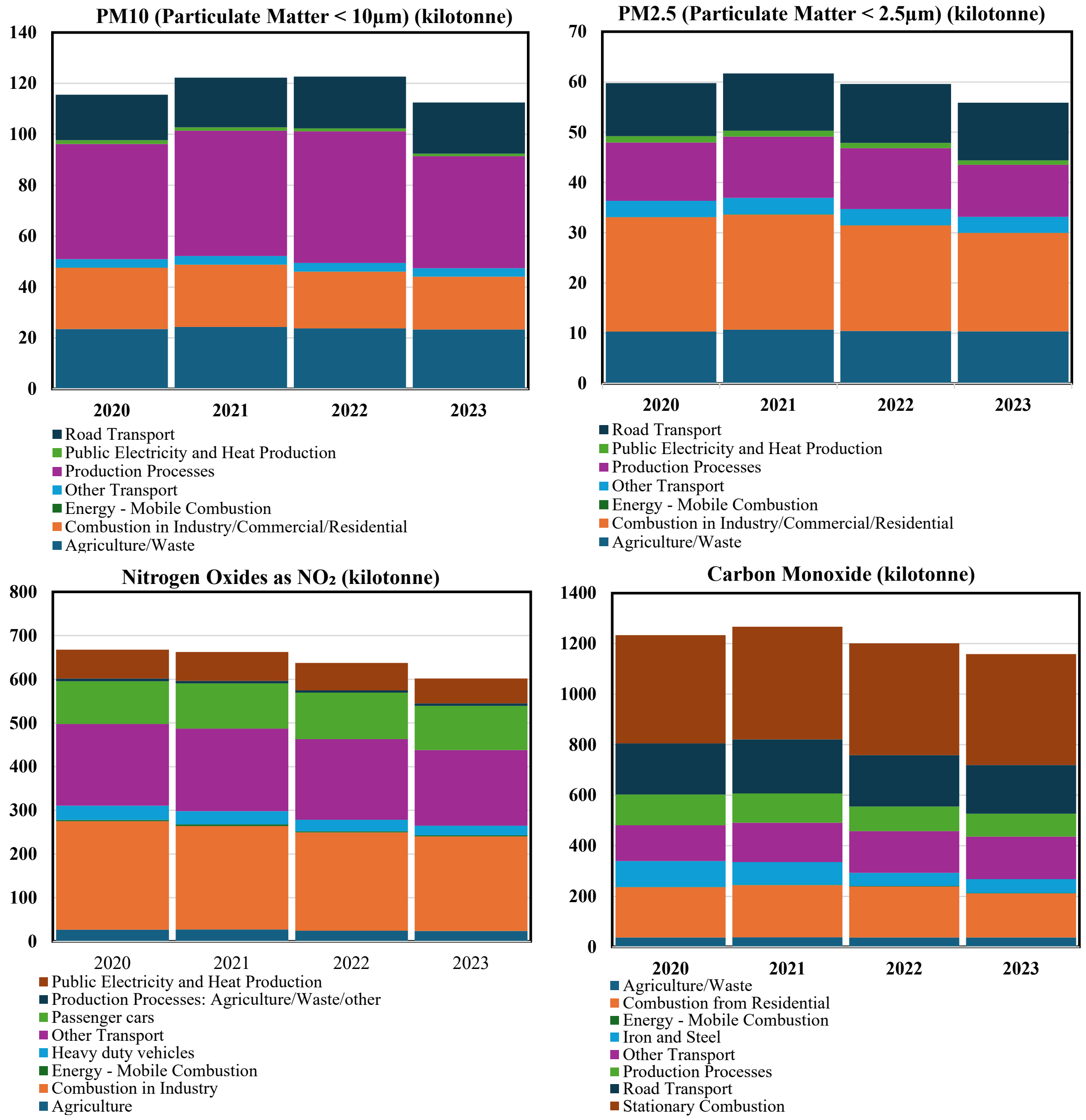

**Fig. 1**. UK emissions of key air pollutants by source sector (2020–2023). Sectoral emissions of nitrogen oxides expressed as $NO_2$, carbon monoxide (CO), PM10, and PM2.5 for the UK over 2020–2023. Data are from the UK National Atmospheric Emissions Inventory (NAEI) air pollutant statistics (latest available update at the time of writing).

The computational bottleneck in practical combustion CFD is not simply grid size; it is the repeated evaluation of tightly coupled, stiff, high-dimensional submodels inside transient solvers [7]. Even with efficient turbulence closures and numeric, detailed chemical kinetics remain computationally demanding, requiring the integration of stiff ordinary differential equations (ODEs) systems across millions of cells and thousands of time steps alongside transport, radiation, sprays, and multiphase effects. The challenge is acute for alternative fuels: ammonia introduces strong slow–fast timescale separation and complex nitrogen pathways, hydrogen shifts the flame speed and stability boundaries, and oxygenated fuels modify the soot-precursor chemistry and oxidation routes. Several recent reviews addressing ammonia chemistry, turbulent combustion modelling for engines, and renewed interest in ignition physics reach the same conclusion: mechanistic fidelity is available, but direct deployment at industrial conditions is frequently infeasible without model reduction or surrogate acceleration [2, 6, 8]. Historically, the field has relied on structured reduction approaches—skeletal mechanisms, tabulation (flamelet/progress-variable, FGM, FPV), presumed PDFs, manifold methods, and targeted closures for turbulence–chemistry interaction. These remain powerful but are increasingly strained by broader fuel flexibility, wider operating envelopes and the demand for rapid multi-objective exploration. This is precisely the space in which scientific machine learning (ML) has gained traction: not as a replacement for physics, but as a practical route to emulate expensive submodels while retaining the end-to-end predictive utility.

Combustion ML is no longer peripheral; it is becoming a workflow paradigm. Ihme et al. [9] framed "Combustion Machine Learning (CombML)" around the maturation of three enablers: abundant data from high-fidelity simulation and experiments, accessible ML tooling, and scalable high-performance computing. That review also highlighted combustion-specific requirements that generic ML does not automatically satisfy—interpretability, conservation consistency, robustness, and benchmark data curation—an agenda that many studies implicitly follow in the last three years [9]. In parallel, combustion ML has expanded from offline property regressions to embedded surrogates that operate inside CFD and engine solvers. This embedded usage is qualitatively different: the surrogate must remain stable under solver feedback, handle out-of-distribution states gracefully, and preserve key invariants (e.g., positivity of species mass fractions and energy consistency) well enough to avoid corrupting the flow solution. The engine community has similarly moved from predictive models toward closed-loop diagnostics, optimization, and control concepts. The review by Aliramezani et al. [5] surveys this shift and emphasizes that reliability and validation often matter as much as raw fit quality.

Among the most visible progress areas is the ML-based acceleration of chemical kinetics. Rather than integrating stiff kinetics directly, a surrogate is trained to map the thermochemical states to reaction rates, source terms, or time-advanced states. Recent open-source frameworks have made this direction practical at scale. For example, the DeepFlame platform introduced a deep learning–empowered pathway for reacting-flow simulation in an open framework, explicitly targeting the "inner-loop" cost of kinetics in CFD [10]. DeepFlame has continued to evolve toward GPU-native, low-Mach reacting-flow workflows [11].

Methodologically, progress has diversified beyond standard feedforward surrogates. Neural-operator approaches have been developed to better handle stiffness and improve generalization across initial conditions and time horizons [12]. Neural ODE formulations offer an alternative philosophy: learn the dynamical system in continuous time in a way that is structurally aligned with the original kinetics [13]. Operator learning and reduced latent dynamics, which are often coupled with autoencoders, aim to reduce the dimensionality burden while preserving the chemistry's temporal structure [14-15]. Multi-fuel generalization is also emerging as a design objective rather than an afterthought, motivated by blended fuels and drop-in replacements. Recent work suggests that surrogates can be trained and validated across distinct fuels with strong a posteriori performance when embedded in reacting-flow simulations [16]. Collectively, these contributions signal a shift from "ML as curve fitting" toward ML as a numerical component within multiscale solvers.

A second major front is learning for turbulence–chemistry interaction and tabulated combustion. Neural replacements for flamelet/FGM databases are attractive because they preserve the legacy conceptual framework (progress variables and manifolds) while reducing the memory and interpolation overhead and enabling smooth differentiability. In another study, a multitask learning strategy replaced flamelet-database lookups with deep neural networks (DNN) and was evaluated in a configuration representative of turbulent flame benchmarking [17]. Meanwhile, data-driven closure concepts have been explored using a posteriori validation framework that emphasize in-solver realism rather than isolated regression success [18].

Beyond tabulation, ML is increasingly used to reconstruct or infer high-fidelity fields from sparse observations, which matters for both experiments and rapid design iteration. Physics-informed neural networks (PINNs) have been applied to recover multiphysics fields in reacting flows without pretraining, leveraging the governing constraints to reduce data demands [19]. In propulsion contexts where experimental access is constrained, deep learning (DL) has also been used for supersonic combustor/scramjet flow-field reconstruction and combustion-model augmentation [20-21]. In LES contexts, convolutional architectures have been explored for learning spatial structure in reaction-rate closures [22], while generative and reduced-order approaches aim to emulate the high-dimensional evolution more efficiently [15, 23].

The alternative-fuel transition is not an “application area” for ML surrogates but is arguably the forcing function that exposes where the current modelling approaches break. Hydrogen’s high reactivity shifts stability limits and can amplify thermoacoustic concerns; ammonia’s low flame speed and NOx sensitivity demand careful staging, blending (often with hydrogen), or plasma assistance; and oxygenated e-fuels modify soot pathways in ways that can challenge reduced mechanisms calibrated to conventional hydrocarbons. Reviews on oxymethylene ethers [4] and ammonia pyrolysis/oxidation chemistry [2], together with recent ammonia combustor synthesis work [24], illustrate that “clean combustion” is not a single-axis problem but is a coupled, kinetics–transport–stability design problem. Accordingly, some of the most compelling ML contributions in alternative fuels are those that connect the surrogate acceleration to design-relevant outcomes: knock limits in engines [25], injection-strategy optimization in ammonia–diesel dual-fuel operation [3, 24, 26], stability and emissions in swirl burners and gas turbines [28], and fast screening of plasma-assisted ammonia ignition strategies [29]. Surrogates for laminar burning velocity and ignition delay remain highly practical because they serve as “inputs” to higher-level models and design rules [30-31]. When such property surrogates are paired with integrated frameworks for control or optimization, they become tools for rapid iteration rather than standalone predictors [32-33].

A persistent gap across the AI combustion literature is not the absence of promising results but inconsistent validation reporting. Across studies, AI/ML models have been validated against the shock-tube ignition delays, burner-stabilized flame species/temperature fields, canonical turbulent jet flames, and engine/combustor tests with optical and emissions diagnostics [31, 34-35]. Thus, the community still lacks standardized protocols for reporting accuracy, extrapolation robustness, and uncertainty. High accuracies are achievable in existing studies, but only a minority report comprehensive quantitative metrics across multiple facilities and operating regimes. This limitation is not editorial but slows progress by making it harder to compare methods, transfer models across configurations, and build shared benchmark datasets, which has been concerned in the CombML agenda [9].

Against this backdrop, this review examines the recent evolution of AI-powered surrogate modelling in multiscale combustion, with particular attention to solver-integrated learning, cross-scale data coupling, and emissions-relevant deployment. The review is organised as follows. Section 2 introduces the foundations of AI for combustion modelling, including learning paradigms, neural architectures, neural operators, physics-constrained learning, and data/training strategies. Section 3 surveys AI applications across combustion scales, spanning the atomistic chemistry, experimental diagnostics, continuum-scale reacting-flow simulation, and an application-level posteriori assessment. Section 4 then provides a quantitative synthesis of surrogate-model performance by comparing reported evidence across accuracy-speed trade-offs, deployment efficiency, robustness, uncertainty, Pareto-optimality, temporal evolution, and multi-metric model-selection criteria. Finally, the cross-cutting limitations, methodological gaps, standardisation needs, and the emerging role of agentic AI in enabling trustworthy, scalable virtual-lab workflows for cleaner combustion are discussed in section 5. By combining a structured narrative review with quantitative cross-study synthesis, this state-of-the-art review aims to distinguish mature advances from still-fragile claims, clarify where AI surrogates already deliver practical value, and identify the benchmark, validation, and reporting practices needed for reliable deployment in multiscale combustion modelling. Fig. 2 summarizes the workflow perspective that organizes this review, linking multiscale combustion physics and high-fidelity data generation to solver-integrated surrogate modelling and emissions-oriented deployment.

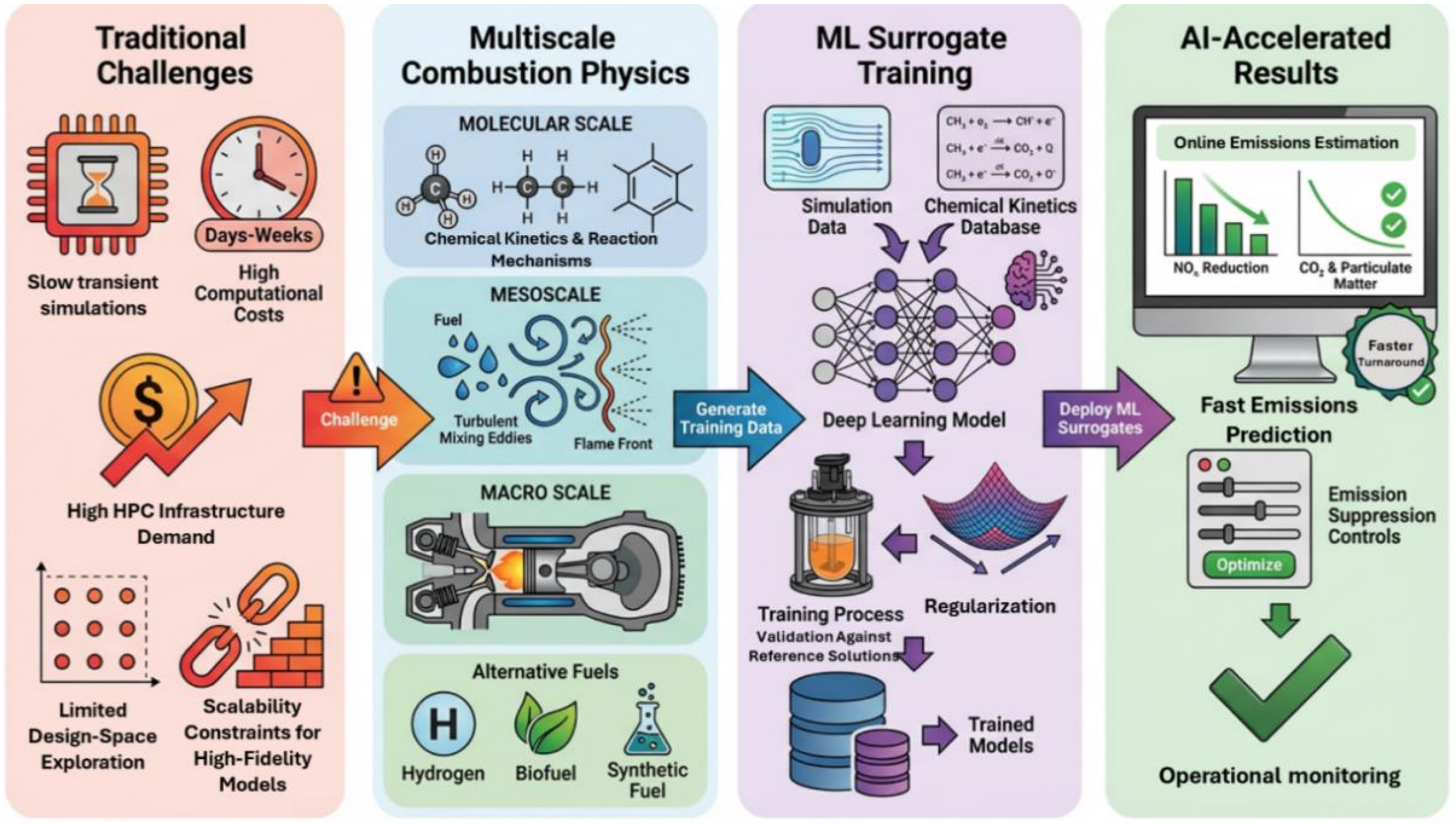

**Fig. 2**. Conceptual overview of the workflow examined in this review. Traditional computational bottlenecks in high-fidelity combustion modelling motivate the multiscale data generation spanning detailed chemical kinetics, turbulence–chemistry interaction, and device-scale reacting flows. High-fidelity simulation and experimental datasets support the training, regularization, and validation of neural surrogate and closure models. When embedded within MD, CFD or engine solvers, these models enable accelerated design-space exploration, fast in-solver emissions estimation, and operational diagnostics under fuel-flexible and emissions-constrained conditions.

## 2. Foundations of AI for Combustion Modelling

Recent advances in AI and ML have introduced new paradigms for modelling complex reacting-flow systems across multiple spatial and temporal scales. Combustion processes are inherently multiscale, spanning atomistic reaction dynamics, mesoscale chemical kinetics, turbulent reacting flows, and device-level performance. The increasing availability of large datasets from high-fidelity simulations (e.g., direct numerical simulation (DNS), large-eddy simulation (LES), molecular dynamics (MD), and density functional theory (DFT)) and experimental diagnostics has created opportunities for data-driven approaches to augment or replace traditional modelling strategies [9, 36].

Within this context, machine learning models can be interpreted as approximators of mappings between physical quantities. For instance, a surrogate model may approximate mappings from thermochemical state variables to reaction source terms, from spatial fields to turbulence closures, or from boundary conditions to full flow solutions. Such mappings may be represented using a parameterized function $f_\theta(x)$, where the model parameters $\theta$ are learned from data. Deep learning architectures provide flexible approximators capable of capturing nonlinear relationships in high-dimensional datasets, making them well suited to the complexity of combustion systems [37].

Despite their flexibility, purely data-driven models often struggle to maintain physical consistency when applied outside their training domain. Consequently, the integration of physical knowledge through constraints, hybrid modelling, or governing equations has emerged as a central theme in scientific machine learning for combustion [38]. This section briefly introduces the principal learning paradigms, model architectures, and physics-constrained approaches that underpin AI-enabled combustion modelling.

### *2.1 Machine Learning Paradigms in Combustion Research*

Machine learning methods used in combustion modelling can generally be grouped into several learning paradigms, each corresponding to a different class of modelling tasks.

*i. Supervised Learning*

Supervised learning is the most widely used paradigm for surrogate modelling in combustion. In this setting, models are trained using labelled datasets consisting of input–output pairs. Typical tasks include predicting reaction source terms, estimating ignition delays, reconstructing flow fields, or approximating turbulence–chemistry closures. Many of these problems can be formulated as regression tasks in which a neural network learns a mapping between thermochemical state variables and target quantities of interest.

Supervised surrogates have demonstrated significant computational advantages when integrated into combustion simulations, particularly for accelerating chemical kinetics calculations. However, their predictive performance is often sensitive to the distribution of the training data. Models trained on limited operating regimes may experience degradation under extrapolation to new conditions such as different fuels, pressures, or equivalence ratios [9]. Hybrid approaches combining neural networks with mechanistic solvers have therefore been proposed to mitigate extrapolation errors. For example, on-the-fly neural-network chemistry models can be coupled with direct integration methods to maintain accuracy when the surrogate encounters previously unseen states [39].

*ii. Unsupervised Learning*

Unsupervised learning focuses on discovering structure in unlabelled datasets. In combustion research, this approach is often used for dimensionality reduction, regime identification, or feature extraction from high-dimensional datasets. Techniques such as principal component analysis (PCA), autoencoders, and manifold learning methods can identify low-dimensional representations of complex thermochemical state spaces.

Deep generative models, including variational autoencoders (VAEs), provide probabilistic frameworks for learning latent representations of high-dimensional data [40]. Such latent spaces can facilitate model compression, reduced-order modelling, and uncertainty quantification in combustion simulations. However, the interpretability of learned latent variables is not always guaranteed, and their physical relevance depends on the alignment between learned representations and the underlying physics.

*iii. Reinforcement Learning*

Reinforcement learning (RL) addresses decision-making problems in which an agent learns control policies through interaction with an environment. In combustion systems, RL has potential applications in active control of combustion devices, fuel injection strategies, and adaptive optimization of engine operating conditions.

Deep reinforcement learning combines neural networks with policy optimization algorithms to enable learning from high-dimensional observations [41]. Although RL offers promising opportunities for combustion control, practical deployment remains challenging due to the high cost of data generation, safety constraints, and the difficulty of ensuring stable operation during exploration.

*iv. Physics-Informed Learning*

Physics-informed machine learning integrates governing equations, physical constraints, and prior knowledge into the training process. Rather than relying solely on data, these approaches incorporate mathematical models to guide the learning process and improve generalization. Physics-informed neural networks (PINNs) represent one prominent example. In PINNs, neural networks are trained to satisfy governing partial differential equations (PDEs) by including residual terms of the equations in the loss function [42]. This approach allows models to incorporate boundary conditions and conservation laws while simultaneously fitting available data.

Physics-informed learning is particularly attractive for combustion modelling because the governing equations of reacting flows are well established. Embedding these equations into the learning process can reduce the required amount of training data and improve the physical consistency of surrogate predictions [38].

*2.2 Neural Network Architectures for Combustion Modelling*

Several neural network architectures have been adopted for combustion applications, each suited to different types of data and modelling tasks.

*i. Feedforward Neural Networks*

Fully connected feedforward neural networks (FCNN), often referred to as multilayer perceptrons (MLPs), remain the most common architecture for surrogate chemical kinetics and algebraic closure models. These networks approximate nonlinear mappings between the input variables and target outputs through a sequence of nonlinear transformations. Their simplicity and computational efficiency make them attractive for embedding within combustion solvers. However, feedforward networks strongly rely on the choice of input representation and may struggle to generalize beyond the training domain without additional constraints or hybridization strategies.

*ii. Convolutional Neural Networks*

Convolutional neural networks (CNNs) are designed to process spatially structured data such as images or grid-based fields. By exploiting the local connectivity and shared filters, CNNs efficiently capture the spatial correlations in high-dimensional datasets. In combustion research, CNNs have been applied to flame imaging, flow-field reconstruction, and turbulence modelling. Their ability to extract hierarchical spatial features makes them particularly useful for interpreting experimental diagnostics and high-resolution simulation outputs [37].

*iii. Graph Neural Networks*

Graph neural networks (GNNs) extend the neural-network learning to graph-structured data. In chemistry and materials science, molecular structures and reaction networks are naturally represented as graphs composed of nodes (atoms) and edges (chemical bonds). GNNs operate on these graphs by iteratively propagating information between neighbouring nodes, allowing them to learn representations of molecular interactions. This framework is particularly relevant for atomistic combustion modelling, where GNN-based models can be used to learn interatomic potentials, reaction pathways, or coarse-grained representations of molecular systems [43].

*iv. Transformer Models*

Transformer architectures employ the self-attention mechanisms to capture long-range dependencies within sequences or sets of inputs. Originally developed for natural language processing (NLP), transformers have since been applied to a variety of scientific modelling tasks [44-45]. In combustion research, attention-based architectures may offer advantages for modelling complex chemical reaction networks or long-range temporal dependencies in reacting flows. However, their application remains an emerging area of investigation.

### *2.3 Neural Operators and Operator Learning*

Many scientific modelling problems are naturally formulated as operators that map functions to functions. For example, solving a set of PDEs can be viewed as learning a mapping from boundary and initial conditions to solution fields. Neural operators aim to learn such mappings directly from data. DeepONet is a neural operator architecture that represents operators using two networks: a branch network that encodes the input function and a trunk network that represents spatial coordinates [46]. Another prominent approach is the Fourier Neural Operator (FNO), which learns the integral kernel operators in Fourier space and has demonstrated strong performance in modelling solution operators for PDEs such as Burgers and Navier–Stokes equations [47]. Neural operators are particularly attractive for surrogate modelling of reacting flows because they can learn solution operators for entire families of PDEs rather than individual simulation instances. This capability opens possibilities for mesh-independent surrogate solvers and digital-twin frameworks for combustion systems.

### *2.4 Physics Constraints and Hybrid Modelling*

A key challenge in data-driven modelling of physical systems is to maintain consistency with known physical laws. In combustion modelling, violations of conservation principles such as elemental mass conservation or energy balance can lead to numerical instability or unphysical predictions. Two primary strategies are commonly employed to incorporate physical constraints into neural-network models. The first involves embedding constraints directly within the model architecture so that conservation laws are satisfied by design. Such approaches can exactly enforce the analytic constraints, ensuring physically consistent predictions [48]. The second strategy involves adding penalty terms to the training objective that penalize violations of physical constraints. Although this approach does not guarantee strict conservation, it provides flexibility when dealing with noisy data or approximate models. Extensions of physics-informed learning have also been developed for conservation laws. Conservative physics-informed neural networks enforce the flux continuity and conservation across domain interfaces, improving the ability of PINNs to represent physical systems governed by conservation equations [49].

### *2.5 Data Sources and Training Strategies*

AI-driven combustion modelling relies on data generated across multiple scales and modalities. High-fidelity simulations such as DNS, LES, and molecular dynamics provide detailed datasets for training surrogate models. Experimental measurements from laser diagnostics, imaging systems, and sensor networks also play a critical role in model calibration and validation [9].

Since generating labelled data can be computationally expensive, strategies such as transfer learning and active learning are often employed to improve the training efficiency. Transfer learning leverages knowledge from related tasks or domains to reduce the amount of data required for new problems [50]. Active learning frameworks iteratively identify regions of the input space, where the additional data would be most informative, allowing models to be trained more efficiently. On-the-fly learning approaches integrate the model training directly within simulation workflows. For example, neural-network surrogates can be trained dynamically during DNS simulations, allowing the surrogate to adapt to the evolving state distribution encountered during the simulation [39].Fig. 3 summarizes how major machine-learning paradigms interact with the multiscale hierarchy of combustion modelling while highlighting the central roles of physics integration, data sources, and adaptive training strategies.

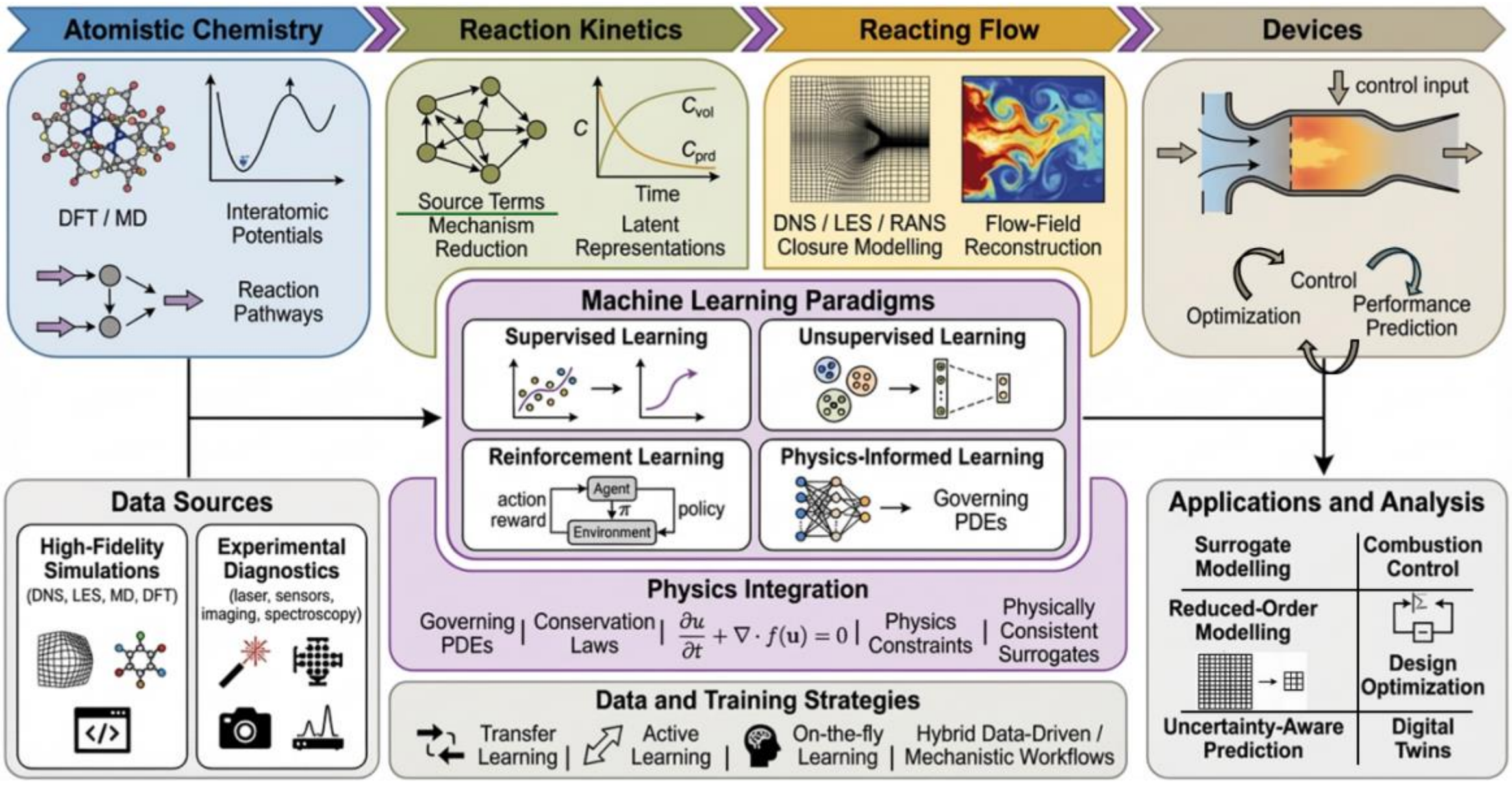

**Fig. 3**. Schematic representation of AI-enabled combustion modelling across scales, from atomistic chemistry to devices, showing major learning paradigms, physics integration, data sources, training strategies, and application pathways.

## 3. AI Across Scales in Combustion Modelling

### *3.1 Molecular/atomistic AI for combustion chemistry and scale bridging*

Multiscale combustion is often discussed in terms of CFD and turbulence–chemistry interaction, but the credibility of any reacting-flow model ultimately rests on chemistry: which pathways are active, how fast they proceed, and how sensitive they are to temperature, pressure, mixture composition, and dilution. At the smallest scales, these questions are governed by the potential-energy surface and the resulting reactive dynamics [51-52]. The past few years have therefore seen rapid growth in AI at the molecular/atomistic level, not as a replacement for combustion modelling, but as a way to (i) approach quantum accuracy at MD cost, and (ii) extract chemically meaningful information (i.e., pathways, intermediates, and rate parameters) that can be carried upward into reduced mechanisms or closures [51].

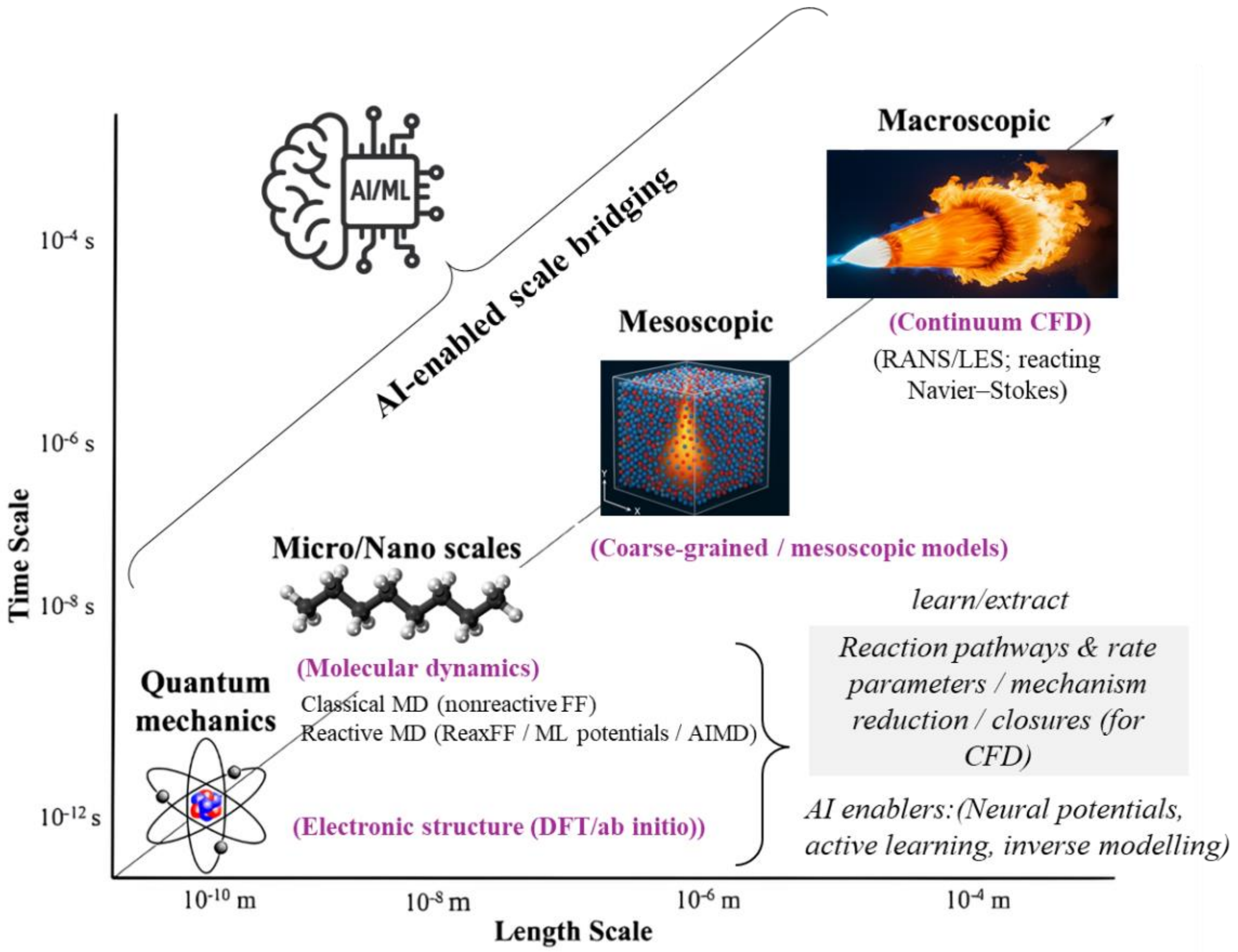

**Fig. 4**. AI-enabled scale-bridging from the electronic structure and molecular dynamics to combustion chemistry closures for reacting-flow simulations. AI is used to learn reactive potentials and extract pathways/rate information that inform reduced mechanisms, tabulated chemistry, or surrogate closures used at larger scales. Adopted from [51, 53].

As schematized in Fig. 4, the molecular/atomistic AI literature that supports the combustion chemistry and scale bridging can be organized into three connected roles: (i) machine-learned interatomic potentials that enable the reactive molecular dynamics with near–DFT accuracy; (ii) reaction discovery and network decoding from reactive trajectories; and (iii) data-driven learning of kinetic ingredients that interface with the mechanism development, reduction, and closure construction.

### *3.1.1 ML interatomic potentials enabling reactive combustion MD*

Reactive molecular dynamics can reveal competing pathways and intermediate populations without prescribing a mechanism a priori, but classical reactive force fields and *ab initio* molecular dynamics represent opposing ends of the cost–accuracy spectrum. Machine-learned interatomic potentials (MLIPs), including neural network potentials (NNPs), aim to narrow this gap by learning high-dimensional potential-energy surfaces from quantum mechanics (QM) reference data. Fig. 5(a) summarizes a typical active-learning workflow for constructing combustion-relevant MLIP training sets, combining broad reactive sampling with iterative QM labelling and model-deviation–guided resampling. Fig. 5(b) illustrates the corresponding NNP model form, in which the local atomic environments are mapped to descriptors, processed by atom-centred networks, and summed to recover the total potential energy used for force evaluation in MD.

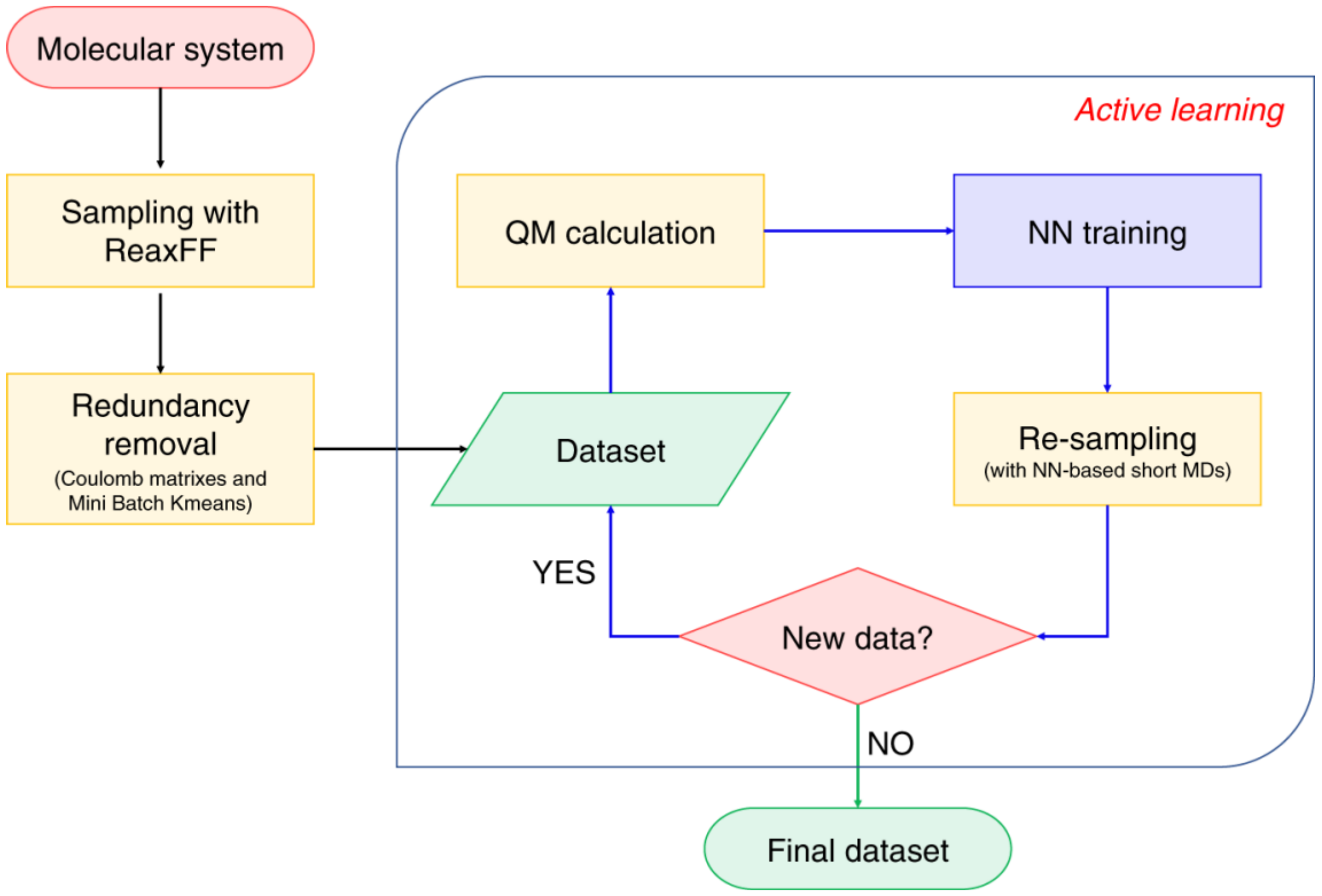

(a)

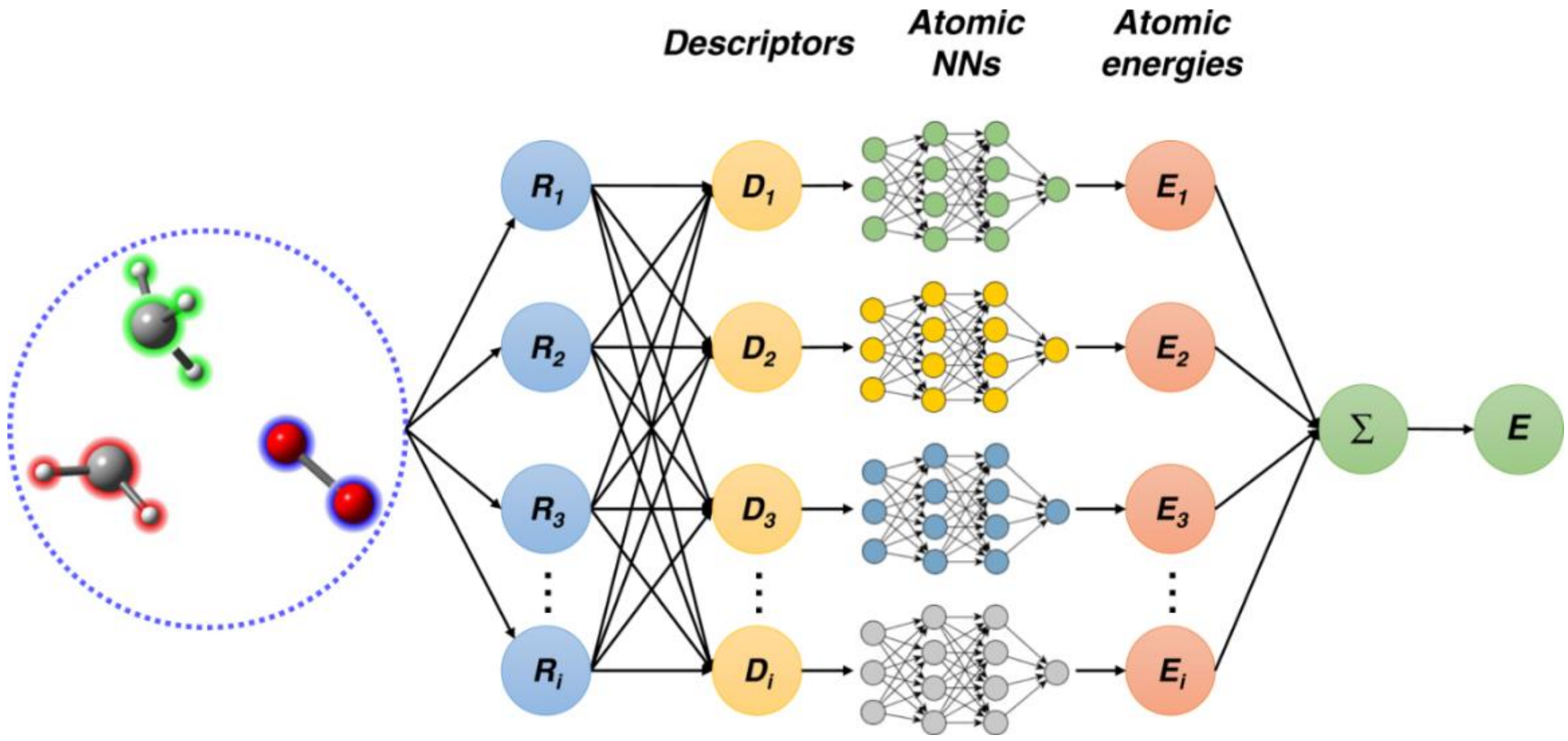

(b)

**Fig. 5**. (a) Active-learning workflow for constructing reactive MLIP/NNP training datasets and (b) atom-centred neural network potential architecture for energy/force prediction in reactive molecular dynamics. Reprinted by permission from [54].

A representative demonstration is neural-network-potential molecular dynamics applied to combustion chemistry, showing that reactive trajectories can recover chemically rich event sequences and reaction pathways that are difficult to enumerate in advance [54]. Extending beyond single-molecule demonstrations, Deep Potential–based workflows have been used to explore the pyrolysis chemistry across linear alkane families, which is important for transferring atomistic insight across fuel classes rather than isolated prototypes [55]. Foundational developments in scalable ML potentials (e.g., Deep Potential Molecular Dynamics) have enabled much of this progress by delivering quantum-accurate forces at scales compatible with chemically meaningful sampling [56]. For nitrogen-containing fuels and blends, molecular-scale AI is particularly valuable because the pollutant formation can sensitively depend on minor channels and the local radical chemistry. Neural-network-potential reactive simulations have been used to interrogate the pollutant formation pathways in ammonia and ammonia–hydrogen combustion, resolving how the equivalence ratio and hydrogen addition influence the $NO/NO_2/N_2O$-relevant chemistry [57-58].

*3.1.2 Reaction-network discovery from ML-driven trajectories*

A second trend has been shifted from "running reactive MD" toward "learning from reactive MD" in a way that produces transferable chemistry knowledge. Instead of treating trajectories as qualitative movies or images, recent studies explicitly decode the reaction networks by identifying elementary events, building graph-based networks of intermediates, and quantifying pathway prevalence under different thermodynamic conditions. A recent example is the reaction-network decoding for nitromethane combustion using neural-network-potential MD, where the extensive sampling is used to map the decomposition networks and reveal how density/conditions alter dominant pathways [59]. Beyond fully learned potentials, hybrid strategies are also emerging in which ML is used to diagnose or mitigate reactive force-field limitations (e.g., ReaxFF transferability and extrapolation), improving reliability of atomistic predictions under combustion-relevant conditions [60-61].

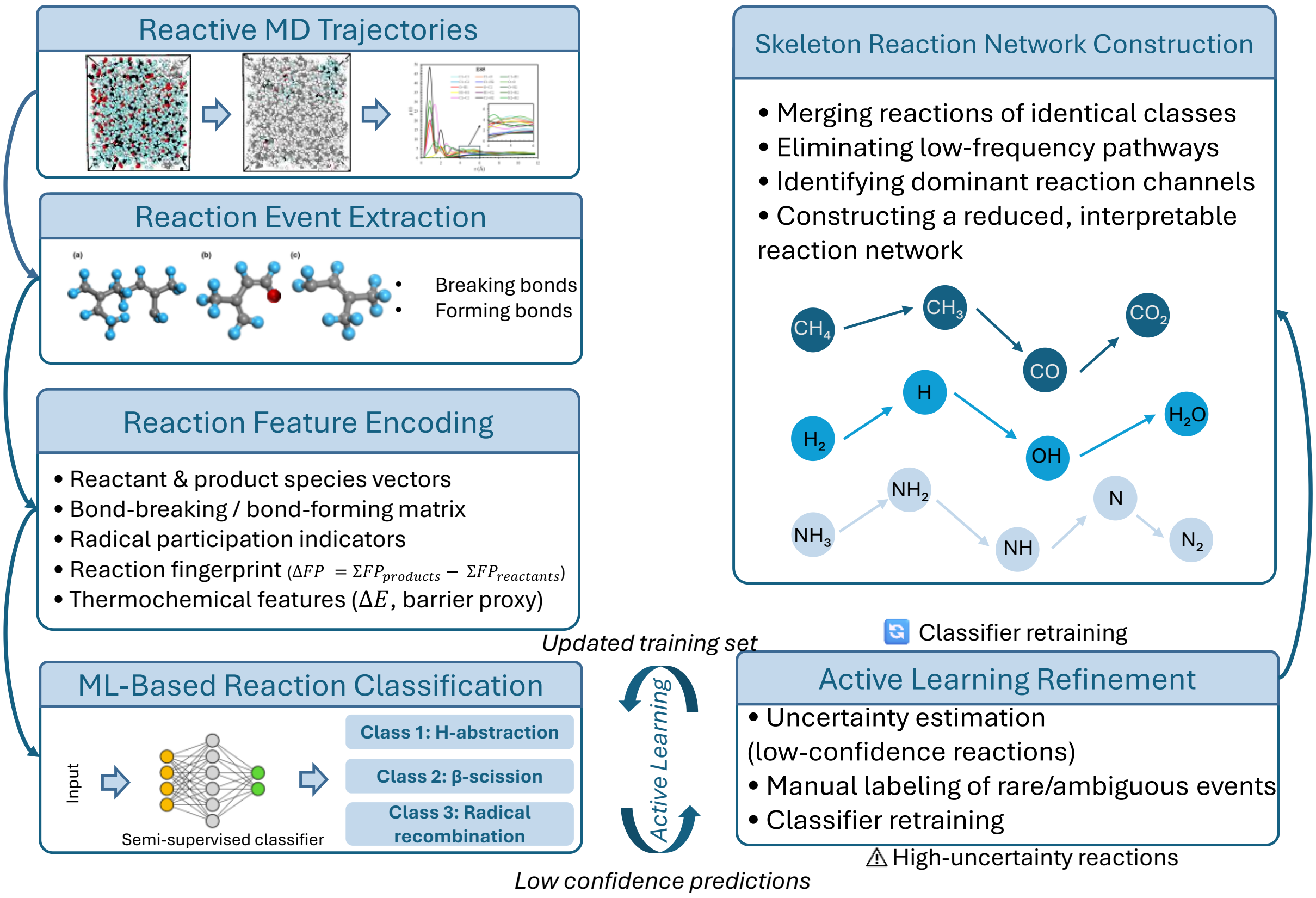

**Fig. 6**. Schematic workflow for reaction-network discovery from ML-driven reactive molecular dynamics. Reactive MD trajectories are processed to extract elementary reaction events, encode reaction features, and classify reactions using machine learning. Active learning iteratively improves the classification reliability, enabling construction of a reduced skeleton reaction network representing dominant hydrocarbon, hydrogen, and nitrogen pathways [45, 54, 60-62].

The general workflow underlying such reaction-network discovery is schematically summarized in Fig. 6. Reactive MD trajectories are first analysed to extract the discrete reaction events, which are then encoded into chemically meaningful descriptors to capture bond rearrangements, species identities, and radical formation. Machine-learning classifiers group elementary events into reaction classes, while active learning strategies iteratively refine the model by targeting uncertain or rare reactions. The resulting classified reactions are merged and reduced to construct a compact skeleton reaction network that highlights the dominant hydrocarbon, hydrogen, and nitrogen pathways relevant to combustion chemistry.

### *3.1.3 From atomistic insight to kinetic parameter learning and closures*

The third role is the most directly "scale-bridging": using ML to learn kinetic ingredients that can be inserted into combustion models. This includes learning rate constants and kinetic parameters to accelerate the mechanism development and calibration. To clarify this atomistic-to-kinetic information flow, Fig. 7 schematically illustrates the scale-bridging pathway.

Reactive molecular dynamics simulations provide access to both static energetic information (e.g., potential-energy surfaces and activation barriers) and dynamic trajectory-level chemistry, from which kinetic ingredients such as activation energies ($E_a a$), reaction enthalpies ($\Delta H$), and characteristic frequencies ($v$) can be extracted. These atomistic observables are subsequently processed through the machine learning-based parameter inference, which ranges from the regression-based Arrhenius fitting to Bayesian calibration and neural-network surrogate modelling, to obtain the temperature-dependent rate expressions $k(T)$. The learned kinetic parameters are then embedded into reduced mechanisms and source-term formulations used at larger scales. For example, ML has been used to predict the rate constants for reactions appearing in combustion kinetic models, supporting faster parameterization and screening of mechanisms [63]. When coupled with pathway evidence and intermediate populations inferred from reactive simulations, such approaches provide a principled route from atomistic evidence to model-form decisions (retained pathways, parameter priors, or uncertainty-aware reduction targets). Recent quasi-classical trajectory studies using *ab initio*–trained ML force fields also show that the kinetic accuracy can be strongly limited by training-set coverage, particularly when multiple barriers and intermediate wells must be represented consistently [64].

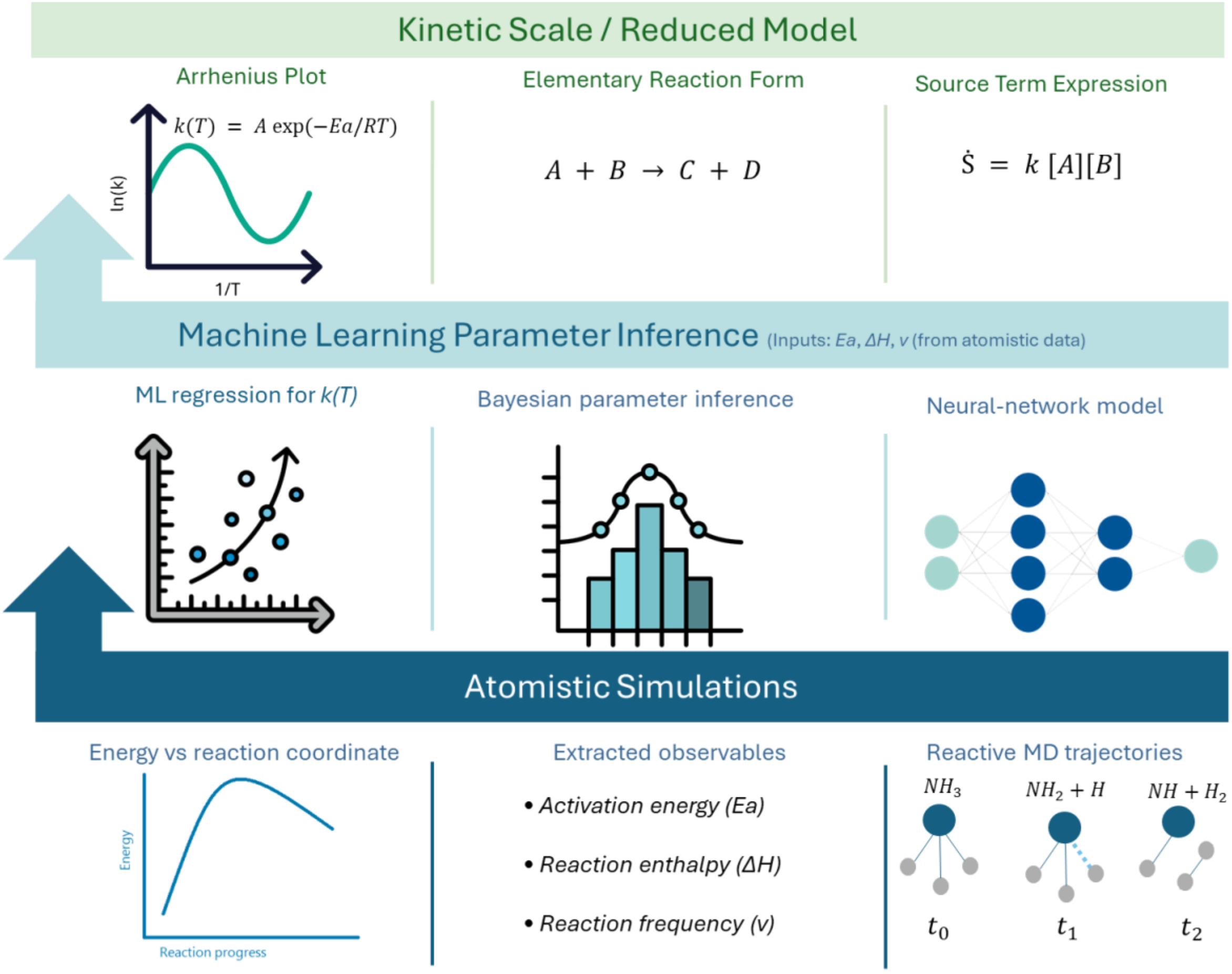

**Fig. 7**. Conceptual illustration of atomistic-to-kinetic scale bridging. Reactive molecular dynamics simulations provide energetic and trajectory-level observables (e.g., $E_a$, $\Delta H$, $v$), which are processed through machine-learning-based parameter inference to construct temperature-dependent rate expressions and reduced kinetic closures.

In parallel, molecular simulation and molecular-scale learning are increasingly used to provide the thermophysical-property support for fuel blends relevant to practical combustion. For example, MD combined with ML regression has been applied to predict thermophysical properties for ethanol–octane blends, producing property trends that matter for spray/mixing boundary conditions and surrogate-fuel formulation in higher-scale models [65]. Table 1 summarizes the representative molecular/atomistic AI studies and highlights their specific scale-bridging contributions, from enabling reactive sampling at near-QM fidelity to extracting pathway- and rate-relevant information that supports reduced modelling.

**Table 1**. Summary of representative molecular/atomistic AI studies and their scale-bridging contributions, from enabling reactive sampling at near-QM fidelity to extracting pathway- and rate-relevant information for reduced modelling and closure development.

| Study | Molecular-scale task | System/fuel | AI method | Scale-bridging value |
|---|---|---|---|---|
| [56] | Scalable ML potential with QM accuracy | General | Deep potential MD | Enables large-scale reactive MD with near-QM fidelity |
| [54] | Reactive chemistry discovery via ML-MD | Combustion chemistry | NNP-MD | Discovers complex pathways/intermediates from trajectories |
| [55] | Chemical-space exploration of pyrolysis | Linear alkanes | DP-GEN + DP-MD | Pathway discovery across a fuel class (transfer beyond single molecules) |
| [66] | Extreme-scale ML-MD | General | ML-MD + HPC scaling | Demonstrates feasibility of very large ML-MD systems |
| [67] | Ab initio soot-formation workflow (“nanoreactor”) | Soot precursors | ML + ab initio workflow | Links molecular growth pathways to soot-model concepts |
| [68] | Pollutant formation pathway analysis | $NH_3$ and $NH_3$-$H_2$ | NNP + reactive MD | Resolves $NO/NO_2/N_2O$-relevant pathways vs $\varphi$ and $H_2$ blending |
| [63] | Kinetic parameter learning | General | ML for rate constants | Accelerates mechanism development / screening |
| [69] | NNPES-assisted reactive sampling of pyrolysis | Spiro-hydrocarbon | NNPES + concurrent learning | Reactive chemistry exploration beyond classical MD limits |
| [70] | Initial pyrolysis behaviour | Spiro-hydrocarbons | DPMD (NN-assisted MD) | High-fidelity trajectory evidence for early pyrolysis channels |
| [71] | General reactive MLIP (condensed phase) | C/H/N/O chemistry | ANI-1xnr + active learning | Transferable reactive exploration across diverse systems |
| [72] | Skeleton reaction network generation | RP-3 aviation fuel pyrolysis | ReaxFF MD + ML reaction classification | Condenses large reactive MD chemistry into interpretable networks |
| [73] | Catalytic mechanism prediction | Metal oxides (CuO and PbO) | NNP + reactive MD/DFT | Atomistic catalytic insight; catalysts improve propellant stability and performance |
| [65] | Thermophysical properties for blends | Ethanol–octane blends | MD + ML regression | Property support for blending/spray/mixing boundary conditions |
| [59] | Reaction-network decoding from NNP-MD | Nitromethane | NNP-MD + network discovery | Maps decomposition network; condition/density effects |
| [60] | NOx suppression & extrapolation from reactive MD | $NH_3$-$CH_4$ + alcohol additives | ReaxFF MD + ensemble ML | Predicts NOx at untested blends; reduces sampling burden |
| [64] | Rate constant calculation using ab initio–trained ML force fields | Hydrogen combustion (elementary kinetics) | aML-MD (DPMD framework) + quasi-classical trajectories | Links QM-trained potentials to kinetic rate constants; highlights the need for energy-diverse training data |

### *3.2 Experimental combustion data and AI-enabled diagnostics*

#### *3.2.1 Role of experiments in AI-enabled multiscale combustion*

Experimental combustion remains the primary arbiter of credibility for multiscale modelling because it supplies the measurable targets, i.e., flame stabilisation limits, ignition delay, burning velocity, pollutant markers, and space–time resolved fields, against which chemistry models and CFD closures are ultimately judged. In practice, however, many quantities of interest are only partially observable: optical access is limited, signals are noisy at high pressure or high speed, multi-parameter laser diagnostics are expensive and difficult to synchronise, and three-dimensional (3D) reconstructions are often ill-posed. Recent work therefore uses machine learning not as a substitute for diagnostics but also as an additional “inference layer” that (i) improves the signal fidelity (denoising/deconvolution), (ii) reconstructs the unmeasured quantities from correlated measurements (virtual sensing), and (iii) accelerates the inverse problems such as tomography so that they become feasible on experimental timescales. These developments are particularly relevant for emerging fuels (e.g., $NH_3/H_2$ blends) because the most decision-relevant outputs can be emissions- or stability-related, where the experimental coverage across operating space is typically sparse [74].

Across the recent literature, three experimental-AI roles appear repeatedly: (1) measurement enhancement (raising signal-to-noise ratio (SNR), correcting artefacts); (2) modality substitution (inferring a hard-to-measure field from an easier-to-measure one); and (3) constrained inversion (embedding measurement physics—line-of-sight integrals, limited-view projections, or conservation laws—into the learning problem so that reconstructions remain physically plausible when data are scarce). Representative studies and their scale-bridging value are summarised in Table 2.

**Table 2**. Experimental combustion AI for diagnostics, inference, and scale bridging.

| Study | Experimental modality/data | System/ configuration | AI model type | Scale-bridging value |
|---|---|---|---|---|
| [75] | Simultaneous OH-PLIF and PIV fields | Swirl-stabilised premixed combustor | Neural networks (field-to-field mapping) | Formalises diagnostic redundancy; supports sensor-fusion design and achievable “virtual sensing” |
| [74] | Flame emission spectroscopy (short- vs long-gated spectra) | Methane–air flat flame, 1–10 bar | Denoising CNN + POD-informed loss; kriging/POD calibration | Extends high-pressure/time-resolved spectroscopy by reducing SNR limitations |
| [76] | Multi-view flame projections (emission-based) | Afterburner flame | Autoencoder–LSTM and deep learning reconstruction | Rapid 3D soot-field estimation to complement or replace slow iterative inversions |
| [77] | CH* chemiluminescence + simultaneous CH-PLIF (training pairs) | Turbulent premixed methane/air | Conditional GAN (ResNet/U-Net generators) | Laser-free approximation of planar diagnostics; enables higher-speed and simpler setups |
| [78] | OH-PLIF + $CH_2O$-PLIF with Rayleigh- | Jet-in-hot-coflow flames (multi-fuel) | Multi-scale U-Net + transfer learning | “Virtual thermometry” adaptable across fuels; supports high-throughput validation datasets |
| | scattering temperature | | | |
| [79] | OH-PLIF + soot volume fraction (PLII) with TLAF temperature | Turbulent sooting flames | CNN image-to-image regression | Extends thermometry into sooting/high-speed regimes where direct methods are limited |
| [80] | Planar laser diagnostics with regions obscured by soot scattering | Acoustically forced laminar sooting flames | PINNs + AE-DeepONet (physics-informed) | Fills diagnostically inaccessible regions while enforcing conservation constraints |
| [81] | Multi-view chemiluminescence projections | Volumetric flame chemiluminescence tomography | Physics-enhanced neural network (PENTAGON) | Brings tomography closer to experimental timescales; improves robustness vs pure data-driven reconstructions |
| [82] | Line-of-sight soot integral radiation | Laminar sooting flames | Equation-informed neural networks (PINNs) | Constrains inversion using measurement physics; reduces ill-posedness sensitivity |
| [35] | Flame images under variable operating conditions | Industrial heavy-oil combustion | Semi-supervised ADAE + GPR | Operational emissions monitoring with uncertainty; supports optimisation under sparse labels |
| [83] | High-speed flame image sequences | Lab-scale combustor with stable/unstable regimes | CNN + LSTM | Early warning/monitoring; links imaging diagnostics to stability control logic |
| [84] | High-speed in-cylinder flame images | Optical GDI engine | EfficientNet/ResNet + saliency | Virtual pressure sensing from imaging; reduces dependence on intrusive transducers |
| [85] | Engine-block accelerometer + crank-angle features | Diesel engine, multi-condition data | ResNet–LSTM | Enables control-oriented combustion phasing inference using low-cost sensors |
| [86] | UV–VIS–IR chemiluminescence spectroscopy | Turbulent premixed $NH_3/H_2$/air swirl burner | Gaussian process regression | Non-intrusive optical sensing coupled to emissions prediction for alternative fuels |
| [87] | Sparse wall-pressure sensing + | Hydrogen-fuelled scramjet | Deep learning reconstruction | Enhances observability in constrained geometries; supports |
| | schlieren imaging (targets) | wind-tunnel tests | (local–global feature fusion) | monitoring and potential active control |
| [19] | Low-cost / partial experimental field data | Gas-phase reacting flows | CRF-PINN (physics-informed) | Data-efficient field recovery for validation and model calibration workflows |

### *3.2.2 Measurement enhancement and diagnostic modality substitution*

A practical entry point for AI in laboratories is direct signal conditioning: short-exposure or high-speed measurements frequently trade temporal resolution for signal-to-noise ratio, and this trade-off is often a hard constraint when probing unsteady flames. Yoon et al. [74] developed a denoising convolutional neural network for fast flame emission spectroscopy, training on paired "short-gated" and high-SNR "long-gated" spectra and introducing a POD-informed loss to preserve physically meaningful spectral structure. The denoised spectra materially improved the downstream regression of pressure and equivalence ratio in high-pressure conditions [74]. Fig. 8(a) illustrates the workflow for high-temporal-resolution flame emission spectroscopy (FES) using CNN-based denoising and data mapping, illustrating the acquisition of high- and low-SNR spectra, POD-based preprocessing, hyperparameter optimization, and gas-property prediction.

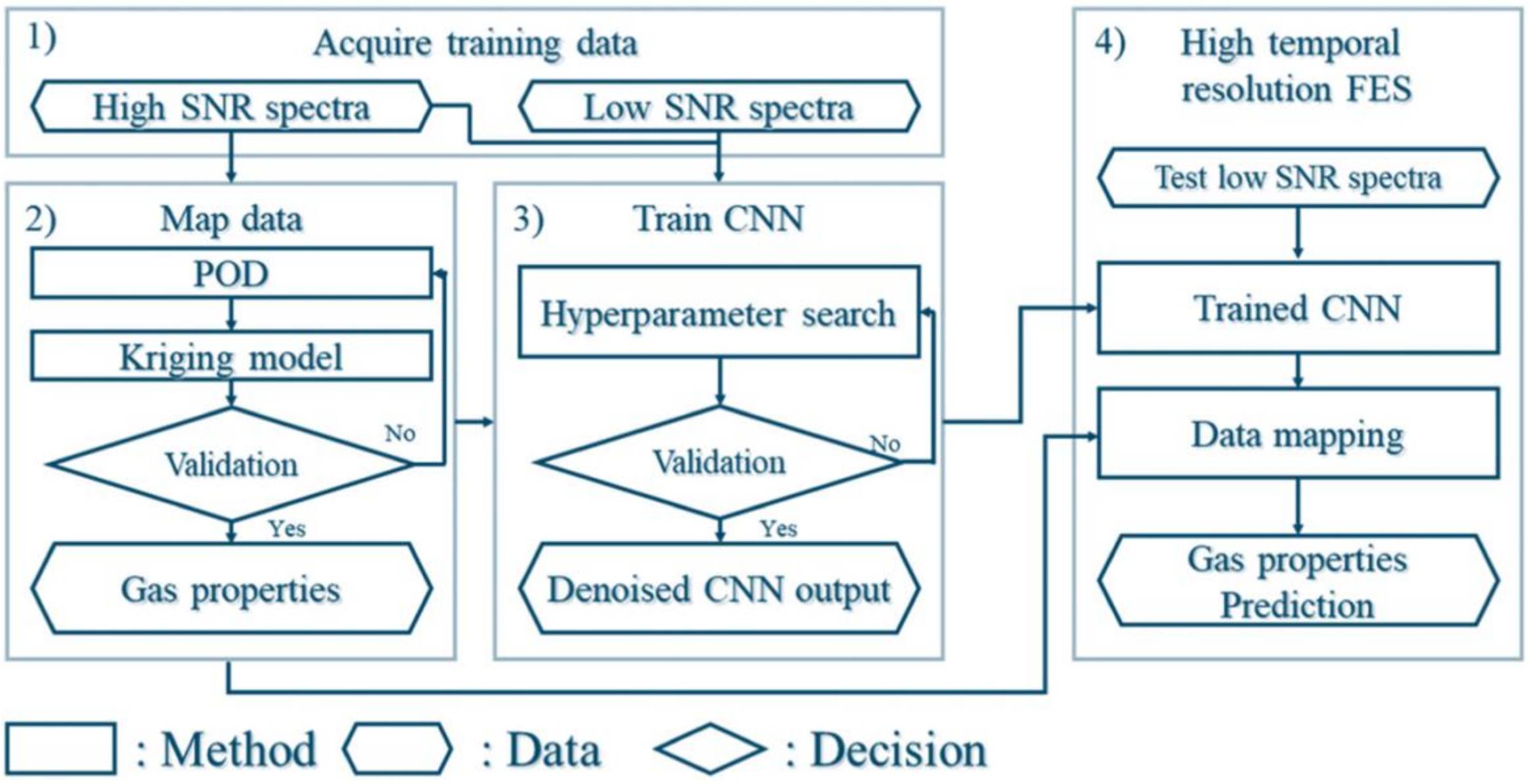

(a)

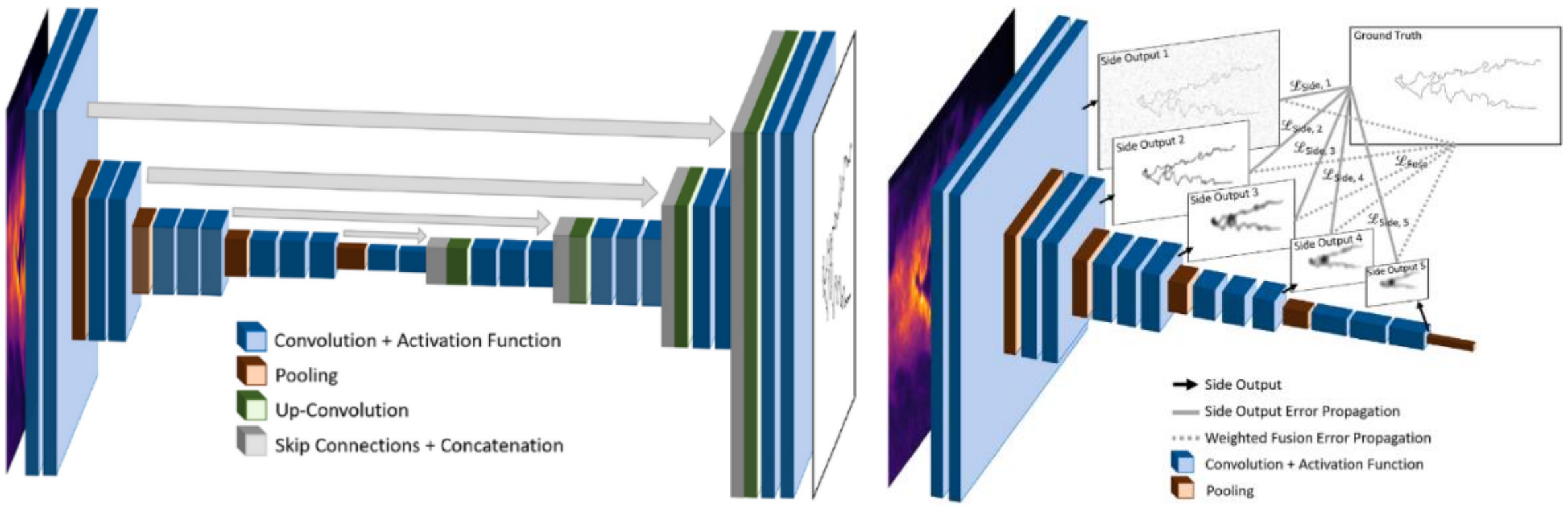

(b) (c)

**Fig. 8**. AI-enabled measurement enhancement and diagnostic modality substitution in experimental combustion. (a) Workflow for high-temporal-resolution flame emission spectroscopy (FES) using CNN-based denoising and data mapping, illustrating the acquisition of high- and low-SNR spectra, POD-based preprocessing, hyperparameter optimization, and gas-property prediction (Reprinted from [74]). (b) U-Net architecture with encoder–decoder structure, convolutional blocks, pooling, up-convolution, and skip connections for image-based combustion diagnostics (Reprinted from [88]). (c) Multi-output convolutional framework with weighted error propagation for physics-consistent feature extraction and diagnostic reconstruction (Reprinted from [88]).

A more consequential use of learning is diagnostic substitution, i.e., using an accessible measurement to infer an inaccessible one, thereby reducing hardware burden or extending temporal resolution. A clear example is the use of multi-input CNN surrogates for temperature imaging. In jet-in-hot-coflow flames, Kildare et al. [78] trained a multi-scale U-Net to map simultaneous OH- and $CH_2O$-PLIF images to Rayleigh-derived temperature fields and then exploited transfer learning to extend the approach across fuels (ethanol and DME) with reduced retraining cost [78]. In turbulent sooting flames, Nie et al. [79] demonstrated that a CNN can reconstruct instantaneous temperature fields from co-measured OH-PLIF and soot volume fraction images, with reported errors that remain bounded even when applied outside the training regime where the direct high-speed thermometry is difficult [79].

The conceptual basis for such substitution is strengthened by studies that explicitly quantify “information overlap” among diagnostics. Barwey et al. [75] used neural networks to assess how much velocity-field information can be inferred from simultaneously acquired OH-PLIF in a swirl-stabilised combustor, formalising a notion that correlated diagnostics can contain recoverable redundancy [75]. The same logic motivates laser-less inference strategies. For example, Han et al. [77] trained a conditional GAN to generate CH-PLIF-like flame-front structure from line-of-sight CH* chemiluminescence, enabling an approximate planar flame-front diagnostic without the experimental complexity of high-repetition laser systems [77]. A related bottleneck is post-processing of low-SNR PLIF at elevated pressures, where the collisional quenching degrades the signal and manual flame-front extraction becomes impractical at scale. Strässle et al. [88] addressed this using CNN-based supervised segmentation, showing improved performance over conventional gradient-based routines across a range of SNR levels representative of pressurised OH-PLIF datasets [88]. Figs. 8(b) and (c) show U-Net architecture with encoder–decoder structure, convolutional blocks, pooling, up-convolution, and skip connections for image-based combustion diagnostics and multi-output convolutional framework with weighted error propagation for physics-consistent feature extraction and diagnostic reconstruction.

*3.2.3 Learning-accelerated reconstruction and tomography from limited measurements*

Many high-value experimental quantities (3D soot fields, volumetric heat-release proxies, or scramjet internal flow structures) are inverse problems: the measurement is indirect (projection, line-of-sight integral, or sparse sensing), and classical reconstructions are either too slow or too sensitive to noise. In this context, machine learning is increasingly used to accelerate inversion while retaining the measurement model in some form.

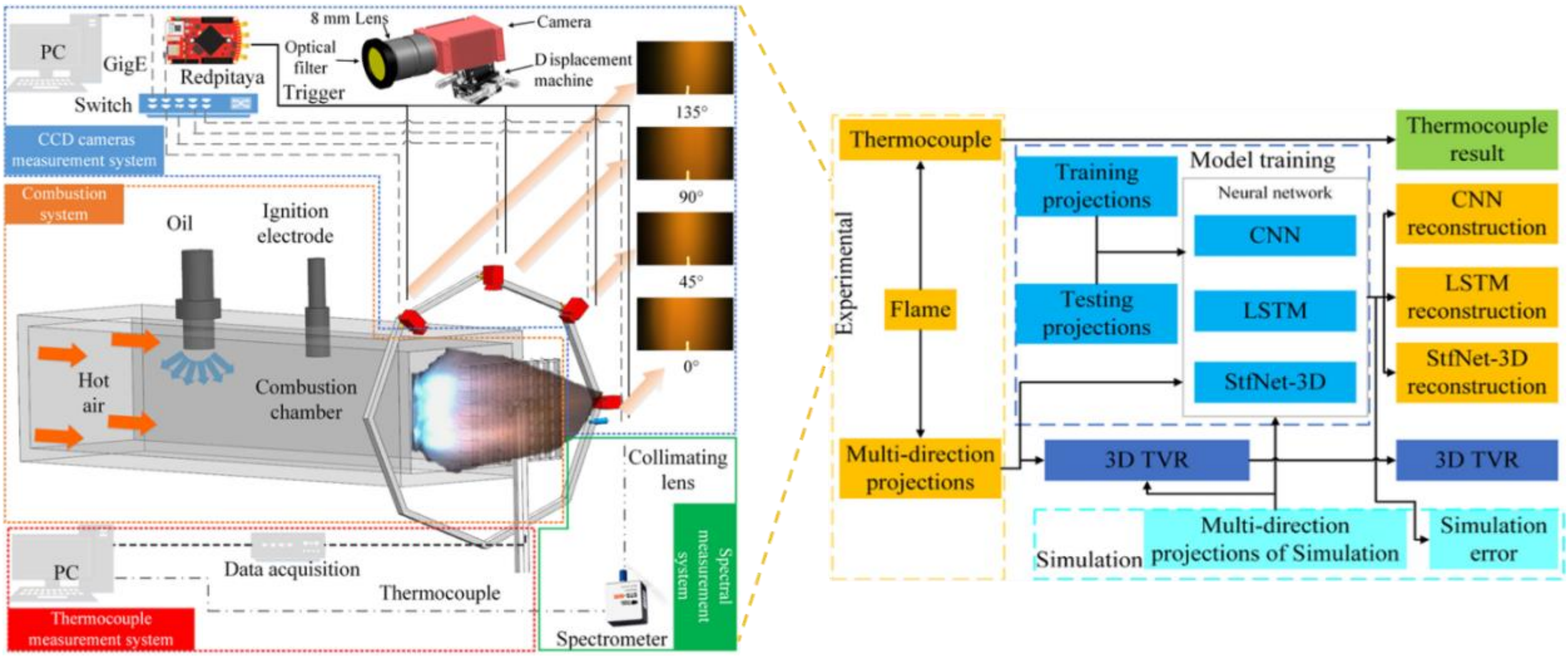

**Fig. 9**. Learning-accelerated three-dimensional (3D) soot-field reconstruction from multi-directional flame projections. Schematic of the experimental configuration and deep-learning-based reconstruction framework for 3D soot temperature and volume fraction estimation. Multi-angle optical projections of the flame are acquired using synchronized CCD cameras and thermocouple measurements, followed by training and testing of neural-network models (CNN, LSTM, and StfNet-3D) for volumetric reconstruction and comparison with traditional 3D TVR methods. The framework integrates experimental measurements and simulation-based projections to enable rapid tomographic inversion. Reprinted from [76].

For soot diagnostics, Dai et al. [76] reconstructed three-dimensional (3D) soot temperature and volume fraction fields in an afterburner flame using deep learning architectures trained on multi-direction projections (with synthetic data generation used to broaden the training set), demonstrating a pathway toward rapid 3D soot-state estimation beyond traditional regularised inversion pipelines (See Fig. 9). More recently, physically constrained variants have gained traction. Wang et al. [82] proposed "equation-informed" neural networks that incorporate the line-of-sight soot radiation integral equation directly into the learning problem, enabling simultaneous inference of soot temperature and soot volume fraction in laminar sooting flames from experimental radiation signals [82]. Volumetric chemiluminescence tomography is a second area where the constrained learning is emerging as a practical enabler. Jin et al. [81] introduced a physics-enhanced neural framework for volumetric flame chemiluminescence tomography that combines data priors with an explicit forward imaging model. The study critically emphasises performance under limited projection views and distribution shift, two failure modes that often undermine purely data-driven reconstructions in laboratory practice [81]. Physics-informed learning also appears in settings where specific regions are "diagnostically inaccessible." In time-varying acoustically forced laminar sooting flames, Liu et al. [80] demonstrated physics-informed machine-learning models (PINNs for velocity/temperature fields and an AE-DeepONet for soot volume fraction) trained on planar laser diagnostics while enforcing conservation constraints Their contribution here is not only the reconstruction accuracy but also the ability to fill regions where the soot scattering makes direct measurement unreliable [80]. Furthermore, the data-driven reconstruction has been extended into high-speed propulsion facilities where the optical access is constrained. In scramjet experiments, Huang et al. [87] reconstructed the schlieren flow-field imagery from sparse wall-pressure sensing using a deep learning architecture designed to preserve weak-gradient features. It is worth noting that the work is positioned explicitly as an aid to state awareness for monitoring and potential active control in narrow combustor geometries [87].

#### *3.2.4 Real-time state estimation and emissions inference in engines and burners*

A distinct stream of experimental-AI work targets real-time inference: estimates must be produced at rates compatible with control or monitoring rather than offline analysis. The models therefore tend to be compact, rely on readily available sensors (accelerometers, low-cost imaging, or chemiluminescence), and must generalise across varying operating conditions.

In optical engines, flame images provide high-dimensional information that is correlated with in-cylinder thermodynamic evolution. Maged and Nour [84] trained deep networks including EfficientNet variants to predict the combustion pressure from flame images in a single-cylinder optical GDI engine dataset and conducted the saliency analysis to interrogate which spatial regions contribute to the prediction, serving as an example of ML being used as a "virtual pressure sensor" when the direct transducer installation is limited [84]. Complementing imaging, non-intrusive vibration sensing offers a route toward low-cost deployment (See Figs. 10(a) and (b). Pushpalayam et al. [85] estimated CA50 cycle-by-cycle from engine-block accelerometers using a ResNet–LSTM architecture, explicitly aiming at real-time combustion phasing inference across varying fuels and conditions.

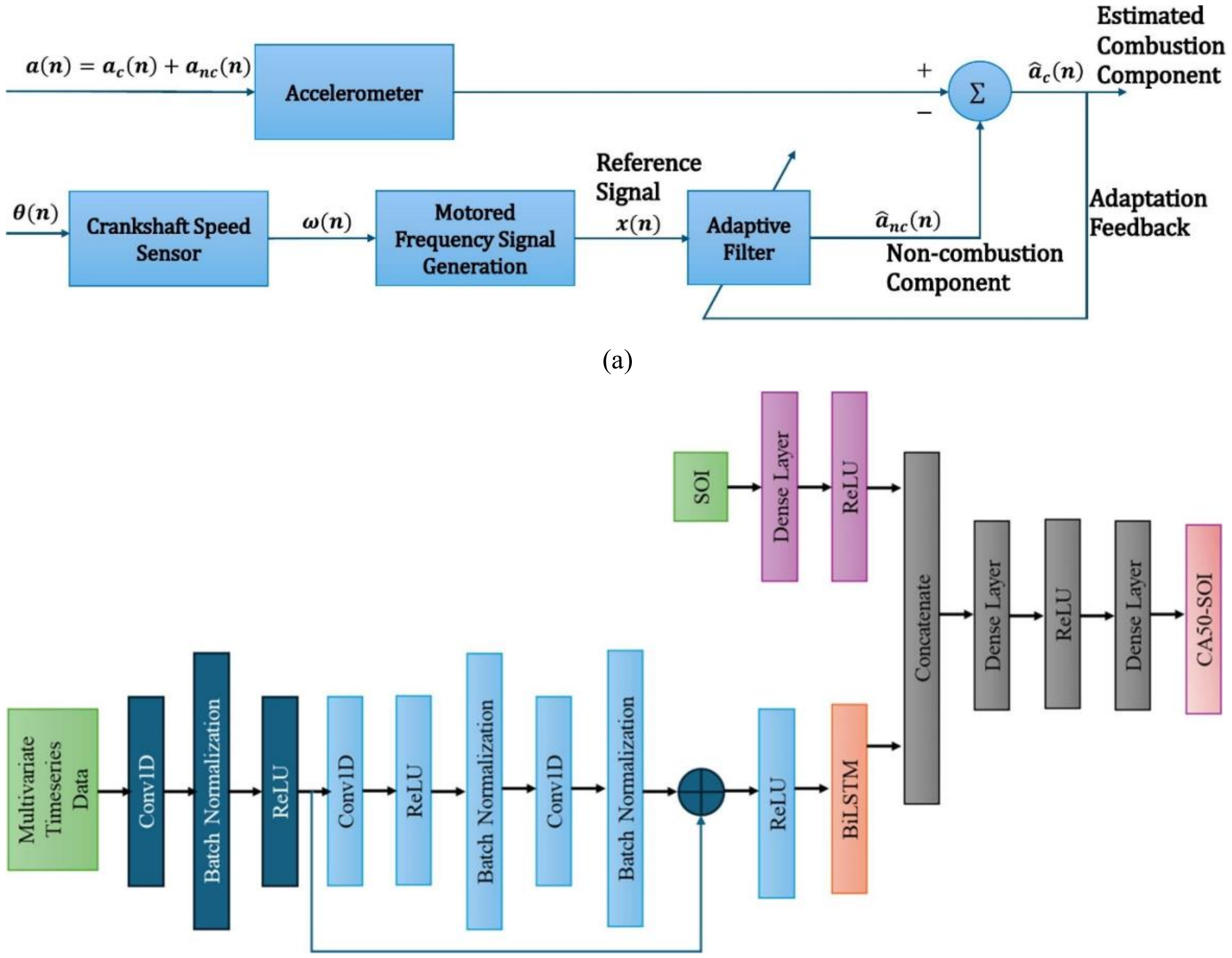

(a)

(b)

**Fig. 10**. Real-time combustion state estimation framework using the vibration-based sensing and adaptive signal processing. (a) Signal-processing architecture for separating combustion-induced and non-combustion vibration components using accelerometer measurements, crankshaft speed sensing, motored-frequency reference generation, and adaptive filtering with feedback correction. (b) Integrated deep-learning framework for real-time combustion progress estimation, combining feature extraction and temporal modelling to infer combustion metrics (e.g., CA50) from engine vibration signals. Reprinted from [85].

Emissions inference from optical signals has also been advanced and motivated by the operational need for fast feedback under variable conditions. Han et al. [35] proposed a semi-supervised adversarial denoising autoencoder coupled with Gaussian process regression (GPR) to predict $CO_2$ and NOx from heavy-oil flame image features. Notably, the employment of Gaussian processes supports confidence intervals, aligning the prediction with decision-making needs rather than point estimates only [35]. For alternative fuels, chemiluminescence-based sensing coupled with regression models is particularly attractive because it is non-intrusive and can be applied in turbulent flames. Mazzotta et al. [89] used GPR to infer equivalence ratio, $NH_3$ fuel fraction, and $NO/NO_2/N_2O$ emissions from multi-band chemiluminescence signals in premixed $NH_3/H_2$/air swirl flames, illustrating a direct route from optical signatures to emissions-relevant metrics. Moreover,, the image-sequence learning has been increasingly used for instability detection, which is experimentally measurable (pressure or imaging) but is operationally costly when it progresses to damaging amplitudes. Lyu et al. [83] combined CNN feature extraction with LSTM temporal modelling to detect the combustion instability from high-speed flame image sequences, demonstrating that the time-resolved imaging can support early warning and monitoring.

*3.2.5 Validation, uncertainty, and transferability*

The principal scientific risk for experimental AI models is not whether they interpolate within a narrow operating envelope, but whether they remain reliable under distribution shift. Differences in facilities, optics, seeding, camera response, fuel composition, pressure, or dilution can alter the measurement manifold even when the underlying physics remains similar. The practical importance of this challenge is illustrated in Fig. 11, which evaluates the robustness of deep-learning-based flame-front segmentation under progressively degraded OH-PLIF signal quality and thus provides a controlled test of model stability under measurement corruption. Recent studies have addressed this problem in three pragmatic ways: (i) transfer learning across fuels and operating conditions, for example in temperature inference for jet-in-hot-coflow flames; (ii) physics-enhanced loss functions that preserve measurement structure, such as POD-informed denoising for emission spectra; and (iii) explicit incorporation of measurement physics to constrain reconstruction, for example through line-of-sight integral equations or forward tomography [78].

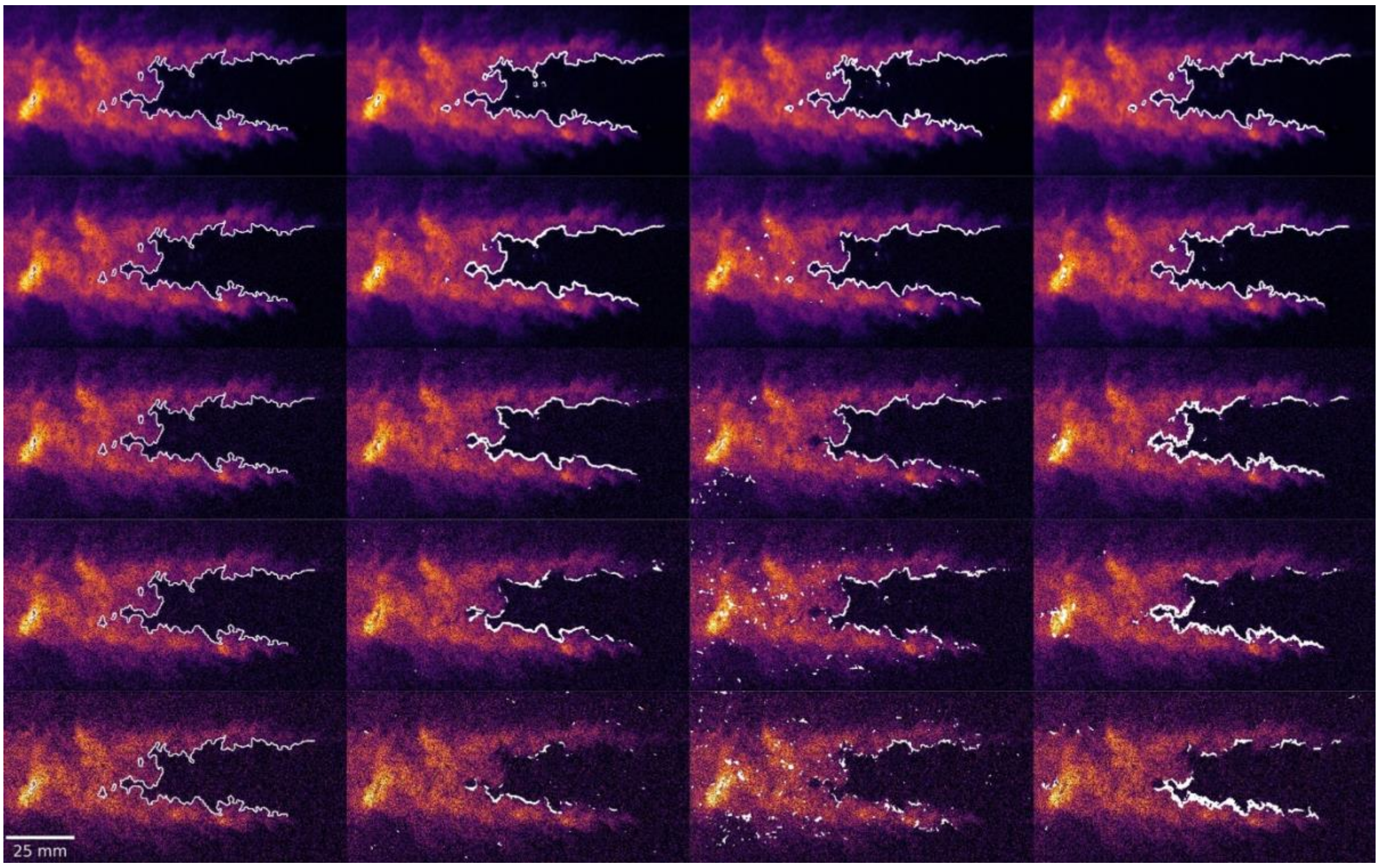

**Fig. 11**. Robustness and uncertainty evaluation of deep-learning-based flame-front segmentation under controlled signal degradation. Representative examples of artificially applied Gaussian and pepper noise to OH-PLIF images at progressively decreasing signal-to-noise ratios (SNR), illustrating model robustness to measurement corruption. Columns (left to right) show the ground truth segmentation and predictions from U-Net VGG16, U-Net EfficientNet-B5, and Attention U-Net VGG16 architectures. Rows correspond to increasing noise levels (SNR = 1.854 to 0.969), enabling the systematic assessment of stability, transferability, and sensitivity of segmentation performance under realistic experimental degradation. Reprinted from [88].

Uncertainty quantification has begun to appear explicitly when the regression model supports it naturally. The heavy-oil emissions study by Han et al. [35] reports prediction intervals from a Gaussian process stage, an approach that is directly relevant if ML outputs are used for operational decisions rather than qualitative interpretation.

A parallel trend is the rise of physics-informed frameworks that attempt to reduce reliance on large, labelled datasets by enforcing the conservation laws and transport equations. For example, CRF-PINN integrates Navier–Stokes, multispecies transport, and global kinetics with low-cost experimental inputs to aid acquisition of high-fidelity multiphysics fields without conventional pre-training [19]. While the maturity of such approaches in routine laboratory workflows is still evolving, they align closely with the broader aim of scale-bridging: producing experimental-grade field information that can be used to validate and calibrate higher-level combustion models in a more systematic and data-efficient manner.

*3.3 Continuum-scale CFD and AI for combustion*

*3.3.1 AI-Augmented Turbulence and Combustion Closures in CFD*

CFD remains the principal predictive framework for combustor and engine design as it resolves the coupled transport of momentum, heat, and chemical species within complex geometries over engineering-relevant time scales. Despite its maturity, two persistent bottlenecks continue to constrain both accuracy and computational efficiency: (i) turbulence–chemistry closure modelling in RANS and LES formulations and (ii) the numerical cost associated with detailed chemical kinetics, particularly stiff ODE integration and large reaction mechanisms. Consequently, recent research has increasingly focused on integrating artificial intelligence techniques into reacting-flow CFD—not as a substitute for conservation-law-based solvers, but as a structured augmentation strategy aimed at improving the closure fidelity, accelerating chemistry evaluation, and enabling rapid-response surrogate models for parametric studies and optimisation tasks [90-92].

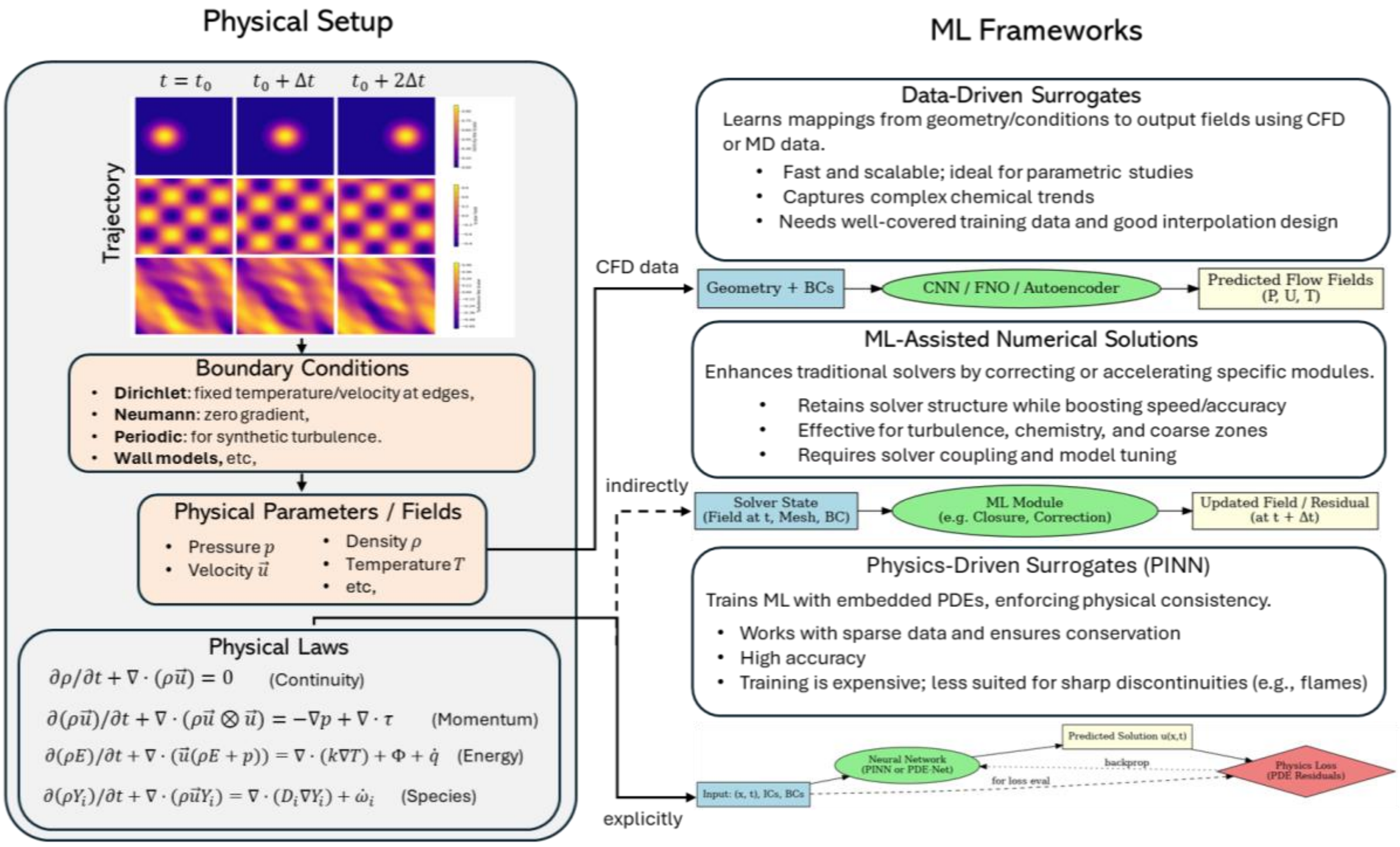

**Fig. 12**. Schematic overview of three primary paradigms for integrating AI with continuum CFD in reacting-flow simulations (top to bottom at the right panel): (i) data-driven surrogate modelling; (ii) ML-assisted solver augmentation; and (iii) physics-informed PDE-constrained learning.

Fig. 12 schematically illustrates three overarching paradigms that currently organise the CFD–AI landscape in combustion research. The first paradigm consists of data-driven surrogates, where neural architectures (e.g., CNNs, FNOs, and autoencoders) are trained to approximate mappings from geometry, boundary conditions, or low-fidelity flow fields to higher-fidelity thermo-fluid solutions. The second paradigm involves ML-assisted solvers, in which machine-learning modules are embedded directly within RANS, LES, or reacting-flow CFD solvers to provide turbulence closures, subfilter-scale combustion closures, or learned residual corrections during time advancement. The third paradigm comprises physics-informed learning frameworks, where governing PDE residuals and conservation constraints are incorporated into the loss function to regularise training and improve the physical consistency. This tripartite classification aligns with recent systematisations in the broader CFD–ML literature and provides a structured lens through which combustion-specific developments can be interpreted [27].

Early and influential work in turbulence modelling established the core design principle that closure models should respect invariances and tensor structure. For example, invariant network architectures were proposed to predict the Reynolds-stress anisotropy while embedding Galilean invariance, which helped frame "physics-aware" ML as more than curve fitting [9, 93]. In parallel, as the most notably gene expression programming (GEP), the symbolic-regression approaches demonstrated that the interpretable algebraic stress–strain corrections can be discovered directly from high-fidelity data, yielding explicit closures that can be implemented in legacy CFD codes [94]. These ideas have been extended from "fit-to-DNS" toward CFD-driven training, where candidate closures are evaluated through integrated RANS calculations during optimisation, thereby targeting a posteriori robustness rather than purely a priori stress matching [95].

In reacting flows, analogous lessons have emerged: closure models must be assessed not only against filtered DNS targets but also for stability and predictive value when inserted into a live LES. Recent examples include machine-learning augmentation of filtered flame-front displacement modelling and flame-surface-density related closures, where the researchers explicitly emphasise a posteriori LES testing and sensitivity to model inputs [96]. In the same spirit, data-centric combustion LES studies have increasingly trained convolutional neural networks on DNS-derived, LES-filtered quantities to learn subfilter burning-rate closures without prescribing the regime-dependent functional forms. Fig. 13 illustrates a representative framework of this approach: a slot-burner DNS configuration is first used to generate a high-fidelity database and is then followed by spatial filtering, downsampling, cube extraction, and supervised CNN training to predict the filtered reaction-rate quantities [22]. The workflow highlights the importance of consistent data generation, filtering operations, and a posteriori validation when deploying ML closures in turbulent combustion LES.

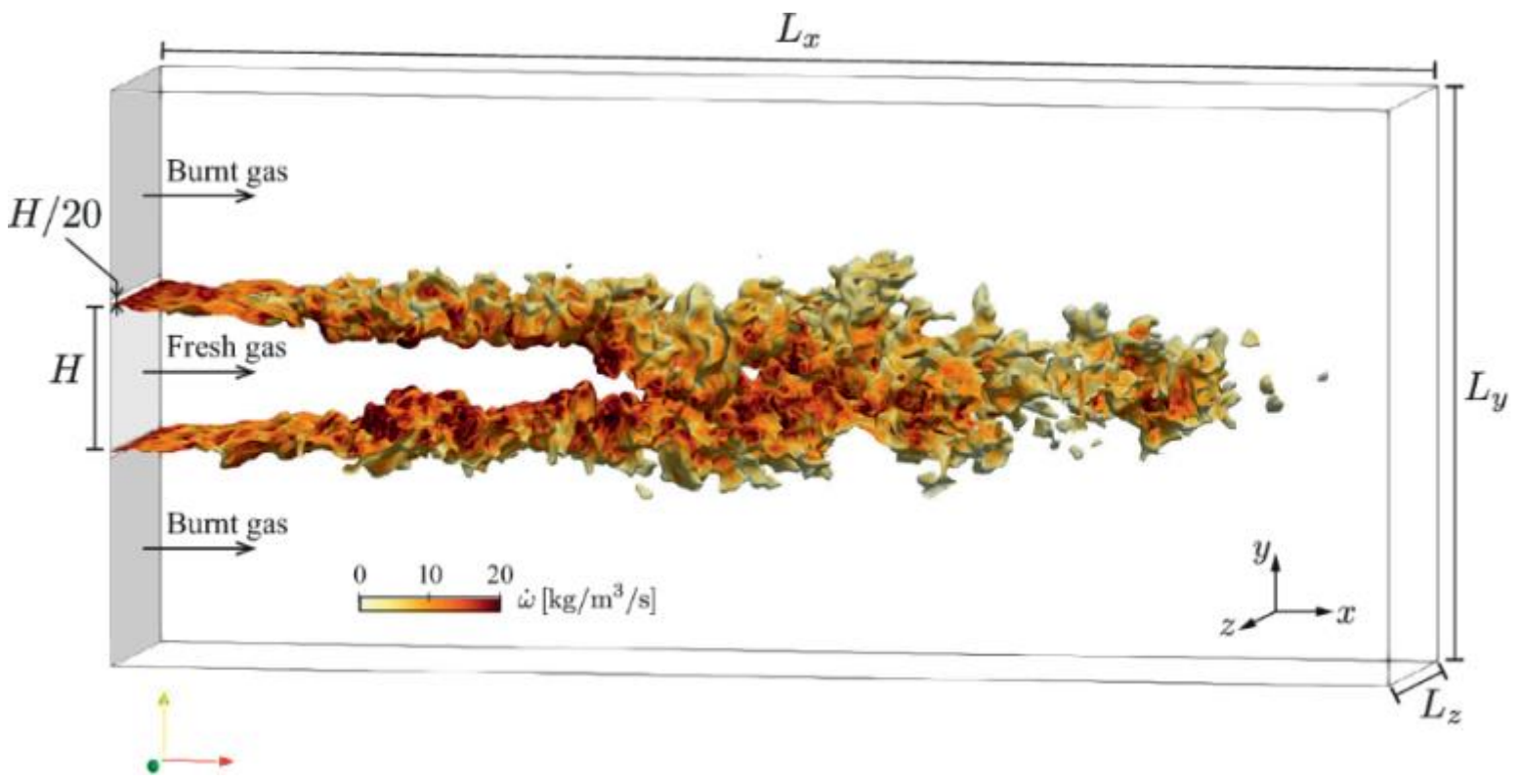

(a)

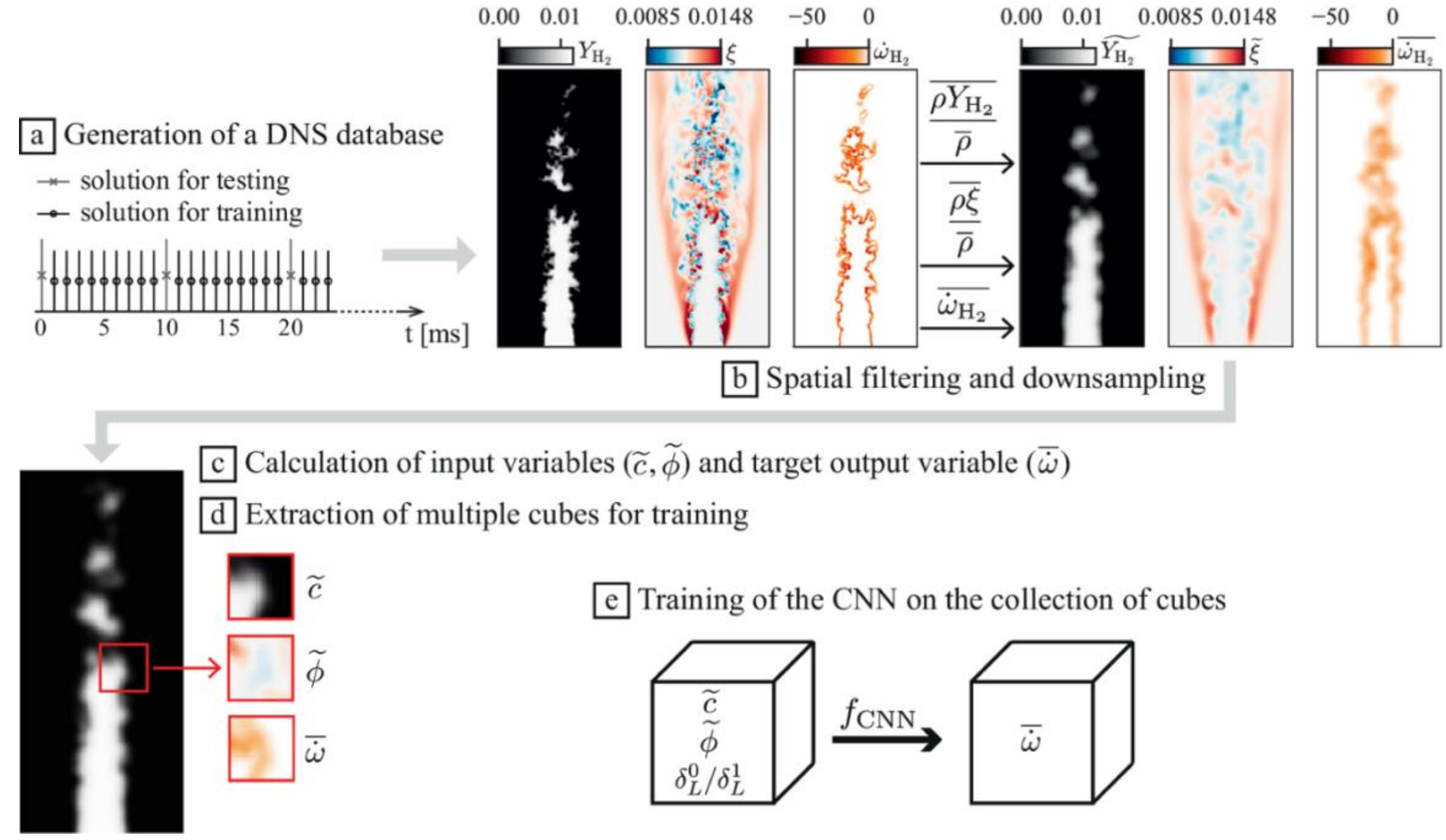

(b)

**Fig. 13**. (a) Slot-burner configuration used to generate the direct numerical simulation (DNS) database for turbulent premixed hydrogen–air combustion and (b) Schematic of the data-generation, spatial filtering, downsampling, and cube-extraction strategy used to train the convolutional neural network (CNN) for subfilter reaction-rate closure modelling. Reprinted from [22].

### *3.3.2 Chemistry Acceleration: Neural Surrogates, Operator Learning, and Conservation Constraints*

For reacting CFD, the dominant cost driver in many practical LES/RANS simulations remains the evaluation of chemical source terms. Consequently, nature of the most mature AI integration in combustion CFD has focused on chemistry acceleration by replacing the stiff ODE integration and memory-intensive tabulation with compact learned surrogates. A representative combustion-facing approach is ML tabulation, where neural networks are trained to reproduce the thermochemical source terms over a targeted composition space (often generated from canonical reactors or flamelets) and subsequently queried during LES–PDF or flamelet-based simulations to avoid repeated expensive chemistry integration. For turbulent DME flames, a hybrid flamelet/random-data strategy coupled with multiple multilayer perceptrons has demonstrated substantial reductions in LES–PDF runtime while preserving agreement with detailed-chemistry baselines [97].

A persistent challenge in such surrogates is to maintain physical consistency, particularly the mass and element conservation during time advancement and in extrapolative regimes. Recent efforts have therefore embedded conservation constraints directly into model design and training. For example, conservation-aware tabulation workflows have been formulated to improve robustness relative to unconstrained regressors [98], while other studies enforce conservation explicitly within neural kinetic surrogates to mitigate the solver-level drift during long integrations [99].

A promising recent direction is the use of operator-learning architectures, such as DeepONets, to emulate the evolution operator of chemical kinetics with adaptive time stepping. Rather than approximating the source terms pointwise, this approach learns the mapping between the thermochemical states across time increments, directly targeting the stiff integration step. A DeepONet-based chemistry-acceleration framework has been demonstrated for complex combustion systems, reporting substantial speedups while maintaining fidelity to detailed kinetics [100]. Fig. 14 illustrates this latent-space evolution strategy: an encoder maps thermochemical scalars into a reduced latent representation, a React-DeepONet operator advances the state within this space, and a decoder reconstructs the full thermochemical field. By evolving the system in a compact latent manifold, the method reduces dimensionality while preserving the dynamic structure. Follow-on work further incorporates auxiliary networks to recover the non-representative species and enforce the conservation constraints in reduced composition space, enabling improved compatibility with reacting-flow CFD solvers [101].

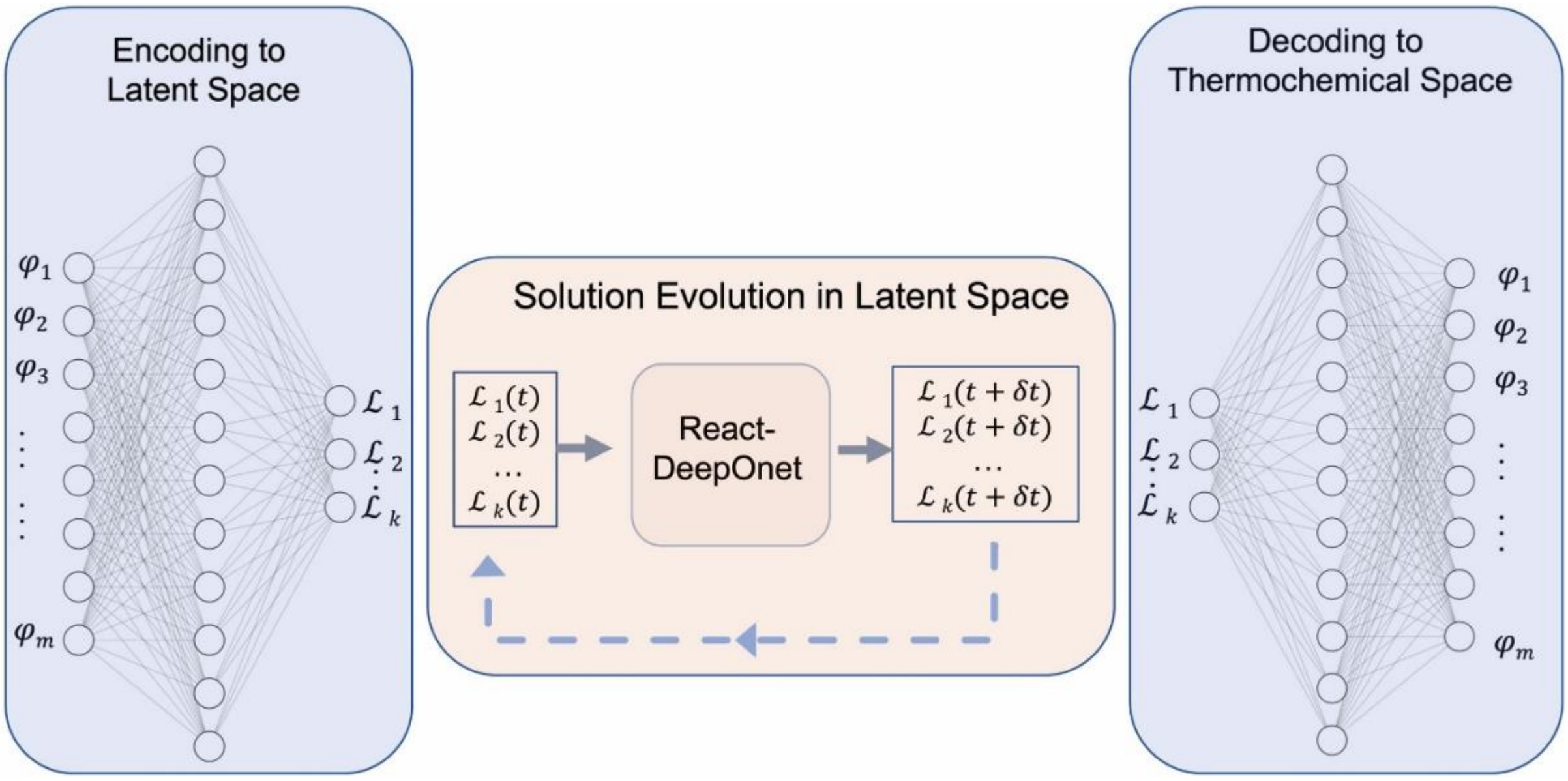

**Fig. 14**. Framework for latent-space kinetics identification and evolution using LS-React-DeepONet. The encoder maps thermochemical scalars to a reduced latent representation, the React-DeepONet operator advances the solution in latent space, and the decoder reconstructs the thermochemical state. Reprinted from [100].

Alongside operator learning, there is renewed interest in generalisation-oriented neural integrators. For example, a multi-scale sampling strategy has been used to build robust DNN surrogates for combustion kinetics that remain stable under temporal evolution and can be implemented across several CFD codes, highlighting that dataset design (coverage of flame branches, intermediate species scales, and perturbations) is as important as architecture choice [102]. More recent work has pushed towards multi-fuel generalisation and explicit a posteriori validation in reacting-flow simulations, reporting large speed-ups for chemistry integration while checking that the key flame characteristics remain consistent with direct integration [16].

### *3.3.3 Data-Driven Flow-Field Surrogates and Digital Twin Applications*

Beyond closures and chemistry, a third family of CFD–AI methods aim to approximate parts—or, in some cases, the entirety—of the CFD forward map, thereby enabling rapid exploration of operating conditions. In combustion-relevant settings, this includes learning mappings from boundary conditions or sparse measurements to full flow-field structures for rapid state estimation and control-oriented modelling. For example, deep-learning-based scramjet flow-field reconstruction has been proposed to infer schlieren-relevant structures from limited wall-pressure measurements under space–time constraints where conventional diagnostics are impractical. Such approaches illustrate the broader shift toward CFD-informed digital twins that are trained offline but deployed as real-time inference models [20]. Fig. 15 presents a representative architecture, in which the multi-scale local–global feature grouping enables reconstruction of 2D combustor flow structures directly from sparse wall-pressure inputs, demonstrating how learned surrogates can approximate the CFD forward operator without iterative solution of the governing equations.

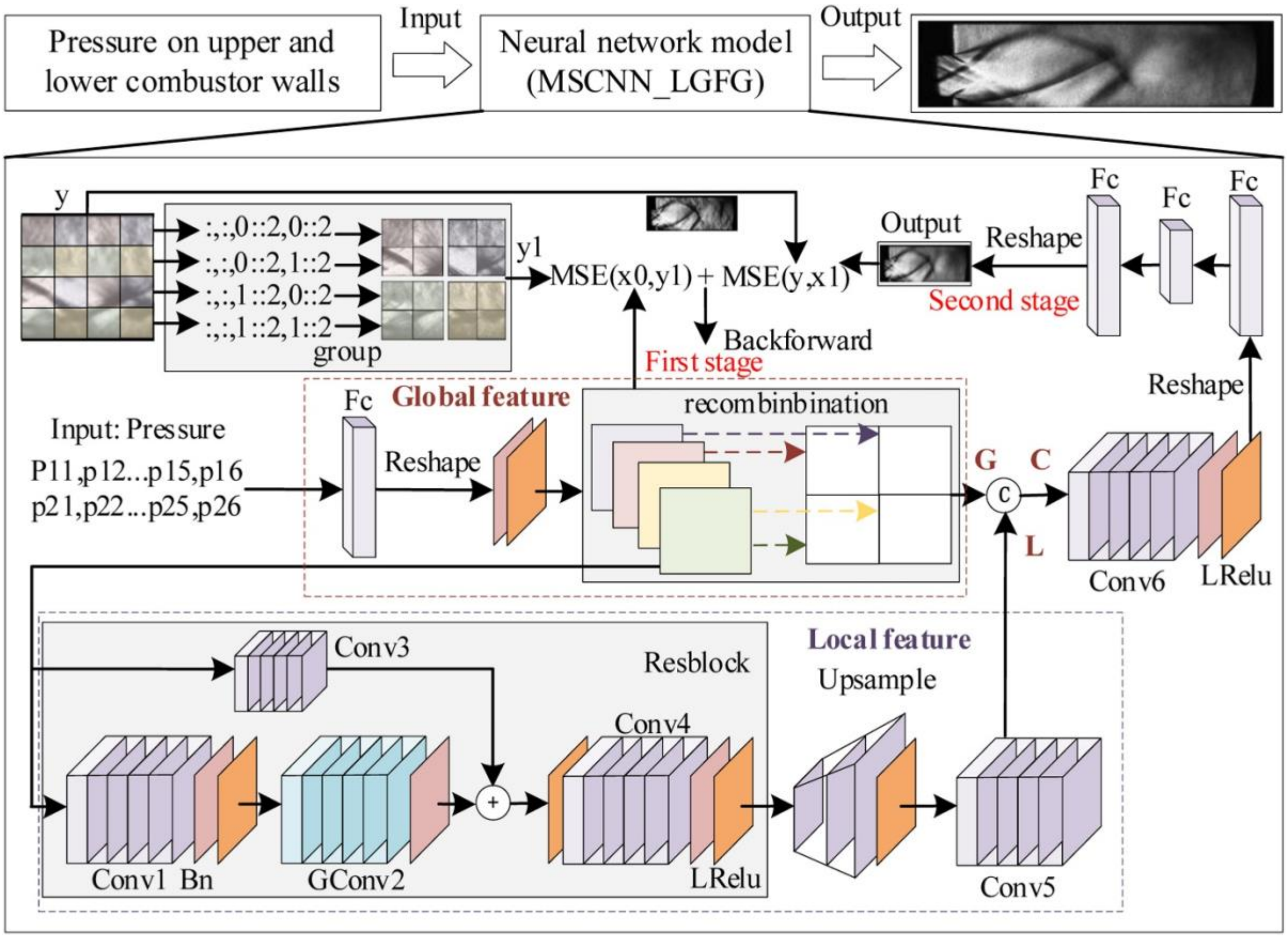

**Fig. 15**. Structure of a multi-scale flow-field reconstruction network based on the local-global feature grouping and fusion. Sparse pressure measurements from combustor walls are encoded and mapped through hierarchical feature extraction to reconstruct the 2D schlieren-relevant flow field. Reprinted from [20].

In turbulent combustion specifically, surrogate modelling must confront strong multi-scale coupling and sensitivity to unresolved scales. One pragmatic response has been the use of neural networks as compact replacements for large flamelet look-up tables, motivated by memory costs and communication overheads in large-scale LES. Flamelet Manifold Neural Networks (FMNN) exemplify this direction by constraining learning to physically meaningful manifold configurations and explicitly reporting both a priori and a posteriori assessment in turbulent premixed simulations [103].

In engine-relevant contexts, data-driven surrogate modelling has also been applied to predict the global performance indicators directly from CFD-informed datasets. Shateri et al. [90] developed and compared neural networks, random forest regression (RFR), and GPR models for predicting dodecane fuel consumption in diesel combustion, with validation against both CFD and experimental data. The GPR model demonstrated superior predictive accuracy and robustness to noise, achieving approximately 1.7× acceleration compared to conventional CFD solvers while preserving consistency in momentum and thermal characteristics. This work illustrates the growing role of ML-based response surfaces as CFD-informed digital twins for rapid design screening and fuel-efficiency optimisation. Extending this paradigm toward design optimisation, Shateri et al. [91] integrated ML surrogates with CFD simulations to predict and optimise the turbulence kinetic energy (TKE) and tumble-y in engine cold-flow conditions. Among the evaluated models, GPR achieved the highest fidelity while enabling a 21.6× reduction in computational time compared with full CFD calculations. By coupling surrogate predictions with Euclidean-distance-based optimisation metrics, the framework enabled rapid identification of engine configurations that enhanced in-cylinder air motion and pre-combustion mixing. Such approaches demonstrate how AI-enhanced CFD workflows can substantially reduce the parametric exploration costs while maintaining physically interpretable flow characteristics.

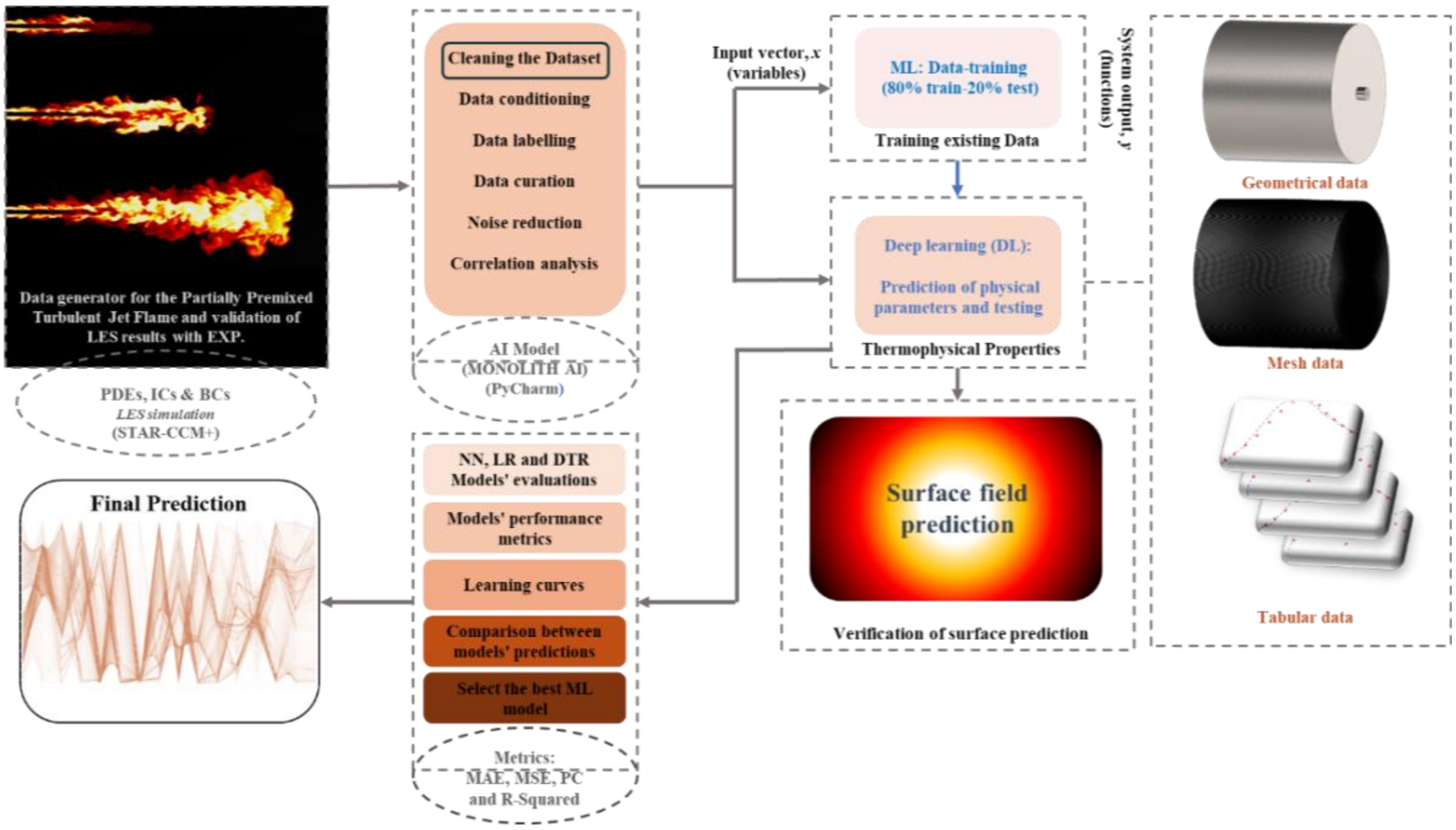

**Fig. 16**. Schematic of a non-intrusive LES-informed surrogate modelling framework. High-fidelity LES data are generated independently, followed by post-processing and development of tabular-input ML surrogates (NN, LR, DTR) and deep-learning surface-field predictors. The ML models operate offline and are not embedded within the CFD solver. Reproduced from Shateri, Yang, & Xie (2025) [92]. Machine learning-based prediction of species mass fraction and flame characteristics in partially premixed turbulent jet flame. Physics of Fluids, 37(7), 075219. https://doi.org/10.1063/5.0277024, with the permission of AIP Publishing.

At the scale of turbulent reacting flows, ML surrogates have also been trained directly on high-fidelity LES datasets. Shateri et al. [92] employed neural networks to predict species mass fractions and flame characteristics in partially premixed turbulent jet flames (Sandia Flame D) using LES-generated flamelet-based data for training. The neural-network surrogate achieved high predictive accuracy and delivered a 17.25× computational speed-up compared to direct LES-based chemistry evaluation. The workflow, which is illustrated in Fig. 16, follows a non-intrusive, a priori strategy, in which the LES data are first generated independently, and after which separate ML pipelines are trained for tabular surrogate prediction and surface-field reconstruction. This separation of data generation and surrogate deployment exemplifies CFD-informed digital-twin development without direct solver coupling while still enabling a posteriori benchmarking against LES baselines.

*3.3.4 Physics-Informed and Hybrid PDE-Constrained Learning for Reacting Flows*

Physics-informed neural networks (PINNs) have been widely proposed to reduce reliance on labelled data by enforcing governing equations in the loss function. While classical PINNs are most mature for lower-dimensional PDE problems, the combustion literature is increasingly exploring hybrid formulations that blend existing reduced-chemistry models with physics-informed training to improve fidelity and robustness. The foundational PINN framework for forward and inverse PDE problems is well established [42].

More combustion-specific work has begun coupling PINNs with flamelet/progress-variable formulations to account for detailed reaction mechanisms while using physics constraints to regularise the learning problem. This direction is notable because it positions PINNs not as "CFD replacement" but as physics-constrained surrogates that can sit between tabulation and full PDE solvers [104]. In practice, the main near-term value of physics-informed approaches in turbulent reacting flows may be less about replacing LES/RANS solvers wholesale and more about improving consistency, aiding data assimilation, and reducing drift in long-time integrations, especially when paired with conservation-constrained chemical surrogates and solver-aware deployment [101]. Across these paradigms, the routes to acceleration are distinct but complementary. Closure augmentation aims to deliver better predictions at a given grid resolution (or comparable accuracy at lower cost). Chemistry acceleration substitutes learned evaluation for stiff ODE solves and/or reduces memory and interpolation costs of tabulated methods. Operator learning offers a particularly direct lever by learning the evolution operator with adaptive time-step handling, whereas conservation-constrained learning is increasingly treated as a prerequisite for stable, long-horizon CFD deployment [100].

**Table 3**. Summary of representative CFD-AI studies that illustrate these themes across turbulence closures, combustion subfilter models, chemistry acceleration, and physics-informed/hybrid frameworks.

| Study | CFD-scale task | Application/ system | AI method | Integration/scale value |
|---|---|---|---|---|
| [90] | Fuel consumption surrogate modelling | Diesel engine combustion (dodecane) | NN, RFR, GPR (GPR optimal) | CFD-informed surrogate; 1.7× speed-up; validated against CFD and experiment |
| [91] | Flow-structure optimisation (TKE, tumble-y) | Engine cold-flow dynamics | NN, RFR, GPR (GPR optimal) | CFD–ML design optimisation; 21.6× acceleration; parametric exploration framework |
| [92] | LES-based species and flame surrogate | Partially premixed turbulent jet flame (Sandia D) | Neural network surrogate | 17.25× speed-up; high-fidelity LES benchmark; uncertainty quantification included |
| [95] | CFD-driven training of RANS closures | Turbomachinery wake mixing | GEP with integrated RANS-in-the-loop training | Targets a posteriori robustness by assessing candidates through CFD during optimisation |
| [105] | Explicit RANS closures via symbolic regression | Turbulent separated flows | Deep-learning-based symbolic regression with non-linear corrections | Generates explicit algebraic closures; balances accuracy and implementability |
| [103] | Replace flamelet LUTs; a priori + a posteriori tests | Premixed turbulent combustion | Flamelet Manifold Neural Networks (FMNN) | Compresses manifold representation; supports turbulent CFD deployment via a posteriori assessment |
| [96] | ML augmentation of combustion-closure terms in LES | Premixed turbulent flames (FSD/FFFD context) | ML regressors for closure terms with a posteriori LES testing | Highlights solver stability and the importance of a posteriori evaluation |
| [106] | Solver-aware coupling of ML with LES code | LES solver integration | Graph + CNN coupling with HPC LES solver | Reduces impedance mismatch between ML models and unstructured CFD workflows |
| [22] | CNN model for filtered burning rates | Lean $H_2$–air premixed combustion (LES) | CNN trained on DNS-to-LES filtered data | Learns subfilter reaction-rate/burning-rate closures without regime-specific assumptions |
| [98] | Conservation-aware chemistry tabulation | Reacting-flow chemistry surrogates | ML tabulation with conservation-law incorporation | Improves robustness/consistency of chemistry surrogates for reacting CFD |
| [97] | ML thermochemistry tabulation to avoid stiff ODE solves | Turbulent DME flames (LES–PDF) | Hybrid flamelet/random data + multi-MLP tabulation | Demonstrates notable wall-time reduction in turbulent combustion CFD with large mechanisms |
| [100] | Operator-learning chemistry acceleration | 0D/1D kinetics for complex fuels (CFD coupling target) | DeepONet-based surrogate (React-DeepONet) | Learns kinetics evolution operator; designed for implementation in reacting-flow solvers |
| [101] | Physics-constrained reduced-composition kinetics for CFD | Reduced composition-space chemistry | Correlation Net + physics-constrained React-DeepONet | Enforces mass/element conservation despite reduced state; improves CFD compatibility |
| [102] | Robust neural kinetics surrogate via dataset design | Multiple canonical flames; CFD-ready training | DNN with multi-scale sampling + preprocessing | Emphasises training-data strategy to improve stability and generalisation under time evolution |
| [99] | Conservation enforcement in neural chemical surrogates | Complex kinetics | Conservation-constrained neural surrogate | Addresses drift/instability risks for long-horizon CFD deployment |
| [107] | Physics-informed ML for dynamic reacting-flow prediction | Hydrogen jet flames | Hybrid physics-informed ML framework | Integrates low-resolution physics models with ML to improve dynamic prediction fidelity |
| [104] | Physics-informed learning coupled to combustion reduced models | Detailed chemistry within flamelet/FPV-style modelling | PINNs + FPV coupling | Illustrates PDE-constrained learning as a physics-regularised surrogate rather than solver replacement |
| [20] | Data-driven flow-field reconstruction for supersonic combustion | Scramjet combustor flow reconstruction | Deep learning reconstruction model | Illustrative digital-twin style surrogate for rapid inference under constrained measurements |

*3.4 A posteriori and application-level assessment of ML-assisted combustion models*

Beyond a priori agreement, the more rigorous test of a machine-learning-assisted combustion model is whether it remains reliable after being embedded in a full reacting-flow solver and applied to realistic combustor configurations. In this transition from offline prediction to solver-level deployment, the study of Malé et al. [22] offers an important bridge because it demonstrates that a convolutional neural network (CNN) can recover filtered burning-rate behaviour more accurately than conventional filtered tabulated-chemistry alternatives in LES-type conditions for lean hydrogen flames. As shown in Fig. 17, the scatter plots compare the CNN-predicted filtered burning rate $\bar{\omega}^{NN}$, the filtered tabulated-chemistry estimates $\bar{\omega}^{F}$ and $\bar{\omega}^{FC}$, and the reference filtered burning rate $\bar{\omega}^{*}$, with all values normalized by $\max(\bar{\omega})$. The four panels correspond to two global equivalence ratios, $\phi_g = 0.35$ and $\phi_g = 0.7$, and two LES parameter sets, $\sigma = 4$, $DSF = 2$ and $\sigma = 8$, $DSF = 4$, where $DSF$ denotes the downsampling factor. The closer clustering of $\bar{\omega}^{NN}$ along the diagonal indicates that the CNN more faithfully reproduces the reference burning-rate distribution, especially under the leaner condition where the thermo-diffusive effects are more severe.

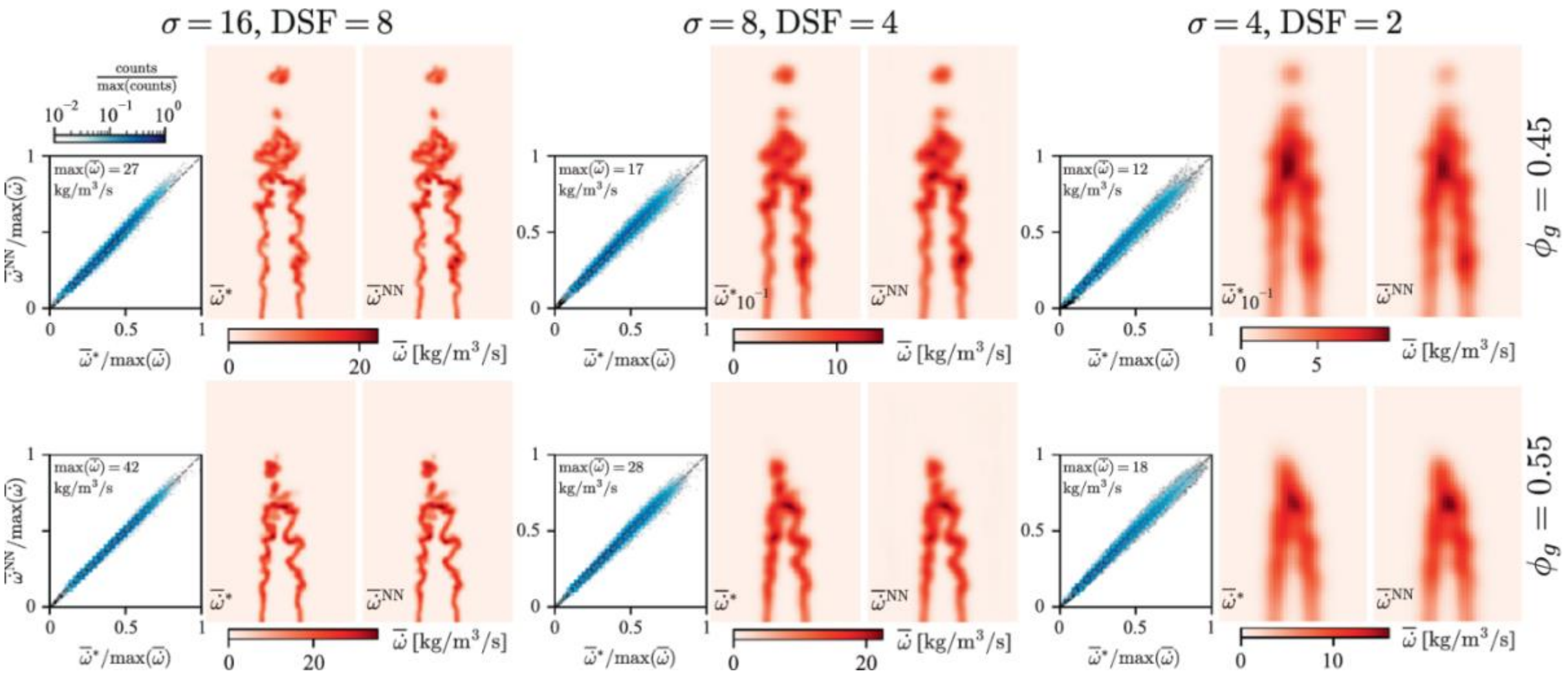

**Fig. 17**. Comparison of convolutional-neural-network-based and filtered-tabulated-chemistry burning-rate predictions for lean hydrogen large eddy simulation cases. Scatter plots with 2D histograms comparing the convolutional neural network modeled burning rate $\bar{\omega}^{NN}$, filtered tabulated-chemistry predictions $\bar{\omega}^{F}$ and $\bar{\omega}^{FC}$, and the reference filtered burning rate $\bar{\omega}^{*}$. Both axes are normalized by $\max(\bar{\omega})$. The columns correspond to $\phi_g = 0.35$ and $\phi_g = 0.7$, while the panels also include the LES parameter sets $\sigma = 4$, $DSF = 2$ and $\sigma = 8$, $DSF = 4$. The gray dashed line denotes the zero-error relation, and the maximum burning rate is reported in kg/$(m^3 \cdot s)$ in each panel. Reprinted from [22].

A more explicit a posteriori demonstration was reported by Chung et al. [108], who incorporated a random-forest classifier into LES of a single-element gaseous-oxygen/gaseous-methane rocket combustor to assign combustion submodels dynamically in space and time. In their framework, finite-rate chemistry, the flamelet/progress variable model, and inert mixing were selected locally according to a user-defined error threshold based on temperature and carbon monoxide mass fraction, denoted by $\theta_{\{T,CO\}}$. Fig. 18 presents a posteriori data-assisted LES results for two threshold values, namely (a) $\theta_{\{T,CO\}} = 0.05$ and (b) $\theta_{\{T,CO\}} = 0.02$. From top to bottom, the results show the temperature $\tilde{T}$, carbon monoxide mass fraction $\tilde{Y}_{CO}$, mixture fraction $\tilde{Z}$, and combustion-submodel assignment fields. In each strip, the upper half shows instantaneous results, whereas the lower half shows time-averaged results. Comparison of panels (a) and (b) indicates that the tighter threshold leads to broader use of the higher-fidelity model and produces fields that more closely resemble the finite-rate-chemistry solution.

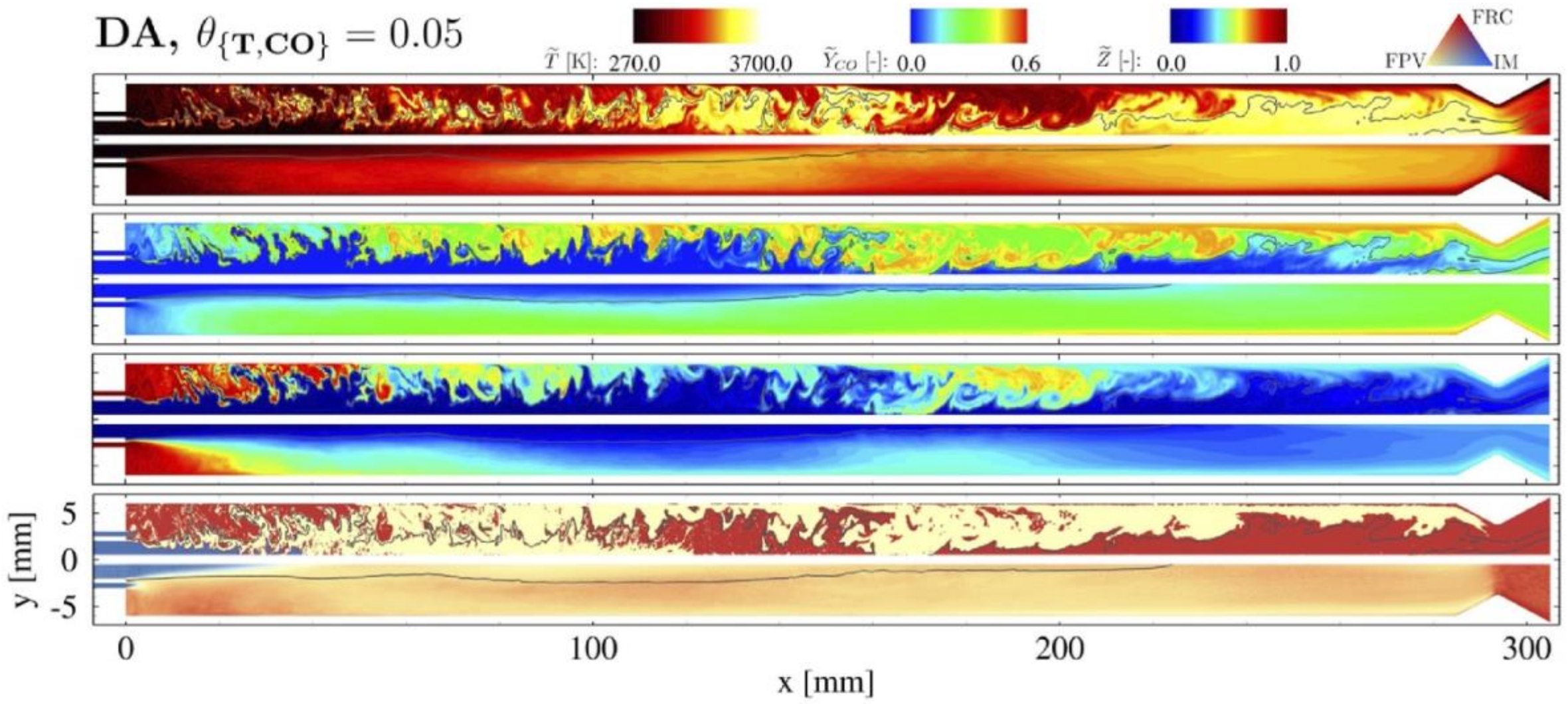

(a) *A posteriori* DA LES with $\theta_{\{T,CO\}} = 0.05$.

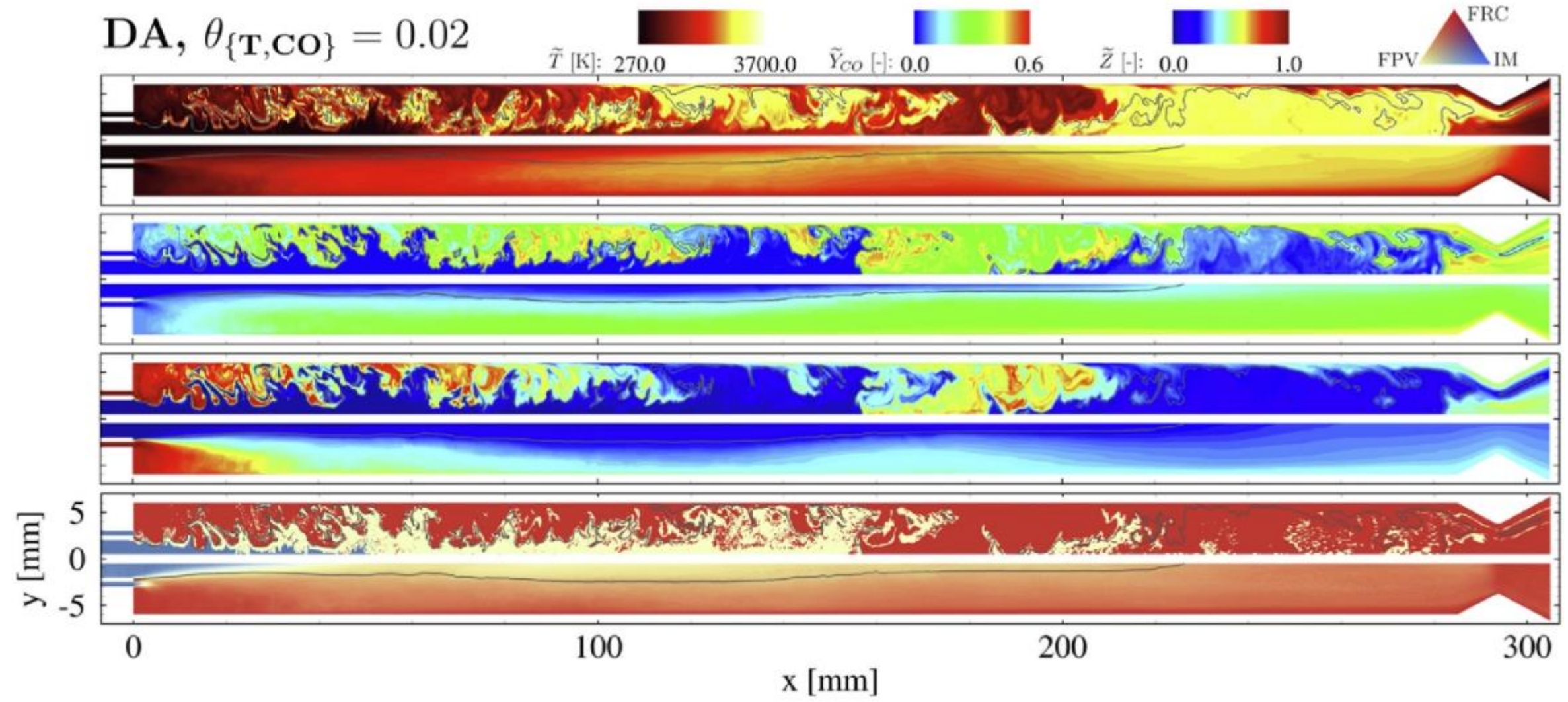

(b) *A posteriori* DA LES with $\theta_{\{T,CO\}} = 0.02$.

**Fig. 18**. A posteriori data-assisted large eddy simulation results for dynamic combustion-submodel assignment in a single-element gaseous-oxygen/gaseous-methane rocket combustor. A posteriori data-assisted large eddy simulation results for (a) $\theta_{\{T,CO\}} = 0.05$ and (b) $\theta_{\{T,CO\}} = 0.02$. From top to bottom, the fields correspond to temperature $\tilde{T}$, carbon monoxide mass fraction $\tilde{Y}_{CO}$, mixture fraction $\tilde{Z}$, and combustion-submodel assignment. In each field, the upper half shows instantaneous results, and the lower half shows time-averaged results. Reprinted from [108].

A complementary application-level validation was presented by Sun et al. [21] for a hydrogen-fueled DLR scramjet, where an artificial-neural-network-assisted compressible flamelet/progress variable solver, denoted AICFPVFoam, was compared with both experiment and the baseline compressible flamelet/progress variable solver, denoted CFPVFoam. Fig. 19 compares the schlieren image from experiment with numerical schlieren fields obtained using AICFPVFoam and CFPVFoam. It shows that the artificial-neural-network-assisted solver reproduces the dominant shockwave and expansion-wave system as well as the major shear-layer structure in a manner that remains consistent with both experiment and the conventional solver. Sun et al. [21] further reported that AICFPVFoam reduced the turbulent flamelet-database memory usage by 69.96% and improved the single-time-step computational efficiency by 7.29% compared to CFPVFoam while maintaining the similar combustion-flow structures.

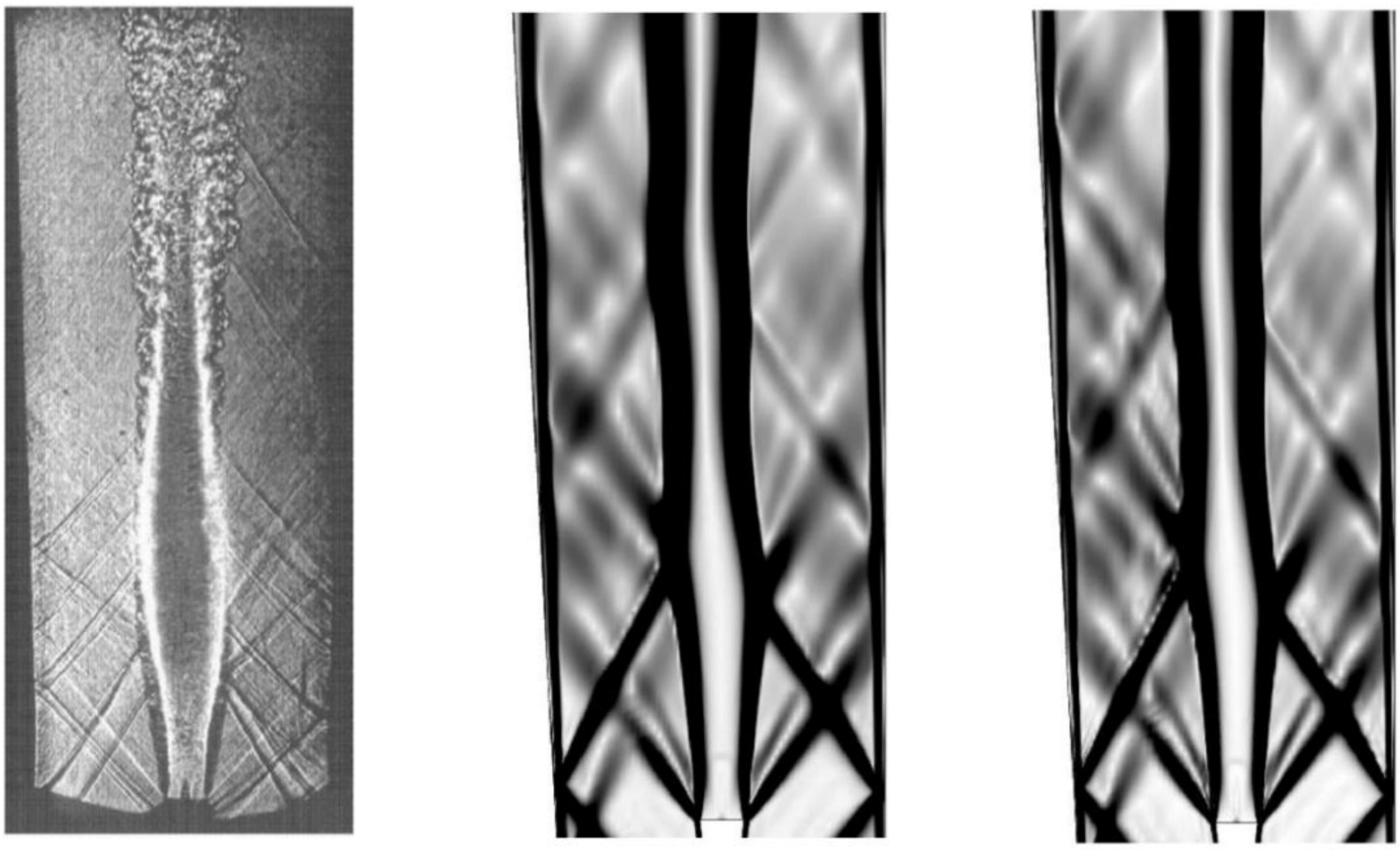

(a) (b) (c)

**Fig. 19**. Schlieren comparison of experiment, artificial-intelligence-assisted compressible flamelet/progress variable solver, and compressible flamelet/progress variable solver for a hydrogen-fueled DLR scramjet: (a) the schlieren image from experiment; (b) the numerical schlieren calculated using AICFPVFoam; and (c) the numerical schlieren calculated using CFPVFoam. The comparison highlights the dominant shockwave/expansion-wave structures and the shear layer generated by the fuel-air mixing process. Reprinted from [21].

Collectively, these studies demonstrate AI's impact on combustion chemistry, diagnostics, and reacting-flow simulation, yet they also highlight a critical challenge: inconsistent benchmarks and tasks make comparing results across the field nearly impossible. This motivates the quantitative synthesis presented in Section 4.

## 4. Quantitative Synthesis of ML Surrogate Performance in Multiscale Combustion

### *4.1 Corpus, Taxonomy, and Evaluation Framework*

This section draws on a curated corpus of 125 peer-reviewed studies published between 2020 and 2026 that investigate machine-learning surrogates for combustion and thermal decomposition modelling across scales. The papers were compiled through manual screening with the aim of capturing both methodological advances and application-driven deployments relevant to multiscale combustion—spanning chemistry and kinetics, turbulence–chemistry interaction, reacting-flow transport, and device-level configurations where performance and emissions are key outcomes. The corpus is therefore intentionally broad: it includes work focused on accelerating detailed-chemistry calculations, learning closures for continuum solvers, building reduced-order surrogates for reacting flows, and enabling near real-time inference for control, diagnostics, or optimisation tasks.

Since the literature uses heterogeneous assumptions, datasets, and validation protocols, the following analysis is framed as a quantitative synthesis rather than a single benchmark comparison. Each paper is assigned a Ref. number (used consistently throughout the tables and figures), and each study is categorised along three complementary dimensions:

i. *Surrogate model family (method taxonomy)*:

To reflect how surrogates are constructed and constrained, studies are grouped into a set of model families that recur across the reviewed literature: physics-informed learning (e.g., PINN-style approaches), operator learning (e.g., DeepONet and related neural-operator formulations), time-continuous models (e.g., Neural ODE variants), deep regressors for emulation (e.g., DNN/ANN and CNN-based surrogates), representation-learning and compression strategies (e.g., autoencoders), classical ML baselines (e.g., Random Forest and SVM), and hybrid clustering-assisted pipelines (e.g., clusterwise models combined with neural predictors). This taxonomy is used throughout Section 4 to organise both performance trends and practical trade-offs.

ii. *Application domain (where the surrogate is used)*:

To preserve the multiscale nature of combustion, each paper is tagged by its primary target application. Broadly, these include: (a) chemistry acceleration and kinetics learning; (b) turbulence–chemistry interaction and flame modelling; (c) reacting-flow field surrogates and digital-twin style emulators; (d) engine/combustor performance prediction; and (e) emissions inference and suppression. Several papers naturally span multiple tags (e.g., chemistry acceleration embedded in LES/RANS workflows), but the categorisation emphasises the dominant modelling objective of each study.

*iii. Evaluation signals (how performance is quantified)*:

Across the reviewed papers, accuracy is reported using a mix of metrics—most commonly $R^2$, RMSE/MAE/MSE/PCC/MAPE, relative error, or task-specific validation measures—while the computational benefit is quantified through speedup factors, wall-clock time reductions, solver-iteration savings, or throughput comparisons that depend on hardware and baseline solvers. For this reason, the figures in Section 4 use reported and, where necessary, normalised measures to support the cross-study synthesis, while treating quantities such as "speedup" as baseline-dependent and "runtime" as hardware-dependent unless a consistent reference is available. The aim is to identify robust patterns (e.g., accuracy–efficiency trade-offs across model families) without overstating precision where the underlying studies are not directly commensurate.

To make this synthesis transparent and traceable, Table 4 provides the evidence map used throughout Section 4. It links each surrogate model family to the dominant application contexts represented in the corpus and lists the corresponding Ref. number supporting each category. This table serves as a compact index from the literature base to the comparative analyses and figures that follow in Section 4.2 and beyond.

**Table 4**. Evidence map of surrogate model families and application domains in the reviewed combustion ML literature (2020–2026).

| **Model family** | **Application** | **Reference** |
|---|---|---|
| PINNs (Physics-informed NNs) | Chemical kinetics/mechanisms (*n=4*), Turbulent flames/TCI (*n=2*), Engines/combustors (*n=2*) | [123, 145, 175, 195, 196, 213] |
| DeepONet / operator learning | Chemical kinetics/mechanisms (*n=4*), Turbulent flames/TCI (*n=2*), Engines/combustors (*n=2*) | [135, 149, 67, 182, 195] |
| Neural ODEs | Chemical kinetics/mechanisms (*n=4*), Ignition/safety (*n=1*) | [133, 154, 193–194, 221] |
| CNNs | Turbulent flames/TCI (*n=9*), Engines/combustors (*n=5*), Mechanism reduction/ROM (*n=2*) | [119, 124, 131, 141, 171, 177, 178, 182, 74, 203, 214, 218] |
| DNN/ANN (general) *(excluding CNN/PINN/DeepONet/ NODE/Autoencoder overlaps)* | Turbulent flames/TCI (*n=47*), Engines/combustors (*n=23*), Chemical kinetics/mechanisms (*n=21*) | [14, 16, 54, 63, 90–92, 116–118, 120–122, 125–130, 136–137, 139–140, 142, 144, 146, 148, 150–153, 155, 156–167, 170, 172, 180–181, 183–189, 192, 199, 201–202, 204, 208, 211, 215–217, 220, 223–224] |
| Autoencoders | Turbulent flames/TCI (*n=8*), Engines/combustors (*n=5*), Mechanism reduction/ROM (*n=4*) | [12, 132, 134, 138, 147, 168–169, 171, 191, 200, 219, 225] |
| Clustering / Cluster + ANN | Turbulent flames/TCI (*n=12*), Chemical kinetics/mechanisms (*n=6*), Engines/combustors (*n=5*) | [118, 121, 125, 140, 156, 160, 163, 176, 179, 193, 200, 205, 210] |
| Random Forest | Engines/combustors (*n=11*), Turbulent flames/TCI (*n=7*), Emissions/pollutants (*n=5*) | [60, 65, 90–91, 108, 143, 164, 167, 169, 173, 178, 198, 206, 217, 222] |
| SVM | Engines/combustors (*n=8*), Turbulent flames/TCI (*n=5*), Emissions/pollutants (*n=3*) | [134, 147, 164, 169, 173–174, 178, 217, 222–223] |
| Other ML | Engines/combustors (*n=3*), Turbulent flames/TCI (*n=2*) | [190, 197, 207, 212, 209] |

Note: *"n" denotes the number of papers within the corresponding application category.*

### *4.2 Accuracy–Speed Trade-offs Across Model Classes*

Combustion surrogate models are rarely evaluated along a single dimension. In most multiscale workflows, surrogates are introduced to reduce the cost of repeatedly evaluating chemistry, closures, or reacting-flow response surfaces inside larger simulation or optimisation loops. Consequently, the model value is determined jointly by predictive fidelity and computational gain, and the literature consistently frames performance as a trade-off between these objectives.

Fig. 20 positions representative surrogate families on a plane of reported/normalized predictive accuracy versus reported/normalized computational speedup (log scale). Classical supervised baselines (e.g., SVMs and Random Forests) occupy the low-speedup/low-to-moderate accuracy region, reflecting their practicality for low-dimensional regression problems but limited expressivity for high-dimensional reacting-flow surrogacy. Representation-learning approaches (e.g., autoencoders) appear in an intermediate regime, consistent with their role as compressive models whose end-to-end accuracy depends strongly on how the latent representations are coupled to downstream predictors. Deep regressors (CNNs and DNNs) populate the higher-accuracy domain, with speedup improving when inference is accelerated (e.g., GPU execution) and when the surrogate replaces a dominant computational bottleneck. Physics-constrained and hybrid strategies, including PINN-style formulations, trend toward the high-accuracy/high-speedup corner, supporting the broader observation that incorporating physical structure can improve robustness and sample efficiency when training data are costly or sparse. Although absolute values remain baseline-dependent across studies, the ordering in Fig. 20 captures a consistent pattern across the reviewed corpus: surrogate families with stronger inductive bias and/or higher representational capacity tend to offer favourable accuracy–efficiency balances for multiscale combustion use cases.

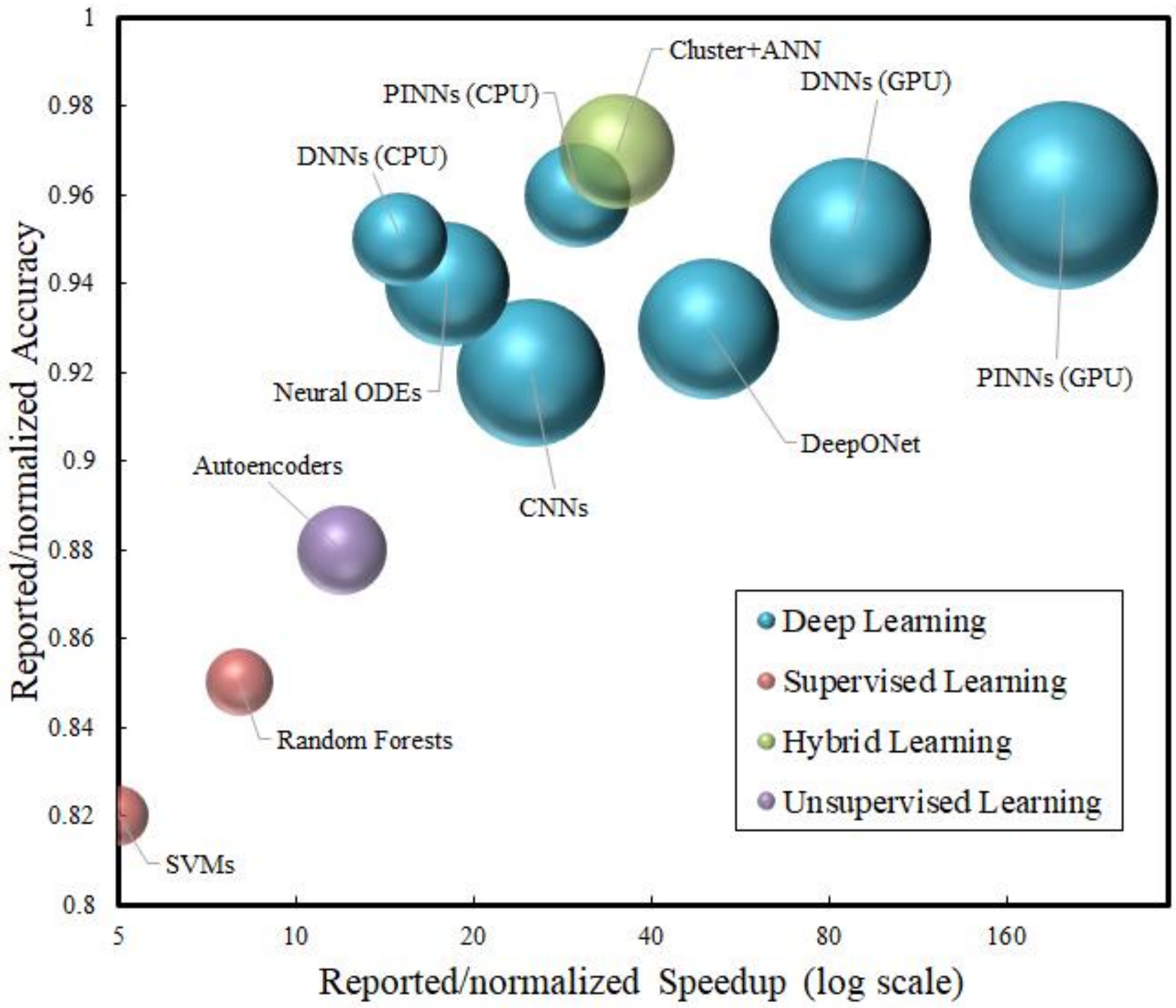

**Fig. 20**. Reported/normalized accuracy–speed trade-off landscape of ML surrogate families. Scatter of model families by reported/normalized predictive accuracy versus reported/normalized speedup (log scale); bubble size encodes the relative resource footprint and colours distinguish learning paradigms.

### *4.3 Deployment Efficiency: Inference Cost Versus Workflow Acceleration*

Reported speedup does not fully characterise deployability. For digital twins, online diagnostics, and closed-loop control, the practical constraint is often the inference latency rather than the offline training cost. A model may deliver large workflow-level acceleration while remaining unsuitable for real-time deployment if the per-query inference is expensive or hardware assumptions are unrealistic.

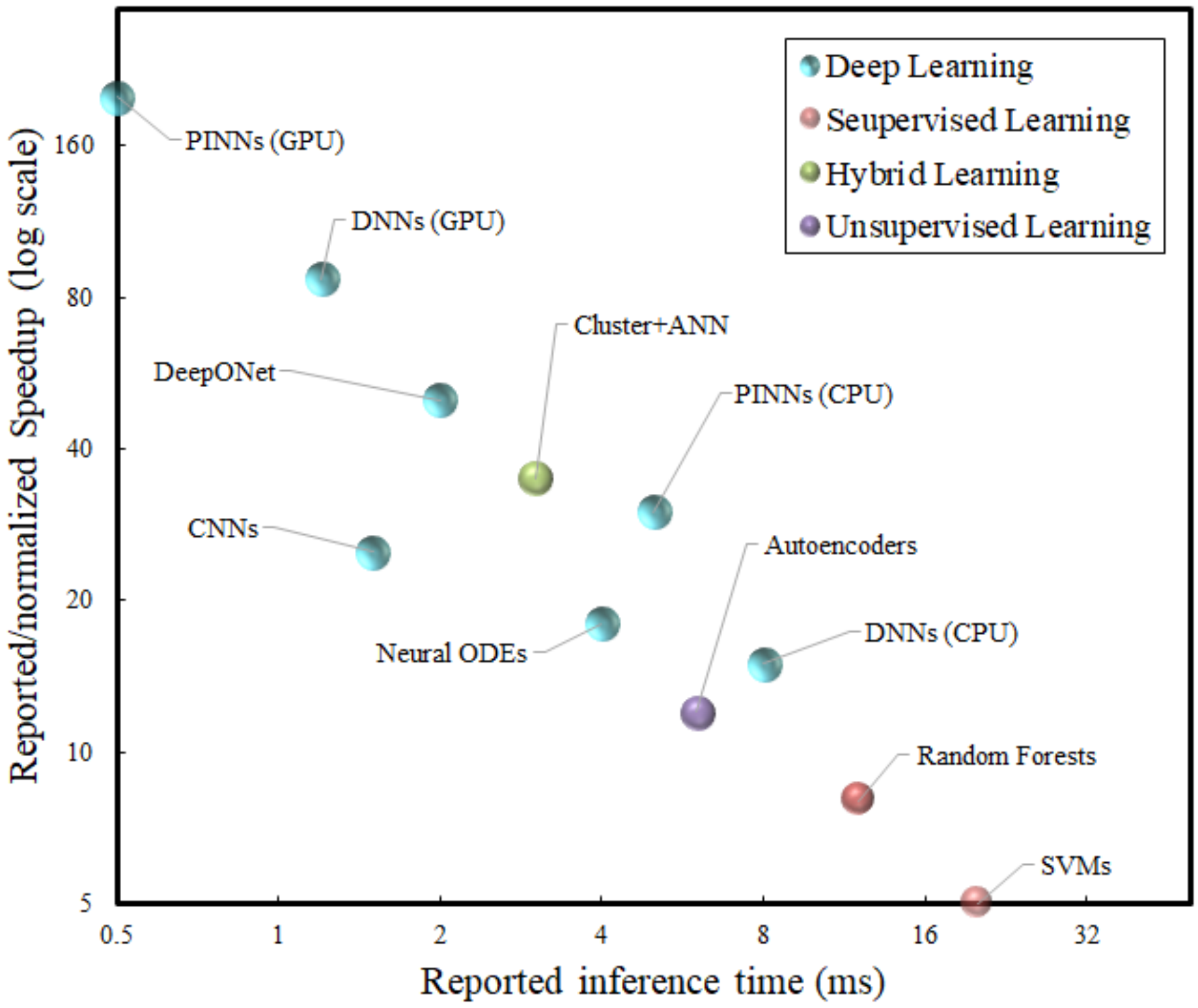

**Fig. 21**. Inference latency versus reported/normalized speedup for deployment-oriented comparison. Inference cost (study-reported latency or a normalized runtime index) is plotted against the reported/normalized speedup to distinguish the workflow-level acceleration from per-query deployment constraints.

Fig. 21 contrasts the inference-time cost against the achieved speedup. The main implication is that the largest computational gains are frequently obtained when the surrogate replaces a dominant solver component and remains inexpensive to evaluate at inference. GPU-enabled deep surrogates and PINN-style models occupy regions of both high speed-up and low inference cost, consistent with their use as fast inner-loop evaluators once trained. In contrast, classical models can be fast at inference yet provide modest workflow-level gains when they are applied only to limited sub-tasks (e.g., regression over summary quantities rather than substituting high-cost chemistry or transport operators). This distinction clarifies why "speedup" values across papers cannot be interpreted in isolation: identical inference costs can correspond to different end-to-end gains depending on what computational bottleneck is being replaced and how the baseline is defined.

*4.4 Pareto-Optimal Model Families and Multi-Objective Decision Structure*

Given the multi-objective nature of surrogate selection, Pareto analysis provides a transparent way to identify the model families that represent the non-dominated trade-offs. This perspective avoids treating surrogate comparison as a single leaderboard and instead highlights viable choices under different fidelity and efficiency requirements.

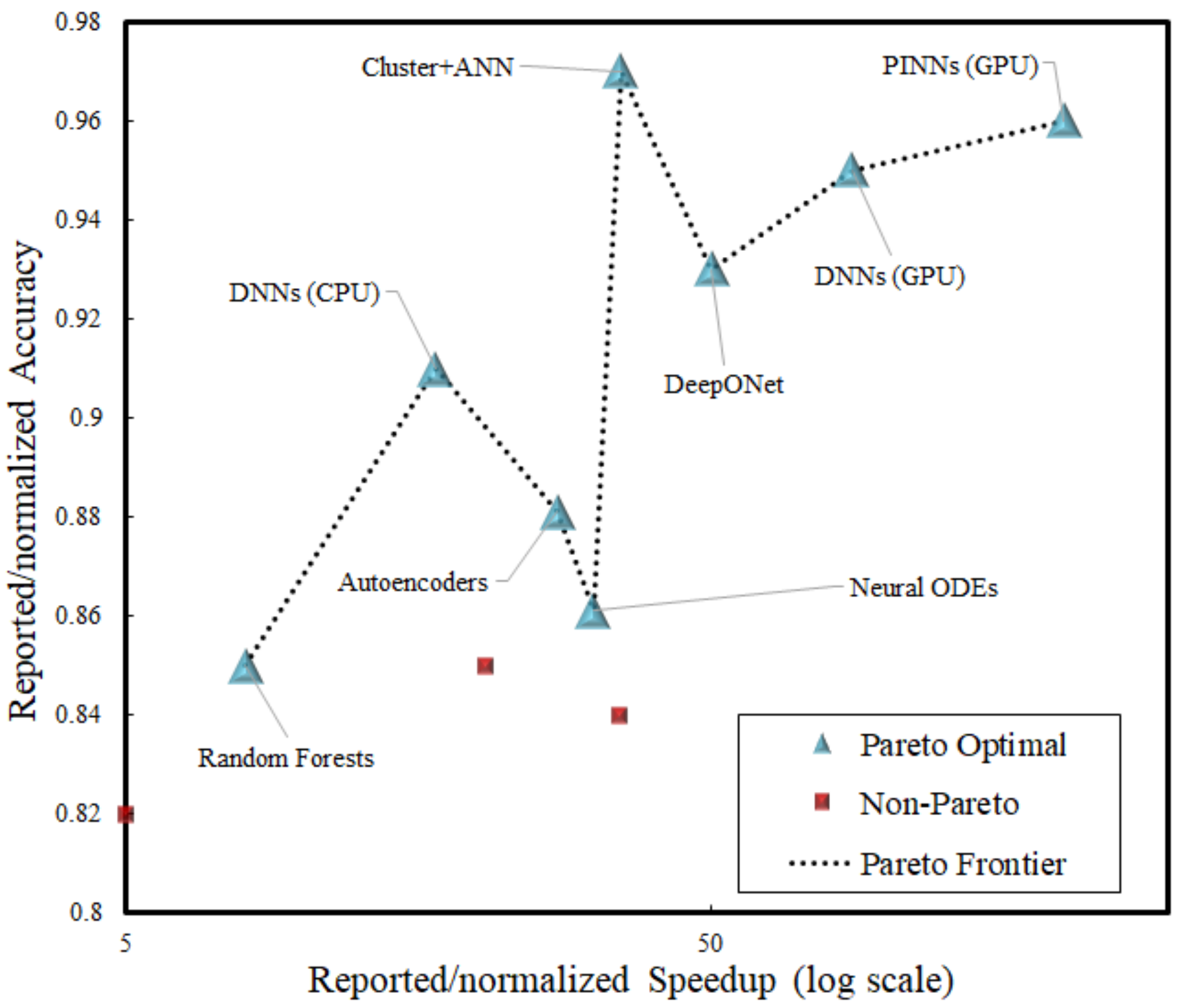

**Fig. 22**. Pareto frontier of reported/normalized accuracy–speed trade-offs. Non-dominated model families are identified under joint objectives of maximizing reported/normalized accuracy and reported/normalized speedup, highlighting candidate choices under different fidelity–efficiency priorities.

Fig. 22 identifies Pareto-optimal model families under the joint objectives of maximising reported/normalized accuracy and reported/normalized speedup. Frontier models represent configurations for which no alternative simultaneously offers higher accuracy and higher speedup under the applied normalisation. Operator-learning approaches (e.g., DeepONet) appear as strong candidates in tasks that resemble operator mapping across conditions, while physics-constrained and hybrid models remain competitive because constraints can stabilise learning and mitigate extrapolation error. Non-Pareto points remain relevant in specific contexts particularly when interpretability, training data limitations, or implementation simplicity dominate, but the Pareto framing provides a concise decision boundary for selecting candidates suitable for integration into multiscale workflows.

*4.5 Temporal Evolution of Reported Performance*

The reviewed literature shows a clear shift from proof-of-concept demonstrations toward models designed for integration into practical workflows. This shift is reflected in both accuracy-focused reporting and a growing emphasis on computational acceleration and deployment considerations.

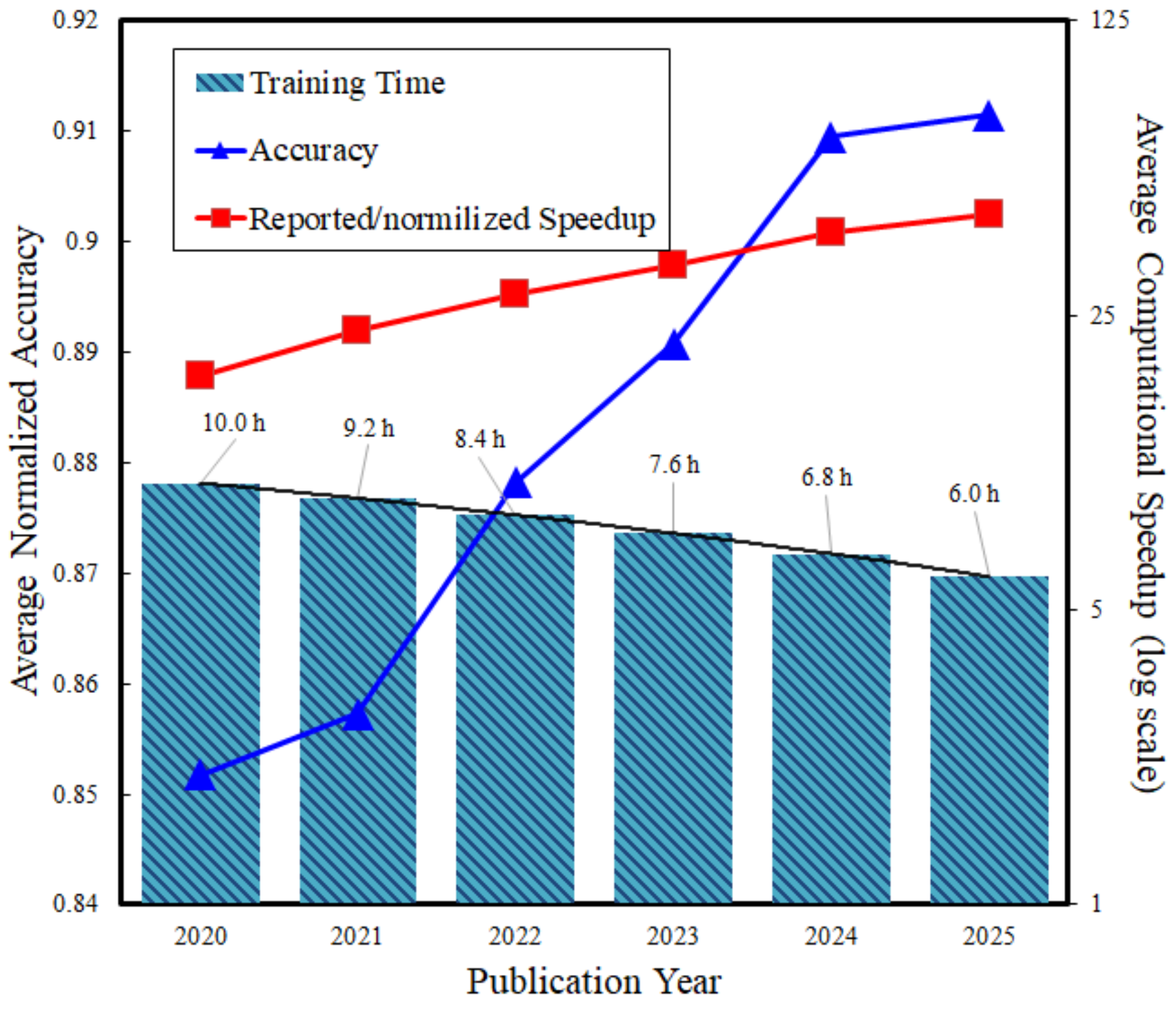

**Fig. 23**. Temporal evolution of reported/normalized surrogate performance (2020–2025). Yearly aggregates track changes in reported/normalized accuracy and reported/normalized speedup (log scale), alongside study-reported training time as an indicative measure of training cost.

Fig. 23 aggregates reported/normalized accuracy and reported/normalized speedup by publication year, together with study-reported training time as a coarse indicator of training cost. Average accuracy increases over the period, suggesting improved validation practices, richer architectures, and broader adoption of physically constrained learning strategies. Reported speedup also trends upward, consistent with the increasing focus on surrogates that replace expensive solver components and leverage accelerator hardware. Training time shows a gradual decline in the plotted averages; however, this signal should be interpreted cautiously because training cost is strongly hardware- and setup-dependent. Nevertheless, the overall direction aligns with a field-wide movement toward more engineering-oriented surrogate pipelines, including transfer learning, more efficient training procedures, and reuse of pretrained representations.

*4.6 Robustness and Uncertainty in Surrogate Combustion Models*

Beyond central accuracy values, the reliability of combustion surrogates is strongly influenced by how performance varies across datasets, operating conditions, and validation protocols. This issue is particularly acute in fuel-flexible modelling and emissions-oriented applications, where the distribution shift, which arises from changes in fuel composition, equivalence ratio, boundary conditions, or geometry, can lead to degradation even when in-domain accuracy appears high. For this reason, robustness-oriented reporting is a necessary complement to single-metric comparisons.

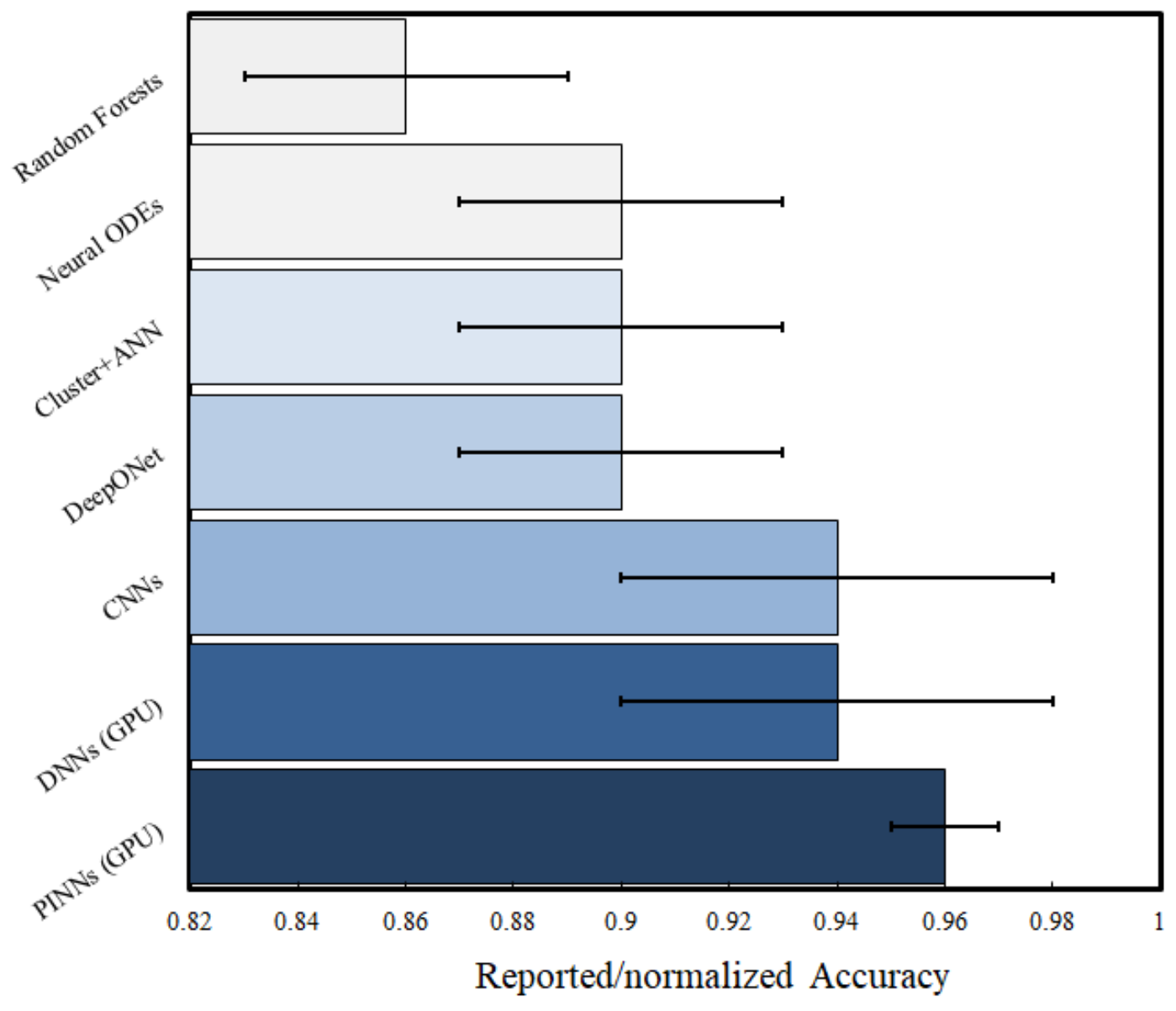

(a)

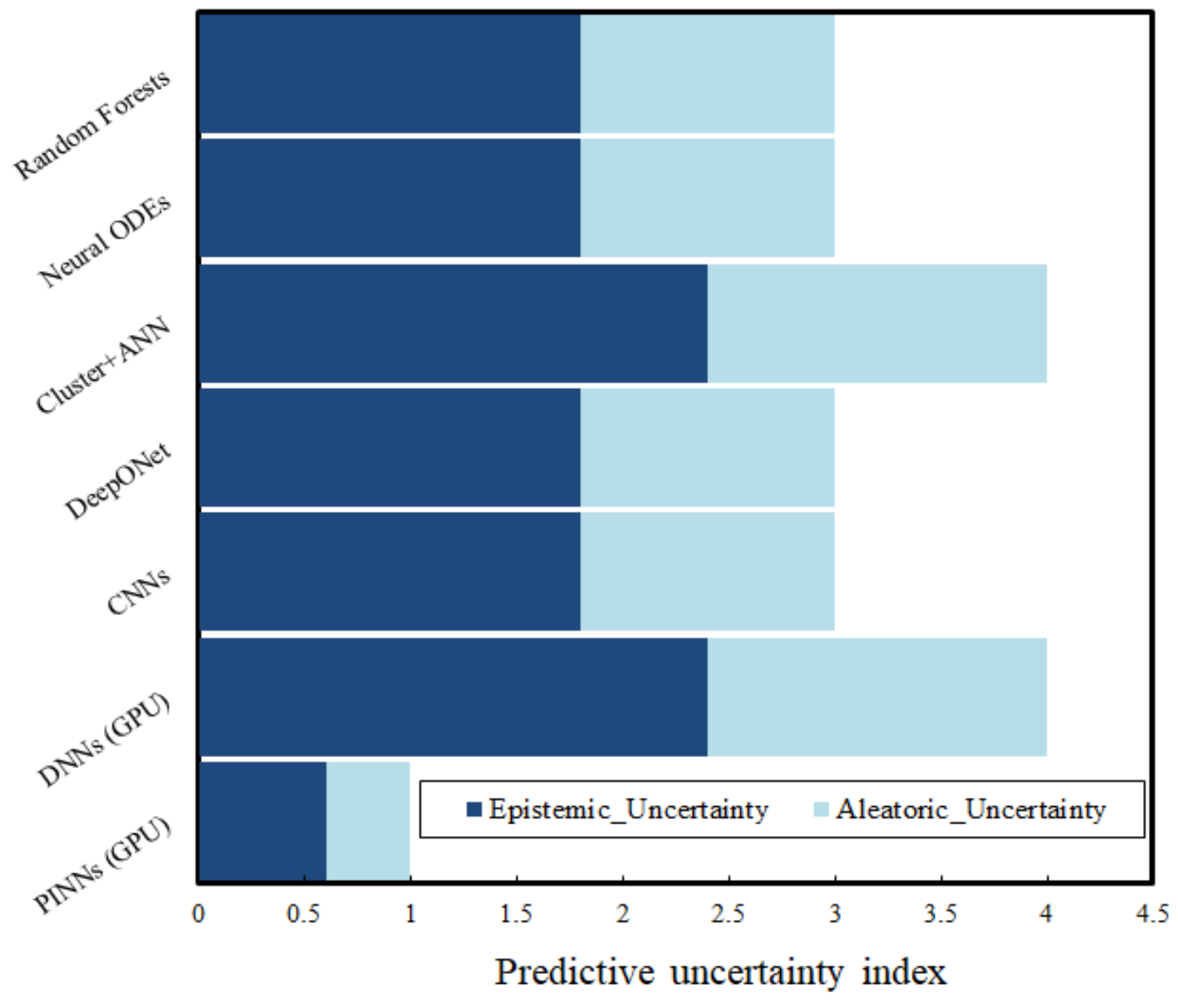

(b)

**Fig. 24**. Robustness and predictive uncertainty across surrogate families: (a) Reported/normalized accuracy with dispersion indicator summarizing variability across the extracted evidence set and (b) Relative decomposition of predictive uncertainty into epistemic and aleatoric components using a consistent index.

Fig. 24(a) summarises representative reported/normalized accuracy across model families along with a dispersion indicator reflecting variability in the extracted evidence set (e.g., across datasets, splits, or study-reported ranges). Physics-constrained and hybrid approaches tend to exhibit high central accuracy with comparatively tighter dispersion, consistent with the role of constraints in limiting degrees of freedom and stabilising extrapolation. Data-driven deep surrogates can also achieve high accuracy, but their dispersion can widen when training data provide narrow coverage of the operating envelope. Classical ML baselines typically show lower accuracy for high-dimensional surrogacy tasks and remain sensitive to the feature design and problem formulation. Collectively, the representative accuracy emphasises that the spread of reported performance can be as decisive as the mean when surrogates are intended for integration into multiscale simulation loops or deployed in inference settings.

To interpret why robustness differs across model classes, Fig. 24(b) presents a relative decomposition of predictive uncertainty into epistemic (model-driven) and aleatoric (data-driven) components, expressed as an index rather than a universal physical unit. The decomposition is useful because epistemic uncertainty is the dominant risk under distribution shift and, in principle, can be reduced through broader training coverage, improved inductive bias, or stronger physical constraints. Across the reviewed surrogate families, physics-constrained approaches often exhibit reduced epistemic contributions in reported settings, consistent with improved identifiability under limited data. In contrast, flexible data-driven models can carry higher epistemic contributions when training data sparsely sample the relevant regime space. Taken together, Figs. 24 (a) and (b) support a consistent inference from the reviewed literature: reliable deployment depends not only on achieving high nominal accuracy, but on controlling performance variability and reporting uncertainty in a way that distinguishes the interpolation-capable surrogates from models that remain fragile under transfer.

### *4.7 Integrated Multi-Metric Scorecard for Model Selection*

Since surrogate selection depends on the multiple practical criteria, a compact scorecard can assist synthesis by presenting a relative ranking under consistent normalisation. This avoids overstating commensurability of raw metrics across studies while still enabling comparative reasoning. Fig. 25 summarises the normalized indicators of accuracy, inference efficiency, and training efficiency on a 0–1 scale. Within this evidence-weighted normalisation, physics-constrained approaches rank highly in combined accuracy and inference efficiency, while training efficiency varies with constraint enforcement and implementation. Operator-learning methods exhibit competitive inference efficiency when the learned mapping replaces repeated operator evaluations. Classical models can be attractive for simplicity and training efficiency in lower-dimensional tasks but often lag in the combined accuracy–efficiency profile required for high-dimensional multiscale surrogacy. The scorecard therefore supports an application-oriented interpretation: model families form distinct performance profiles rather than a single universally dominant choice, and selection should be guided by the governing bottleneck and the required robustness to operating-condition variation.

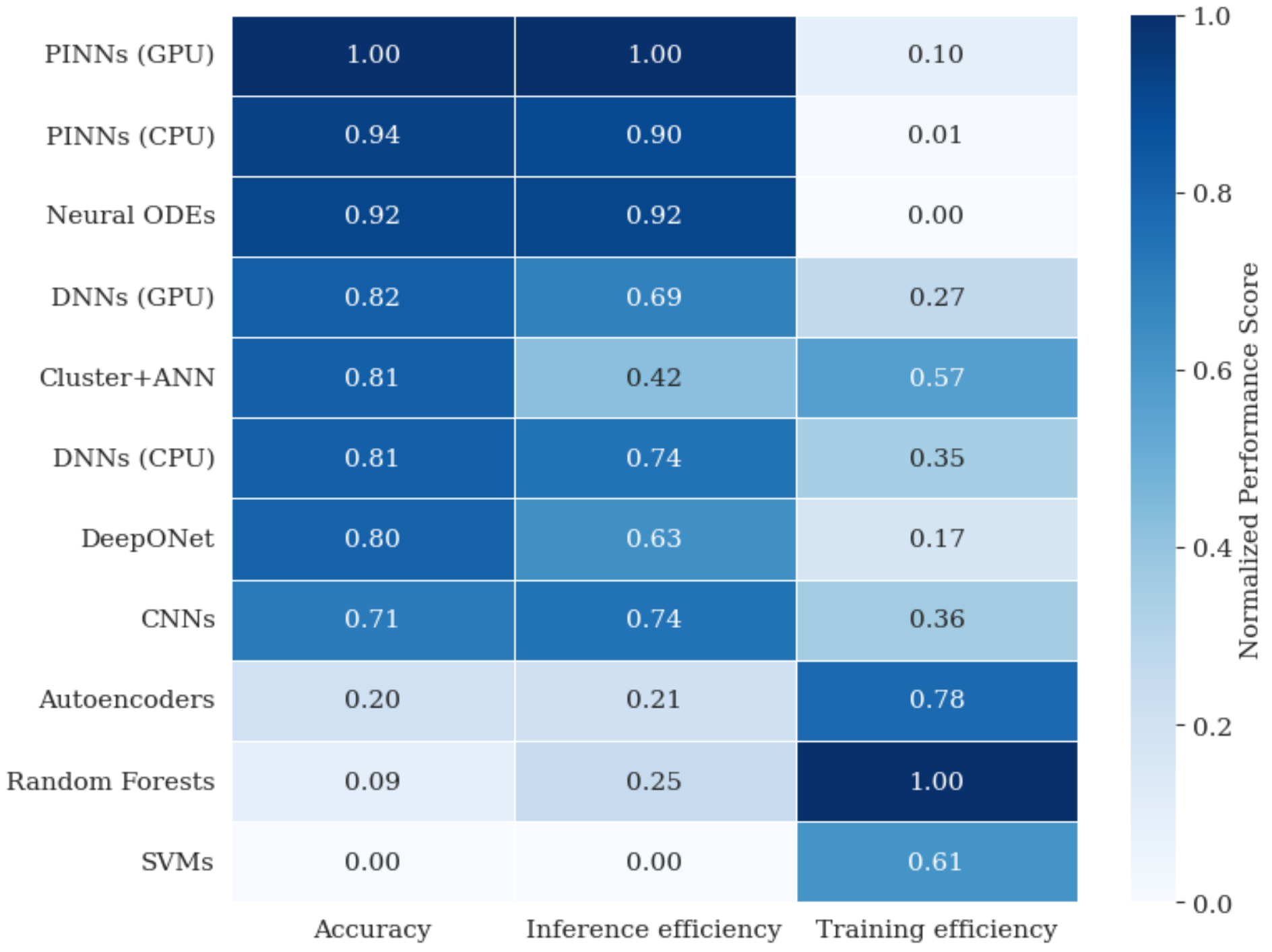

**Fig. 25**. Normalized multi-metric scorecard for surrogate selection. Heatmap summarizing normalized indicators of accuracy, inference efficiency, and training efficiency (0–1 scale) to support integrated, evidence-weighted comparison across model families.

## 5. Current limitations and future directions

Despite the rapid progress surveyed in Sections 3 and 4, the current AI–combustion ecosystem remains fragmented, with many of the most eye-catching speed-ups achieved under narrow conditions, non-standard baselines, or limited validation envelopes. The next step is therefore not only to invent new models but to address the structural limitations in data, metrics, workflows, and trust so that AI becomes a reliable, scale-bridging component of combustion science rather than an *ad-hoc* accelerator [109].

### *5.1 Cross-cutting limitations*

The corpus analysis in Section 4 highlights that “speed-up” is often defined heterogeneously, with studies variously quoting wall-clock ratios to detailed ODE solvers, comparisons to reduced-mechanism tables, or gains relative to low-order surrogates, frequently on different hardware and without counting training cost. As a result, nominally similar numbers can reflect profoundly different trade-offs, making it difficult to construct a robust Pareto frontier of accuracy versus efficiency across AI architectures and application regimes.

A second limitation is that most benchmarks remain task-local: models are typically tuned and evaluated on a single canonical flame, reactor configuration, or fuel blend, with only modest probing of distribution shift across facilities, diagnostics, fuels, or operating envelopes. Sections 3.2 and 3.3 already illustrate how the measurement manifolds and solver behaviour can change under soot-loading, pressure, or geometry variation, but only few studies in the corpus explicitly quantify robustness to these shifts or report uncertainty estimates aligned with operational decision-making.

Third, workflow-level frictions are rarely reported but are central in practice. Much of the current literature still assumes labour-intensive manual pre-processing (geometry clean-up, mesh generation, diagnostic denoising, and *ad-hoc* feature engineering) and equally manual post-processing (scripted field extraction, bespoke plotting, or hand-crafted pathway analysis), which can dominate the person-hours even when the learned surrogate itself is fast. Since these steps are often undocumented and unlogged, they become brittle failure points and major barriers to reproducibility and reuse across laboratories.

Finally, reproducibility and auditability remain uneven: code and data availability are inconsistent, metadata on training sets and preprocessing are often incomplete, and only a minority of studies provide the information needed to reconstruct end-to-end pipelines [110].

### *5.2 Methodological gaps and standardisation needs*

Addressing these limitations requires methodological conventions that go beyond individual papers. At minimum, chemistry-surrogate and CFD-integrated studies should report: (i) a clearly defined baseline including mechanism size, solver tolerances, and hardware; (ii) separate training, inference, and end-to-end workflow timings; and (iii) error metrics linked to decision-relevant quantities (e.g., ignition delay, NOx, blow-off margins) rather than only pointwise scalar discrepancies. Similarly, experimental AI studies should more routinely demonstrate transferability across facilities and operating regimes, quantify sensitivity to measurement noise and variations in optical diagnostics, and provide explicit uncertainty estimates through ensemble methods, Bayesian models, or calibrated predictive intervals.

The corpus in Section 4 also underscores the need for shared benchmarks and standardized comparisons. A small number of well-designed, open benchmark suites covering. For example, the turbulent premixed flames with alternative fuels, canonical engine-relevant operating points, and representative high-pressure diagnostics would enable fairer comparison of AI models and reduce the tendency to evaluate them only under favourable conditions. Equally important are reporting standards for physics consistency (e.g., conservation diagnostics and long-time integration stability) and for deployment constraints such as memory footprint, inference latency, and hardware portability, which remain under-emphasised in the current literature.

### *5.3 Agentic AI as a route to trustworthy, scalable workflows*

The limitations above are not unique to combustion: parallel discussions are emerging in materials and chemistry, where "agentic AI" has been proposed to turn large language models and task-specific predictors into autonomous, tool-using research agents that plan, execute, and analyse experiments under explicit constraints [109]. In this paradigm, agents couple reasoning, memory, planning, and tool orchestration to run closed-loop discovery campaigns, selecting candidate simulations or experiments, calling external codes or robots, and updating internal models based on outcomes, often with uncertainty-aware decision logic and human-in-the-loop checkpoints [109-110].

To illustrate the conceptual progression of AI technologies, Fig. 26 presents a hierarchical view of the evolution from traditional AI and ML to agentic AI systems. At the foundation, AI/ML includes fundamental learning paradigms such as supervised, unsupervised, and reinforcement learning. Building upon this layer, deep learning introduces advanced neural architectures including CNNs, recurrent neural networks (RNNs), and transformers. The next stage, Generative AI, enables the creation of content through technologies such as large language models (LLM), multimodal generation, and retrieval-augmented generation (RAG). On top of this, AI agents incorporate additional capabilities including planning, tool usage, memory, and task orchestration to execute complex tasks. Finally, Agentic AI represents the highest level of autonomy, characterized by goal-driven behaviour, autonomous decision-making, and multi-agent collaboration. As illustrated in Fig. 26, system autonomy and complexity increase progressively across these layers.

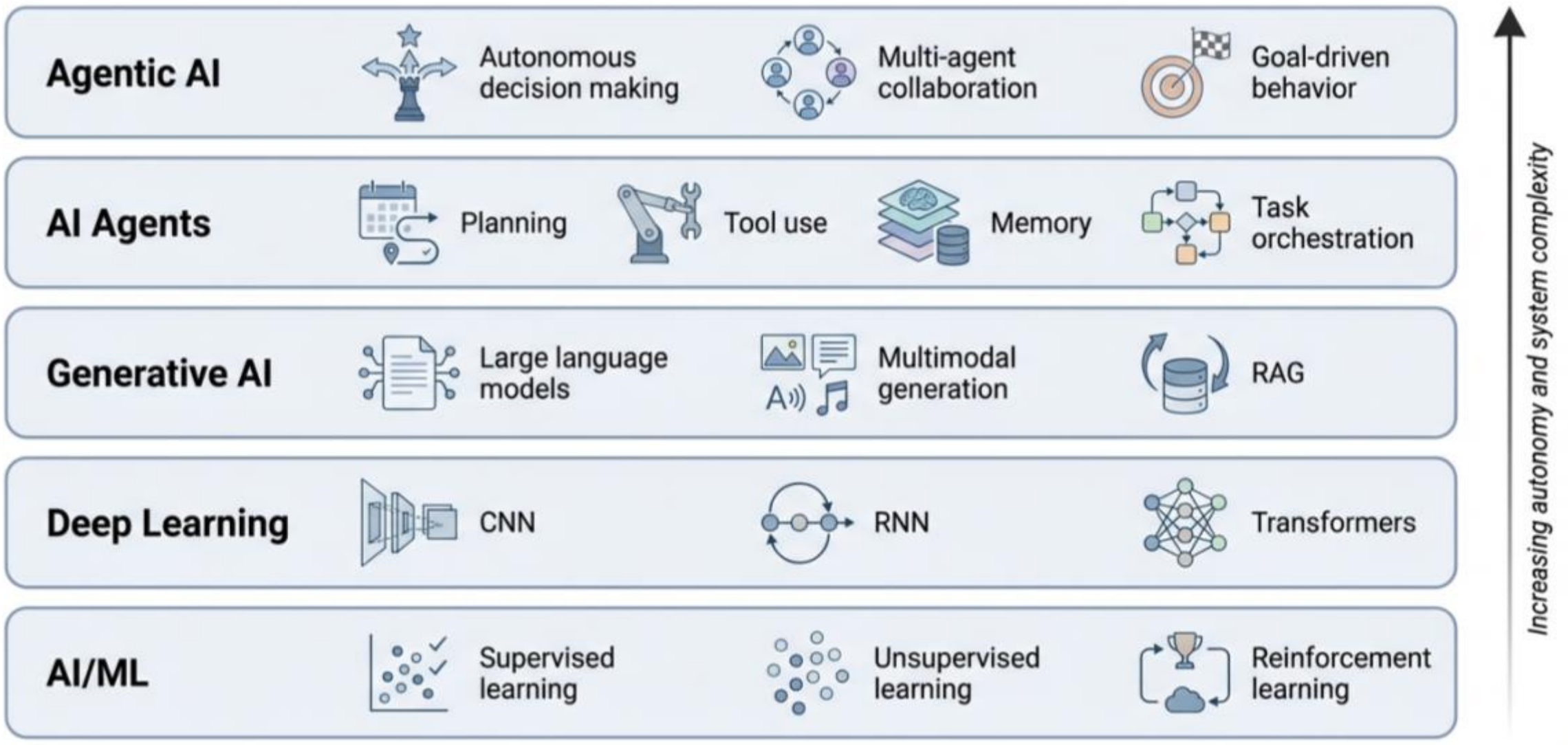

**Fig. 26**. Hierarchical progression of artificial intelligence technologies from traditional AI/ML and deep learning to Generative AI, AI agents, and Agentic AI. The roadmap highlights how system capabilities, autonomy and complexity increase across these layers.

Concrete implementations in energy materials and chemistry provide instructive precedents. Recent work on AI agents for energy-materials discovery traces the evolution from knowledge-assistant LLMs to autonomous multi-agent systems that generate material hypotheses, plan synthesis routes, and coordinate automated laboratory execution in self-driving labs [111]. Autonomous intelligent agents have likewise demonstrated end-to-end control of high-throughput DFT workflows, where agents select the next calculations based on surrogate models, thermodynamic priors, and logic rules, achieving substantial gains in sample efficiency and uncovering hundreds of new stable compounds with minimal human intervention [110]. Cross-domain reviews of agentic AI and lab automation emphasise that such systems can, in principle, manage full "design–synthesise–test–learn" loops, provided that safety, calibration, and governance are treated as first-class design constraints. In particular, recent surveys highlight the need for cross-layer approaches that connect agent architectures, threat models, and governance strategies, arguing that agentic AI demands new forms of lifecycle control, observability and defence compared with static prediction models [111].

*5.4 An agentic virtual lab for cleaner combustion*

One concrete instantiation of these ideas is an agentic AI virtual lab built around the reactive-molecular-dynamics (R-MD) pipeline for alternative fuels and NOx suppression. In such a lab, autonomous, auditable agents would span the full MD lifecycle: assembling chemically plausible fuel–oxidiser–diluent systems, orchestrating large ensembles of ReaxFF-or ML-potential simulations under physics-aware safety critics, and performing standardised post-processing for pathway analysis and emissions quantification.

By automating not only individual tasks but also the coordinating logic between them, including system validation, parameter sweeps, convergence checks, and provenance tracking, the virtual lab seeks to convert a currently time-consuming and largely manual blend-screening campaign into a repeatable one-month workflow with traceable decision-making and calibrated uncertainty. In the pre-simulation stage, agentic pre-processing would wrap tools such as PACKMOL [112], Moltemplate [113], and ReaxFF coverage scripts in agents that propose, validate, and document starting configurations, checking charge neutrality, density targets, and force-field coverage before any long trajectory is launched. During simulation, orchestration agents interfaced with LAMMPS [114] or similar engines would adapt timesteps, thermostats, and restart strategies in response to live diagnostics from a safety critic monitoring energy drift, temperature, density, and suspicious force-field excursions, with human-in-the-loop gates for escalation and revert actions when thresholds are exceeded. In the post-simulation stage, analytics agents would standardise bond-event parsing, reaction-network extraction, and NOx/NOy time-series analysis, attaching bootstrap- or ensemble-based uncertainty estimates and emitting FAIR, versioned datasets suitable for mechanism reduction or surrogate training. An active-learning layer would close the loop by training surrogate models on MD-derived descriptors and using uncertainty-aware acquisition functions to propose the next simulations, prioritising conditions that reduce NOx prediction uncertainty or probe extrapolative regimes (e.g., new alcohol fractions, pressures, or temperatures) most efficiently [110].

The evaluation of such virtual-lab agents can itself be structured systematically. Fig. 27 sketches an A/B protocol, in which two R-MD agents are run on the same benchmark suite of $NH_3/CH_4$ and $NH_3/CH_4$/Ethanol scenarios, with performance assessed by a code-based evaluator (e.g., trajectory validity, steps-to-stability, NOx/NOy RMSE, wall-time per GPU, and revert rate) and an LLM-as-a-judge evaluator scoring trace clarity, plausibility, and documentation quality.

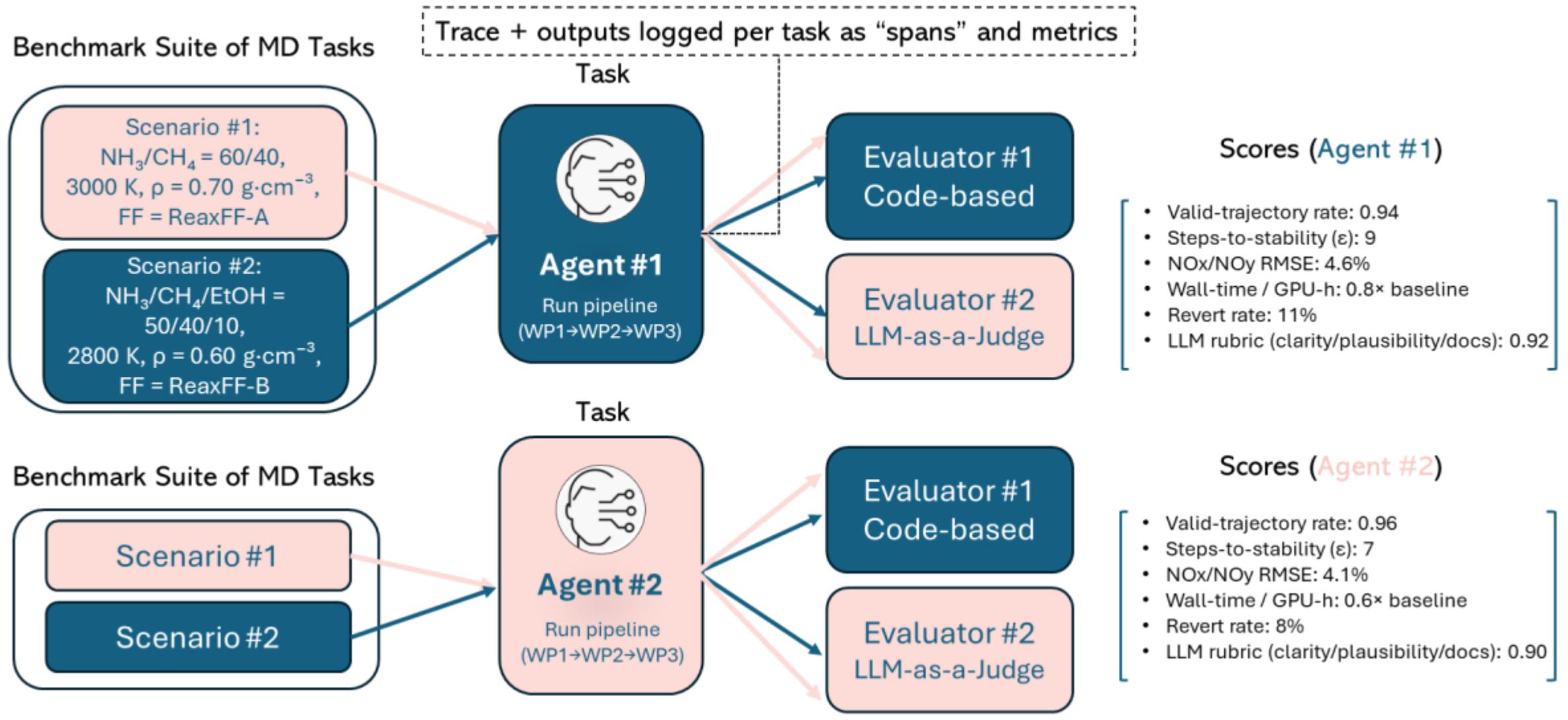

**Fig. 27**. Conceptual A/B evaluation of two R-MD agents on a benchmark suite of combustion-relevant MD tasks. Each agent runs the same scenarios, and independent evaluators compute code-based metrics and LLM-based rubric scores along with all traces span-logged for audit.

Complementing this, Fig. 28 illustrates the end-to-end agentic routing for the R-MD virtual lab, in which an LLM-based router dispatches tasks to three main work-packages: WP1 for molecular setup and validation, WP2 for simulation orchestration under a safety critic, and WP3 for post-processing and emissions analytics. Crucially, the diagram emphasises retrial loops under uncertainty, human-mentor checkpoints, and structured span logging for every action, design principles that directly address the manual pre/post-processing and reproducibility limitations identified earlier.

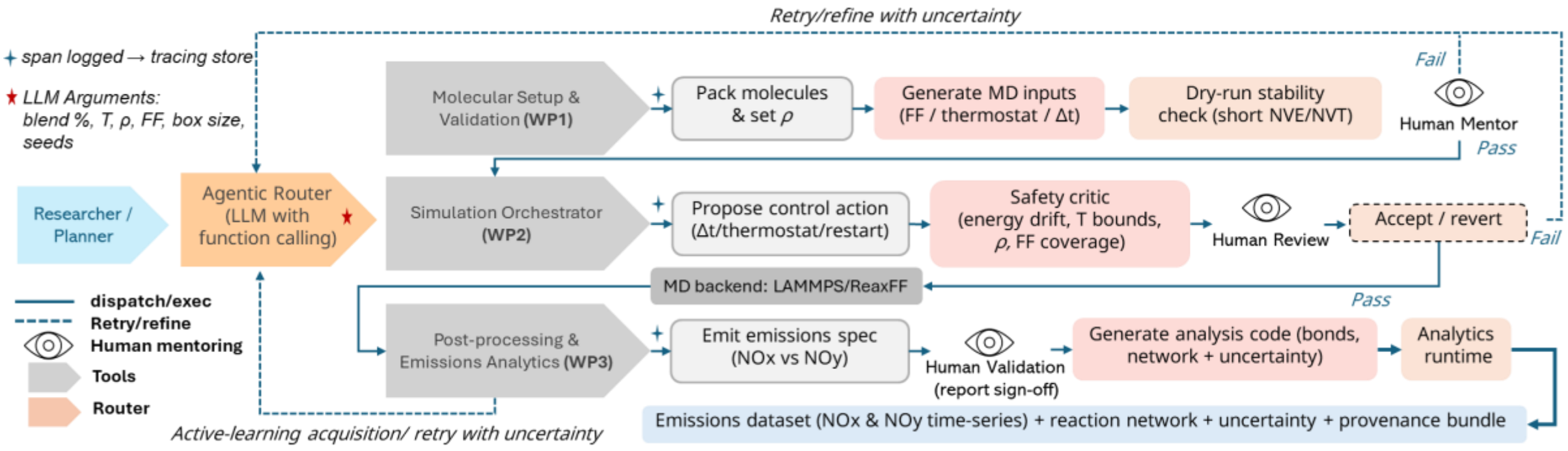

**Fig. 28**. Agentic router for the R-MD virtual lab. An LLM with function-calling capabilities routes tasks among autonomous setup, orchestration, and analytics agents, with safety critics, human validation, and active-learning acquisition loops built into the workflow.

To summarise, the virtual-lab concept takes inspiration from existing self-driving-lab and autonomous-agent architectures in chemistry and materials [109-111, 115] but adapts them to combustion-specific objectives such as high-temperature NOx suppression, ReaxFF coverage validation, and emissions-focused analytics. Although developed in a combustion context, the agentic virtual lab is fundamentally domain-agnostic and can be adapted to other scientific problems by modifying the molecular representation, operating conditions, and task-specific constraints. In this sense, the same framework could be extended to applications such as drug discovery, computational biology, and broader molecular design workflows.

**6. Conclusions**

This review has examined how AI is reshaping combustion research across molecular, experimental, and continuum scales, with a particular focus on surrogate modelling, chemistry acceleration, diagnostics, and data-driven closure development. Collectively, the literature shows that AI has evolved from a peripheral post-processing tool into the core of multiscale combustion workflows, capable of accelerating kinetic evaluation, enriching experimental inference, and improving the tractability of complex reacting-flow simulations.

In addition, the quantitative evidence assembled in this review indicates that reported gains must be interpreted carefully. Although many studies report substantial speed-up, these values are often derived under different baselines, hardware settings, datasets, and validation conditions, which limits direct comparability across model classes and application domains. A central conclusion of the present synthesis is therefore that progress in AI for combustion should be judged not only by raw acceleration but by the combined criteria of accuracy, robustness, physical consistency, uncertainty awareness, and deployability within realistic scientific and engineering workflows.

The review also highlights that current limitations are often workflow-level rather than purely algorithmic. Manual pre-processing, manual post-processing, fragmented data pipelines, weak reproducibility, and limited out-of-distribution validation continue to slow translation from promising demonstrations to dependable practice. Consequently, the next phase of the field is unlikely to be driven by isolated architectural refinements only, but by the development of integrated, benchmarked, and auditable AI ecosystems that can support reliable decision-making across scales.

Looking forward, one of the most promising directions is the emergence of agentic and virtual-laboratory paradigms for combustion science, in which autonomous but supervised AI systems help design simulations, monitor execution, analyse outputs, quantify uncertainty, and propose the next most informative cases. If developed with strong physics constraints, transparent logging, and human oversight, such frameworks could help address many of the current bottlenecks identified in this review, including inconsistent workflows, slow campaign turnaround, and limited reproducibility. Ultimately, the real promise of AI in combustion lies not simply in making existing simulations faster but in enabling a more rigorous, reproducible, and predictive science of clean reactive systems across scales.